\documentclass{aa}  

\usepackage{graphicx}
\usepackage{txfonts}
\usepackage[utf8]{inputenc}
\usepackage{hyperref}
\usepackage{tabularx} 
\usepackage{caption}
\usepackage{subcaption}
\usepackage[table]{xcolor}
\usepackage[flushleft]{threeparttable}
\usepackage{multirow}
\usepackage{orcidlink}
\usepackage[rightcaption]{sidecap}
\usepackage{placeins}

\hypersetup{
  colorlinks=true,   
  urlcolor=blue,     
  linkcolor=blue,     
  citecolor=blue
}

\begin{document}

   \title{Establishing a mass-loss rate relation for red supergiants in the Large Magellanic Cloud\thanks{Full Table 2 is only available in electronic form at the CDS via anonymous ftp to \url{cdsarc.u-strasbg.fr} (\url{130.79.128.5}) or via \url{http://cdsweb.u-strasbg.fr/cgi-bin/qcat?J/A+A/}}}

   \author{K. Antoniadis
          \inst{\ref{noa}, \ref{nkua}}\orcidlink{0000-0002-3454-7958}, 
          A.Z. Bonanos \inst{\ref{noa}}\orcidlink{0000-0003-2851-1905},
          S. de Wit\inst{\ref{noa}, \ref{nkua}}\orcidlink{0000-0002-9818-4877}, 
          E. Zapartas\inst{\ref{crete},\ref{noa}}\orcidlink{0000-0002-7464-498X},
          G. Munoz-Sanchez\inst{\ref{noa}, \ref{nkua}}\orcidlink{0000-0002-9179-6918},  
          G. Maravelias\inst{\ref{noa},\ref{crete}}\orcidlink{0000-0002-0891-7564}
          }

   \institute{IAASARS, National Observatory of Athens, 15236 Penteli, Greece. \\
              \email{k.antoniadis@noa.gr} \label{noa}
         \and
             National and Kapodistrian University of Athens, 15784 Athens, Greece \label{nkua}
        \and
            Institute of Astrophysics FORTH, GR-71110, Heraklion, Greece \label{crete}
            }


 
  \abstract
   {The high mass-loss rates of red supergiants (RSGs) drastically affect their evolution and final fate, but their mass-loss mechanism remains poorly understood. Various empirical prescriptions scaled with luminosity have been derived in the literature, yielding results with a dispersion of two to three orders of magnitude.}
   {We determine an accurate mass-loss rate relation with luminosity and other parameters using a large, clean sample of RSGs. In this way, we shed light into the underlying physical mechanism and explain the discrepancy between previous works.}
   {We assembled a sample of 2,219 RSG candidates in the Large Magellanic Cloud, with ultraviolet to mid-infrared photometry in up to 49 filters. We determined the luminosity of each RSG by integrating the spectral energy distribution and the mass-loss rate using the radiative transfer code \texttt{DUSTY}.}
   {Our derived RSG mass-loss rates range from approximately $10^{-9} \ M_{\odot}$ yr$^{-1}$ to $10^{-5} \ M_{\odot}$ yr$^{-1}$, mainly depending on the luminosity. The average mass-loss rate is $9.3\times10^{-7} \ M_{\odot}$ yr$^{-1}$ for $\log{(L/L_\odot)}>4$, corresponding to a dust-production rate of $\sim3.6\times10^{-9} \ M_{\odot}$~yr$^{-1}$. We established a mass-loss rate relation as a function of luminosity and effective temperature. Furthermore, we found a turning point in the relation of mass-loss rate versus luminosity at approximately $\log{(L/L_\odot)} = 4.4$, indicating enhanced rates beyond this limit. We show that this enhancement correlates with photometric variability. We compared our results with prescriptions from the literature, finding an agreement with works assuming steady-state winds. Additionally, we examined the effect of different assumptions on our models and found that radiatively driven winds result in mass-loss rates higher by two to three orders of magnitude, which is unrealistically high for RSGs. For grain sizes $<0.1 \ \mu$m, the predicted mass-loss rates are higher by a factor of 25-30 than larger grain sizes. Finally, we found that 21\% of our sample constitute current binary candidates. This has a minor effect on our mass-loss relation.}
   {}

    \keywords{stars: massive -- stars: supergiants -- stars: mass-loss -- stars: late-type -- stars: evolution -- circumstellar matter}

   \titlerunning{Mass-loss rates of red supergiants in the Large Magellanic Cloud}
   \authorrunning{Antoniadis et al.}

   \maketitle
%
    
\section{Introduction}

The winds of massive stars play a significant role in stellar and galactic evolution by enriching a galaxy with gas and dust, which in turn contributes to star formation. Stars with initial masses $8\lesssim M_\mathrm{init}\lesssim30 \ M_{\odot}$ \citep{Meynet_2003, Heger_2003} evolve to the red supergiant (RSG) phase, which exhibits enhanced mass loss compared to the main sequence (MS). The mass loss of RSGs can impact their evolution, end fate, and the type of the resulting core-collapse supernovae (CCSNe). When the mass-loss rate is substantial, the star can be stripped of the hydrogen-rich envelope, return to hotter temperatures across the Hertzsprung–Russell (HR) diagram, and eventually explode as Type IIb, Ib, or Ic supernovae (SNe) \citep{Eldridge_2013, Yoon_2017, Zapartas_2017}. Some studies showed that RSGs with such winds are rare, suggesting that the stripping mainly occurs through binary evolution or episodic mass loss \citep{Beasor_2022, Beasor_2023, Decin_2023}. Strong winds could also contribute to solving the so-called red supergiant problem, which means the absence of observed RSG Type II SN progenitors with $M \gtrsim 20 \ M_{\odot}$ \citep{Smartt_2009, Davies_2020}. RSGs at the upper mass limit with strong winds would not only be able to strip off the envelope, but would also create a dust shell that could obscure the star \citep{Beasor_2022}. However, the underlying mechanism driving the RSG wind is still unknown.

Determining the mass-loss rates of RSGs is crucial as an input for stellar evolution models. These models mainly rely on empirically derived prescriptions \citep[e.g.][]{deJager_1988}. Many hypotheses have been put forward about the physical process that might dominate the mass-loss mechanism. The highly convective envelope may lift off outer parts of the stellar atmosphere to a distance at which dust can condense. Radial pulsations also occur in RSGs and may additionally have an impact on this mechanism \citep{Yoon_2010}, similar to asymptotic giant branch (AGB) stars \citep{Hofner_2003, Neilson_2008}. However, the case of RSGs is more complicated, as they have higher effective temperatures, lower pulsational amplitudes \citep{Wood_1983}, and larger stellar radii than AGB stars. Therefore, gas needs to be driven to higher altitudes to reach the dust condensation limit. \cite{Arroyo-Torres_2015} modelled the extensions of RSG atmospheres and showed that neither pulsations nor convection alone can lift enough material for dust condensation. \cite{Kee_2021} suggested that atmospheric turbulence is the dominant factor for the mass loss of RSGs, supporting previous studies and observations \citep[e.g.][]{Josselin_2007, Ohnaka_2017}. Under this assumption, \cite{Kee_2021} derived the first analytic mass-loss rate relation. Finally, radiation pressure on the newly formed dust grains can further enhance the driving wind \citep{vLoon_2000}. Different assumptions on the dominant mechanism and properties of the RSG atmospheres and circumstellar matter (CSM) have led to a significant dispersion in the results between different studies and samples, with mass-loss rates ranging between $10^{-7}-10^{-3} \ M_{\odot}$ yr$^{-1}$ \citep{deJager_1988, vLoon_2005, Goldman_2017, Wang_2021, Beasor_2022, Beasor_2023, Yang_2023}.

The empirical derivation of mass-loss rates is often based on modelling circumstellar dust \citep{deJager_1988, vLoon_2005, Goldman_2017, Beasor_2020, Yang_2023}. An observation, however, is a snapshot'of the RSG, and deriving a mass-loss rate implies assuming a constant mass loss throughout this phase. Episodic mass loss can also occur in RSGs in a superwind phase or an outburst \citep{Decin_2006, Montarges_2021, Dupree_2022}. There were attempts to confirm the existence of episodic mass loss or an outburst event by assuming that this would change the density profile of the dust shell \citep{Shenoy_2016, Gordon_2018, Humphreys_2020}. \cite{Beasor_2016} claimed that the result of these works could be reproduced by varying the dust temperature in the model; thus, it is uncertain whether the effect on an observed spectral energy distribution (SED) originates from an episodic event. Alternatively, \cite{Massey_2023} used the luminosity function to derive a time-averaged mass-loss rate of RSGs.

Motivated by the study of \cite{Yang_2023} in the Small Magellanic Cloud (SMC), we extended this work to an environment with a different metallicity, the Large Magellanic Cloud (LMC). This paper aims to determine the mass-loss rates of a large, clean, and complete sample of RSGs and to establish a relation between the mass-loss rate, the luminosity, and the effective temperature. In Sect.~\ref{sec:sample} we describe the sample selection and data analysis, Sect.~\ref{sec:DSmodels} presents the models and fitting method, and Sect.~\ref{sec:results} offers the results. In Sect.~\ref{sec:disc} we discuss our mass-loss rate results and compare them to previous works. Finally, we present a summary and our conclusions in Sect.~\ref{sec:summary}. In the final stages of this paper, we became aware of the similar work by \cite{Wen_2024}, which we briefly discuss in Sect.~\ref{sec:disc}.


\section{Sample} \label{sec:sample}

We compiled the sample of RSG candidates in the LMC from the catalogues of \cite{Neugent_2012}, \cite{Yang_2021lmc}, and \cite{Ren_2021}, who defined RSG candidates by photometric limits in colour-magnitude diagrams (CMDs). We merged the three source catalogues and removed any duplicates by cross-matching their coordinates within a 1" radius, which resulted in 5601 RSG candidates. We used photometric data ranging from the ultraviolet (UV) to the mid-infrared (mid-IR), as in \cite{Yang_2021lmc}, to ensure uniform coverage and a sufficient range to model the observed SED. Furthermore, we updated or added any missing photometric data to create the final catalogue with the required photometry for our study.

We replaced the \textit{Spitzer} Enhanced Imaging Products (SEIP), which included \textit{WISE} and \textit{Spitzer} photometric data, with SAGE-LMC \textit{Spitzer} data \citep{sage_lmc}, which have an improved photometry. We replaced the \textit{WISE} data with data from the AllWISE catalogue \citep{allwise}, applying photometric quality criteria with $ccf=0, \ ex=0$ and the following signal-to-noise ratios $\mathrm{S/N}_{W1}>5,\ \mathrm{S/N}_{W2}>5,\ \mathrm{S/N}_{W3}>7, and\ \mathrm{S/N}_{W4}>10$. We also updated the \textit{Gaia} DR2 and EDR3 data to DR3 \citep{gaia1, gaia2}. When data were missing, we added or updated photometric data from the following source catalogues: SkyMapper DR2 \citep{Onken_2019} with $flags$\_$psf\leq4$; NOAO Source Catalog DR2 \citep[NSC DR2;][]{Nidever_2021} with $flags<4$; Vista Magellanic Cloud Survey \citep[VMC;][]{vmc}, excluding photometry brighter than the detection limit, $J\leq12$ mag and $K_\mathrm{S}\leq11.5$ mag; InfraRed Survey Facility \citep[IRSF;][]{IRSF}, from which we kept the photometric data with quality flag 1 (point-like source); AKARI \citep{akari}, from which we removed photometric data that have flags for contamination by multiplexer bleeding trails (MUX-bleeding), by 'column pulldown' or by artefacts and saturated sources; photometry from \cite{Massey_2002}; the Galaxy Evolution Explorer \citep[GALEX;][]{galex}; and the Optical Gravitational Lensing Experiment \citep[OGLE-III;][]{ogleIII}. We present the list of the surveys and the filters used in \autoref{tab:filters}. Finally, we collected mid-IR epoch photometry from the Near-Earth Object Wide-field Infrared Survey Explorer \citep[NEOWISE;][]{Mainzer_2011, Mainzer_2014} to examine the intrinsic variability of RSGs.

\begin{table}[h]
    \centering
    \caption{Photometric surveys and filters used for the spectral energy distributions.}
    \renewcommand{\arraystretch}{1.3}
    \begin{tabular}{l c}
        \hline\hline
       Survey  & Filters \\
         \hline
       GALEX  & FUV, NUV \\
       SkyMapper DR2 & $u, v, g, r, i, z$ \\
       \citet{Massey_2002}         & $U, B, V, R$ \\
       NSC DR2  & $u, g, r, i, z, Y$ \\
       \textit{Gaia} DR3  & $G_{BP}, \ G, \ G_{RP}$ \\
       OGLE-III & $V, I$ \\
       VMC & $Y, J, K\mathrm{_S}$ \\
       2MASS  & $J, H, K\mathrm{_S}$ \\
       IRSF  & $J, H, K\mathrm{_S}$ \\
       AKARI  & N3, N4, S7, S11, L15, L24 \\
       AllWISE & [3.4], [4.6], [12], [22] \\
       \textit{Spitzer}   & [3.6], [4.5], [5.8], [8.0], [24] \\
       \hline \\     
    \end{tabular}
    \
    \label{tab:filters}
\end{table}

\begin{figure}[h]
    \centering
    \includegraphics[width=\linewidth]{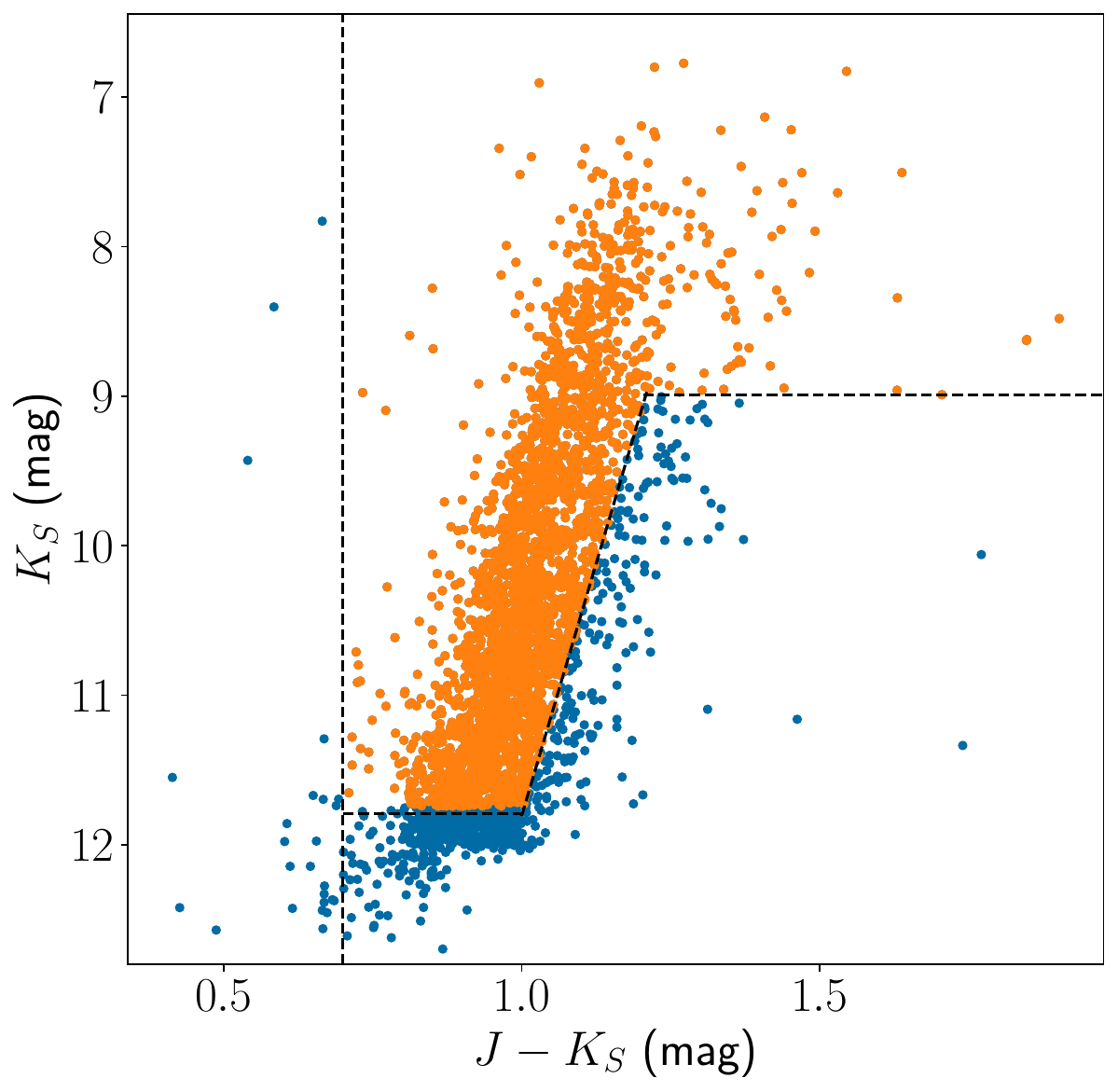}
    \caption{Colour-magnitude diagram (CMD) of the RSGs in the LMC using 2MASS photometric data. The dashed lines define the photometric RSG limits from \cite{Yang_2021lmc}, modified by increasing the slope of the right diagonal line. The orange points show the sources we kept in our final sample.}
    \label{fig:JK_CMD}
\end{figure}

We applied the Two Micron All Sky Survey (2MASS) $J-K_s$ photometric cuts to all our targets as defined in \cite{Yang_2021lmc} for RSG candidates. We used a stricter constraint on the right side of the CMD to avoid AGB contamination, as shown in {Fig.}~\ref{fig:JK_CMD}, {since O-rich AGB stars are located right of the RSG population in these CMDs \citep[e.g.][]{Yang_2019, Ren_2021}}. There may still be some AGB contamination, even though not many AGB stars occupy the brighter end of the CMD due to their lower intrinsic luminosity \citep{Groenewegen_2018}. 

\begin{figure*}[h]
    \centering
    \includegraphics[width=\linewidth]{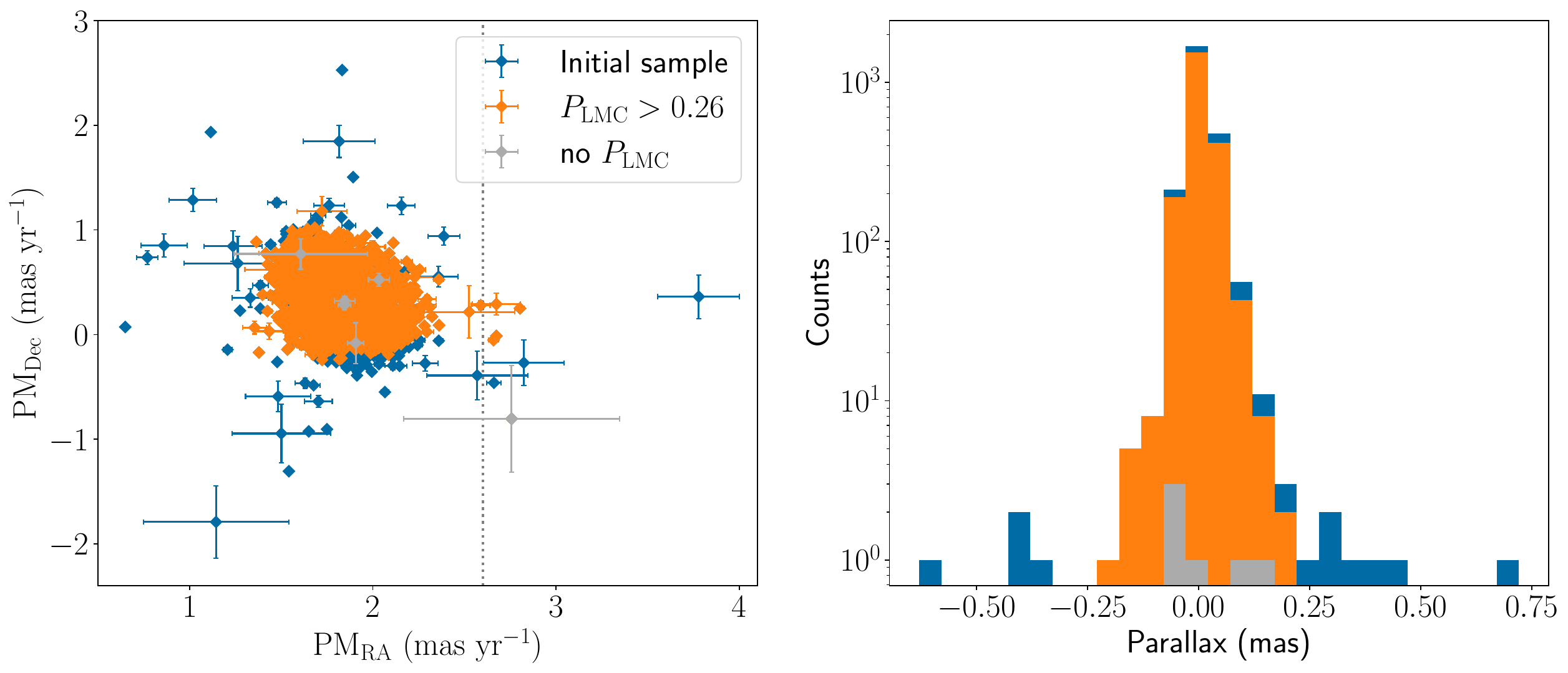}
    \caption{Proper motions in Dec. and R.A. (left) and distribution of parallaxes (right). The blue points show the initial RSG sample, and the orange points show the LMC members after applying the probability criteria from \citet{Gaia_LMC} with $P_\mathrm{LMC}>0.26$. The grey points indicate sources without a probability measurement from \citet{Gaia_LMC}. Sources located right of the dotted vertical line were considered to be foreground.}
    \label{fig:PM}
\end{figure*}

\begin{figure}[h]
    \includegraphics[width=\linewidth]{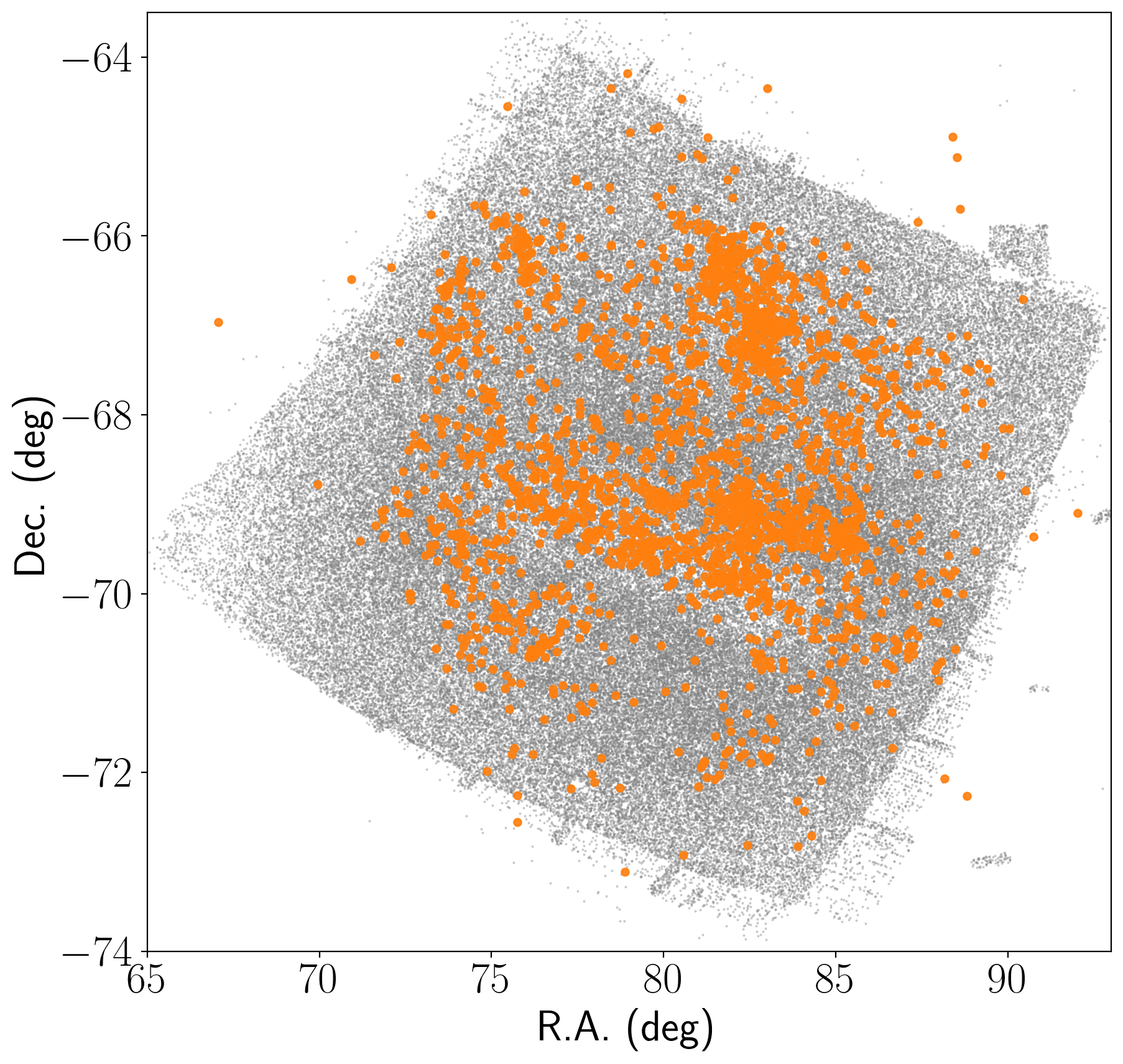}
    \caption{Spatial distribution of the final RSG sample in the LMC (orange points). The background targets (grey) are the complete sample of all sources from \cite{Yang_2021lmc}.}
    \label{fig:spatial}
\end{figure}

We removed foreground sources using the LMC membership criteria from \cite{Gaia_LMC}, in which they present a kinematic analysis using \textit{Gaia} DR3 data, with a probability $P_\mathrm{LMC}>0.26$. Six of our sources did not exist in the probability catalogue of \cite{Gaia_LMC}. We checked whether the proper motions and parallax of the targets followed the distribution of probable members and removed 5 targets with $\text{PM}_\text{RA}>2.6$ mas yr$^{-1}$. \autoref{fig:PM} shows the distribution of proper motions and parallax, and the vertical dotted line shows the mentioned limit in $\text{PM}_\text{RA}$. We also kept 13 targets without \textit{Gaia} DR3 data (since it is uncertain whether they are foreground). We required that all sources have photometry in at least one band at $\lambda>8 \ \mu$m to properly model the infrared excess of the dust. The majority of our sources have at least two observations in that range. For those that had only one, we required that they have photometry from \textit{Spitzer} [8.0] or W3. Figure \ref{fig:spatial} shows the spatial distribution of the final and clean sample of 2,219 RSG candidates, including the dust-enshrouded RSG WOH G64 \citep{Levesque_2009}, which was explicitly added, as it did not exist in our source catalogues.

\begin{figure}[h]
    \centering
    \includegraphics[width=\columnwidth]{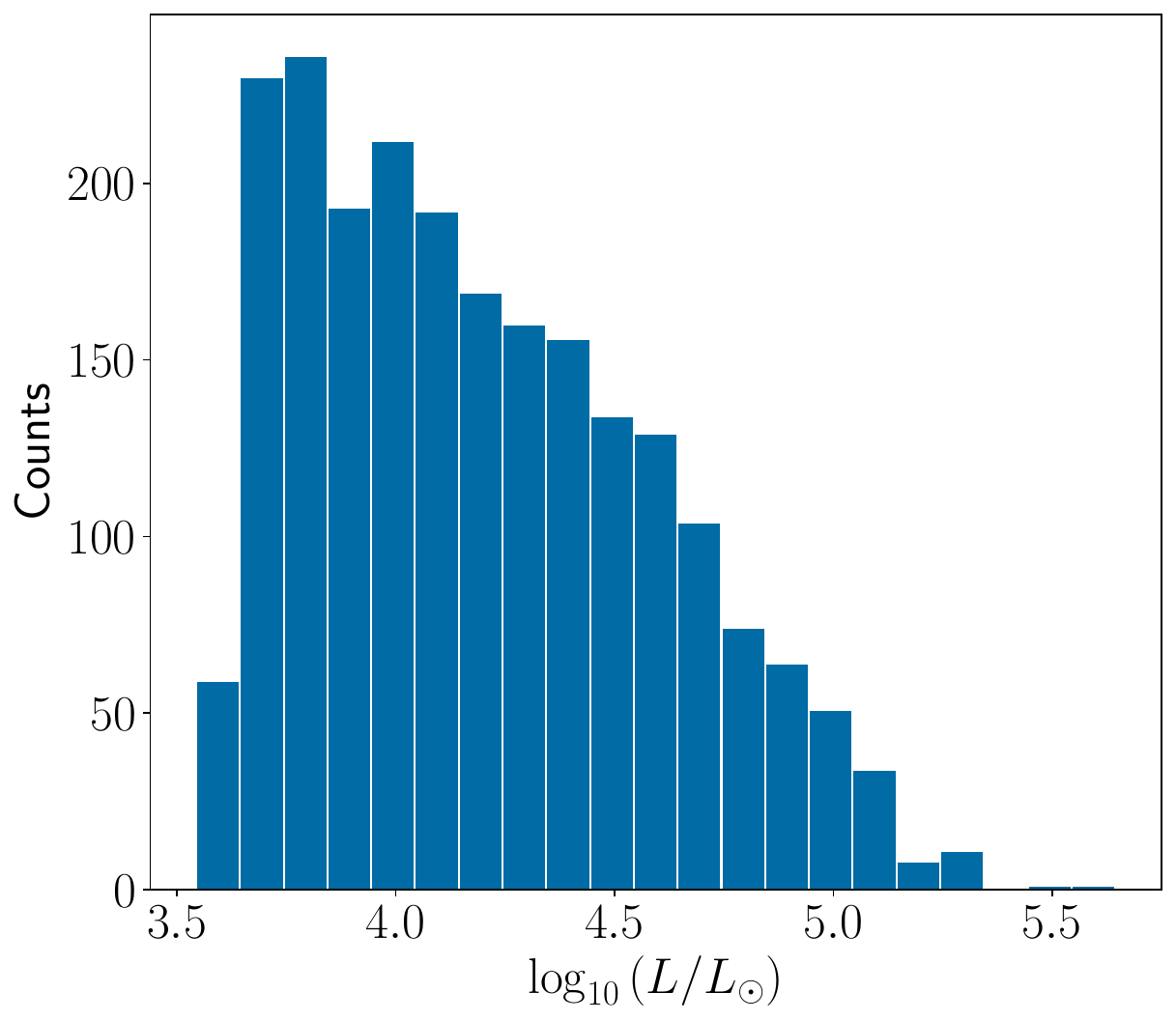}
    \caption{Luminosity distribution of the RSG candidate sample.}
    \label{fig:L}
\end{figure}

We corrected for foreground Galactic extinction using the extinction law and values from \cite{Wang_2019} with $R_V=3.1$ and $E(B-V)=0.08$ mag for the LMC \citep{Massey_2007}. For the OGLE \textit{I} band, we used the individually determined extinction values provided by the OGLE-III shallow survey for each target \citep{ogleIII}. We calculated the luminosity of each RSG by integrating the observed spectral energy distribution (SED), $L = 4\pi d^2\int F_\lambda \mathrm{d}\lambda$, with a distance to the LMC $d= 49.59\pm0.63$ kpc \citep{lmc_distance}. We used the trapezium method to integrate the SED. Even though the observational data did not cover the whole spectrum, the SED beyond 24~$\mu$m does not significantly contribute to the total flux. Additionally, we computed the luminosity by integrating the whole SED of the best-fit model (described in the following section). The results agreed well ($\lesssim1\%$ difference, reaching up to around 2\% for the most luminous sources). Combination of the photometric and distance errors resulted in an error in luminosity below 2\%. \autoref{fig:L} shows the luminosity distribution. We should note that the tip of the red giant branch (TRGB) defines the lower limit in $K_s$ mag \citep{Yang_2021lmc}, which results in low-luminosity sources. The least luminous RSG in our sample, confirmed by spectroscopy \citep{Gonzalez_2015}, has $\log{(L/L_\odot)}\simeq 4$. We used this additional cut in our study of RSG mass-loss rates to avoid contamination by lower-mass stars, such as red helium-burning stars.

We calculated the effective temperatures, $T_{\mathrm{eff}}$ in $K$, using the empirical relation for RSGs in the LMC from \cite{Britavskiy_2019},
\begin{equation}
    T_\mathrm{eff} = 4741 - 791 \times (J-K_s)_0, \label{eq:Teff}
\end{equation}
with the observed $J-K_s$ colour corrected for a total extinction of $E(B-V)=0.13$ mag \citep[internal and foreground;][]{Massey_2007}, with an additional assumed circumstellar extinction of $A_V=0.1$~mag. Sources with low optical depth ($\tau_V<0.1$) have a corresponding CSM $A_V\simeq 0.01$ mag \citep[e.g.][]{Kochanek_2012}; in more extreme cases, $A_V\simeq1$ mag. We checked the effect of these two values on $T_\mathrm{eff}$, and the difference was $\leq3\%$. Equation (\ref{eq:Teff}) is valid for $0.8<(J-K_s)_0<1.4$~mag; therefore, we did not calculate temperatures for sources with redder colours. In our sample, 118 sources had bluer colours ($0.7<(J-K_s)_0\leq 0.8$~mag), only 12 of which had $\log(L/L_\odot)>4$. We therefore did not exclude them as they do not affect our results.


\section{Dust shell models} \label{sec:DSmodels}
\subsection{Model parameters}

We used the 1D radiative transfer code \texttt{DUSTY} (V4)\footnote{\hyperlink{https://github.com/ivezic/dusty}{https://github.com/ivezic/dusty}} from \cite{Ivezic_1997} to model the SEDs of our sources. \texttt{DUSTY} solves the radiative transfer equations for a central source inside a dust shell of certain optical depth, $\tau_{V}$ at $\lambda=0.55 \ \mu \mathrm{m}$ in our case. We assumed a spherically symmetric dust shell extending to $10^4$ times the inner radius, $R_\mathrm{in}$. This value is commonly adopted so that the dust density is low enough at the outer limit so that it does not affect the spectrum. 

\subsubsection{Density distribution}

We assumed steady-state winds with a constant terminal outflow velocity, $v_\infty$, and a density distribution of the dust shell falling off as $\rho\propto r^{-2}$. Outflow wind speeds of RSGs are estimated to be $\sim 20-30$ km s$^{-1}$ based on maser emission measurements \citep{Richards&Yates_1998, vLoon_2001, Marshall_2004}. According to radiatively driven wind theory, the terminal outflow velocity scales with luminosity as $v_\infty \propto L^{1/4}$ \citep{vLoon_2000}. We assumed 30 km s$^{-1}$ for the most luminous RSG of our sample and rescaled accordingly for the wind speeds of the rest of the RSGs, $v_\infty = 30 \ (L/L_{\mathrm{max}})^{1/4}$. For the LMC, \citet{Goldman_2017} also derived a scaling relation between wind speed and luminosity by measuring the wind speeds from maser emission and found $v_\infty=0.118L^{0.4}$. However, this relation yields low values, $v_\infty \lesssim 10$ km s$^{-1}$ for stars with $L<10^5 L_\odot$. In addition, \citet{Beasor_2022} suggested that some of the stars studied by \cite{Goldman_2017} are luminous AGB stars and not RSGs. We chose to use the scaling by \cite{vLoon_2000}, even though the dust-driven wind theory may not explain the mechanism in RSGs, since the empirical correlation of $v_\infty$ with $L$ by \cite{Goldman_2017} needs improvement. Considering the linear relation of the mass-loss rate with the outflow velocity, this assumption does not affect our results significantly. We also note that \cite{Goldman_2017} provided a metallicity dependence on their mentioned relation, which \cite{Wang_2021} then applied the low-metallicity LMC relation to their high-metallicity sample in M31 and M33.

Then, we calculated the mass-loss rate as \citep{Beasor_2016}
\begin{equation}
    \dot{M} = \frac{16\pi}{3}\frac{a}{Q_V} R_{\mathrm{in}}\tau_V\rho_d v_\infty r_{\mathrm{gd}} , \label{eq:dotM}
\end{equation}
where $a/Q_V$ is the ratio of the dust grain radius over the extinction efficiency at the $V$ band, $\rho_d$ is the bulk density, $R_\mathrm{in}$ is the inner shell radius, and $r_\mathrm{gd}$ is the gas-to-dust ratio. We note that \texttt{DUSTY} has $R_\mathrm{in}$ as an output for a given luminosity of $10^4 \ L_\odot$, and the inner radius is scaled as $R_\mathrm{in} \propto L^{1/2}$ (see \texttt{DUSTY} manual Section 4.1\footnote{\hyperlink{https://github.com/ivezic/dusty/blob/master/release/dusty/docs/manual.pdf}{\texttt{DUSTY} manual V4}}); thus, we rescaled each best-fit model radius as $R_\mathrm{in} = R_\mathrm{in}^\mathrm{DUSTY} [{L}/(10^4L_\odot)]^{1/2}$. We used an average value for the gas-to-dust ratio of the LMC, $r_\mathrm{gd}\simeq245.5^{+255}_{-80}$ \citep{Clark_2023}.

We also opted for the numerical solution for radiatively driven winds (RDW) to test our results and compared it with the steady-state wind assumption. The full dynamics calculations can be found in \cite{Ivezic_1995}, \cite{Elitzur_2001} and references therein. The analytic approximation is shown below for reference,
\begin{equation}
    \eta(y) \propto \frac{1}{y^2}\left[\frac{y}{y-1+\left(v_1 / v_e\right)^2}\right]^{1 / 2},
\end{equation}
where $\eta$ is the dimensionless density profile normalised according to $\int\eta dy=1$, $y\equiv r/R_\mathrm{in}$ with $R_\mathrm{in}$ the inner shell radius, and $v_1$ and $v_e$ the initial and final expansion velocities, respectively. The mass-loss rate for RDW is given as an output from \texttt{DUSTY} for $L=10^4 \ L_\odot$, $r_\mathrm{gd}=200$ and $\rho_d=3 \ \mathrm{g} \ \mathrm{cm}^{-3}$, demanding a rescaling according to the dust-driven wind relation by \cite{vLoon_2000},
\begin{equation}
    \dot{M} = \dot{M}_\mathrm{DUSTY}\left(\frac{L}{10^4}\right)^{3/4}\left(\frac{r_\mathrm{gd}}{200}\frac{\rho_d}{3}\right)^{1/2}.
\end{equation}
We should note that \cite{McDonald_2011} have an error in their $\dot{M}$ relation (Eq. (1) in their paper): They divide the dust-to-gas ratio $\psi$ by 200 instead of 1/200. This erroneous equation was also applied by \cite{Yang_2023}, with an index of $-$1/2 instead of 1/2 for the gas-to-dust ratio. Application of the correct relation would yield higher mass-loss rates by a factor of 5 or 0.7 dex in logarithmic space in the \cite{Yang_2023} result.

\subsubsection{Dust composition}
The dust grain composition affects the interaction of radiation with the dust shell and is observed in the output SED. Observations of RSGs indicate O-rich silicate features at roughly $10 \ \mu \mathrm{m}$ and $18 \ \mu \mathrm{m}$, as shown in previous works \citep[e.g.][]{vLoon_2005, Groenewegen_2009, Verhoelst_2009, Sargent_2011, Beasor_2016, Goldman_2017}. We opted for the astronomical silicate of \cite{DraineLee_1984} with a dust bulk density of $\rho_d=3.3 \ \mathrm{g} \ \mathrm{cm}^{-3}$.

\subsubsection{Grain size distribution}
We used the modified Mathis-Rumpl-Nordsieck (MRN) distribution $n(a)\propto a^{-q}$ for $a_{min}\leq a \leq a_{max}$ \citep{MRN}. Observations of VY CMa, a dust-enshrouded RSG, yielded an average grain radius of $0.5 \ \mu\mathrm{m}$ \citep{Scicluna_2015}, while other studies showed that grain sizes are in the range of $0.1-1 \ \mu\mathrm{m}$ \citep{Smith_2001, Groenewegen_2009, Haubois_2019}. Thus, we opted for $a_{min}=0.1 \ \mu\mathrm{m}$, $a_{max}=1 \ \mu\mathrm{m}$ and $q=3$, even though the grain size in the range of [0.1, 0.5] does not significantly affect the mass-loss rate \citep{Beasor_2016}.

A broader MRN distribution with a grain size in the range $0.01 - 1 \ \mu$m and $q=3.5$ has been applied for different populations of RSGs by \citet{Verhoelst_2009, Liu_2017, Wang_2021}. Moreover, \citet{Ohnaka_2008} determined grain sizes in that range from their best-fit model of WOH G64. This profile has an average grain size of approximately $0.02-0.03$ $\mu$m, which affects the extinction efficiency, $Q_V$ \citep[see Fig. 4 in][]{Wang_2021}. Since it is difficult to infer a grain size from photometric data, we used this broader range of the modified MRN distribution to compare with our main assumption and to examine its effect on the mass-loss rate.

\subsubsection{Stellar atmosphere models, $T_\mathrm{eff}$}
We used the \textsc{MARCS} model atmospheres \citep{Gustafsson_2008} as input SEDs for the central source. We chose typical RSG parameters: a stellar mass of $15 \ M_\odot$, a surface gravity $\log g=0$, interpolated models with metallicity $[Z]=-0.38$, consistent with the metallicity of RSGs in the LMC, $[Z]=-0.37\pm 0.14$ \citep{Davies_2015}, and effective temperatures, $T_\mathrm{eff}$, in the range of [3300, 4500] with a step of 100 K. We interpolated the models between 4000 and 4500 K to create models with a step of 100 K, which are not available by default. We chose the commonly adopted value for RSGs of $\log g=0$, although it can be as low as approximately $-0.5$ \citep[e.g.][]{Davies_2015}. However, this affects the peak of the SED due to line blanketing, shifting it to appear slightly cooler for higher $\log g$, and it is degenerate with the effective temperature. Since we could not constrain both parameters, we chose to fix the value for $\log g$. This can affect the derived $T_\mathrm{eff}$ from \texttt{DUSTY}, but we used Eq. (\ref{eq:Teff}) to calculate the $T_\mathrm{eff}$ of our sources. A varied $T_\mathrm{eff}$ in \texttt{DUSTY} leaves the mass-loss rate almost unaffected \citep{Beasor_2016}. We then resampled the \textsc{MARCS} model surface fluxes by decreasing the resolution, and we matched them with the custom \texttt{DUSTY} wavelength grid. Since \textsc{MARCS} models extend up to $20 \ \mu\mathrm{m}$, we extrapolated longer wavelength ranges using a blackbody.

\subsubsection{Inner dust shell temperature $T_\mathrm{in}$ and optical depth $\tau_V$}

We set the dust temperature at the inner radius of the dust shell, $T_\mathrm{in}$, in the range of $[400, 1200]$ with a step of $100 \ \mathrm{K}$. We also set the sublimation temperature to 1200 K, which is considered to be an approximation for silicate dust \citep{Gail_1984, Gail_1999, Gail_2020}.

The optical depth, $\tau_V$, was varied using 30 logarithmic steps in the range of [0.001, 0.1], 70 steps in the range of [0.1, 1], and 99 in [1, 10].

In total, we created 23,049 models in which we varied the mentioned parameters to optimise the data fitting.

\subsection{Spectral energy distribution fitting} \label{sec:sed_fitting}

First, we created synthetic fluxes by convolving the model spectrum with the filter profiles of each observational instrument\footnote{The filter profiles were retrieved from \url{http://svo2.cab.inta-csic.es/theory/fps/}.}. We then calculated the minimum modified $\chi^2$, $\chi^{2}_{mod}$, as defined below \citep{Yang_2023},
\begin{equation}
    \chi^2_{mod}=\frac{1}{N-p-1} \sum{\frac{[1-F(Model, \lambda)/F(Obs, \lambda)]^2}{F(Model, \lambda)/F(Obs, \lambda)}},
\end{equation}
where $F(\lambda)$ is the flux at a specific wavelength, $N$ is the number of photometric data of each source, and $p$ is the free parameters. The output flux of \texttt{DUSTY} is given at the outer dust shell radius, $R_\mathrm{out} = 10^4R_\mathrm{in}$, so to compare with the observed flux on Earth, 
\begin{equation}
    F(Model, \lambda) = F_\mathrm{DUSTY}(\lambda)(10^4R_\mathrm{in}/d)^2 ,
\end{equation}
where $d$ is the distance to the source-galaxy ($d=d_\mathrm{LMC}$).

We used this definition of minimum $\chi^2$ to balance the contribution of fluxes from both the optical and the infrared. A regular $\chi^2$ would barely account for the infrared flux, which is crucial for determining the dust properties, and it would weight the optical photometry much more. The modified $\chi^2$ treats all wavelengths equally since it depends on the ratio of the model over the observation instead of the difference \citep[see Section 3.2 in][for a more detailed explanation]{Yang_2023}.

We determined the errors on the best-fit parameters as in \cite{Beasor_2016}, but since we have a different definition of the minimum $\chi^2$, we defined them by the models within the minimum $\chi_{mod}^2+0.002$. This is a crude estimation of the errors and does not account for all the uncertainties that lie in every assumption that is made.

\begin{figure*}[h]
     \centering
    \begin{subfigure}[t]{0.49\textwidth}
        \raisebox{-\height}{\includegraphics[width=\textwidth]{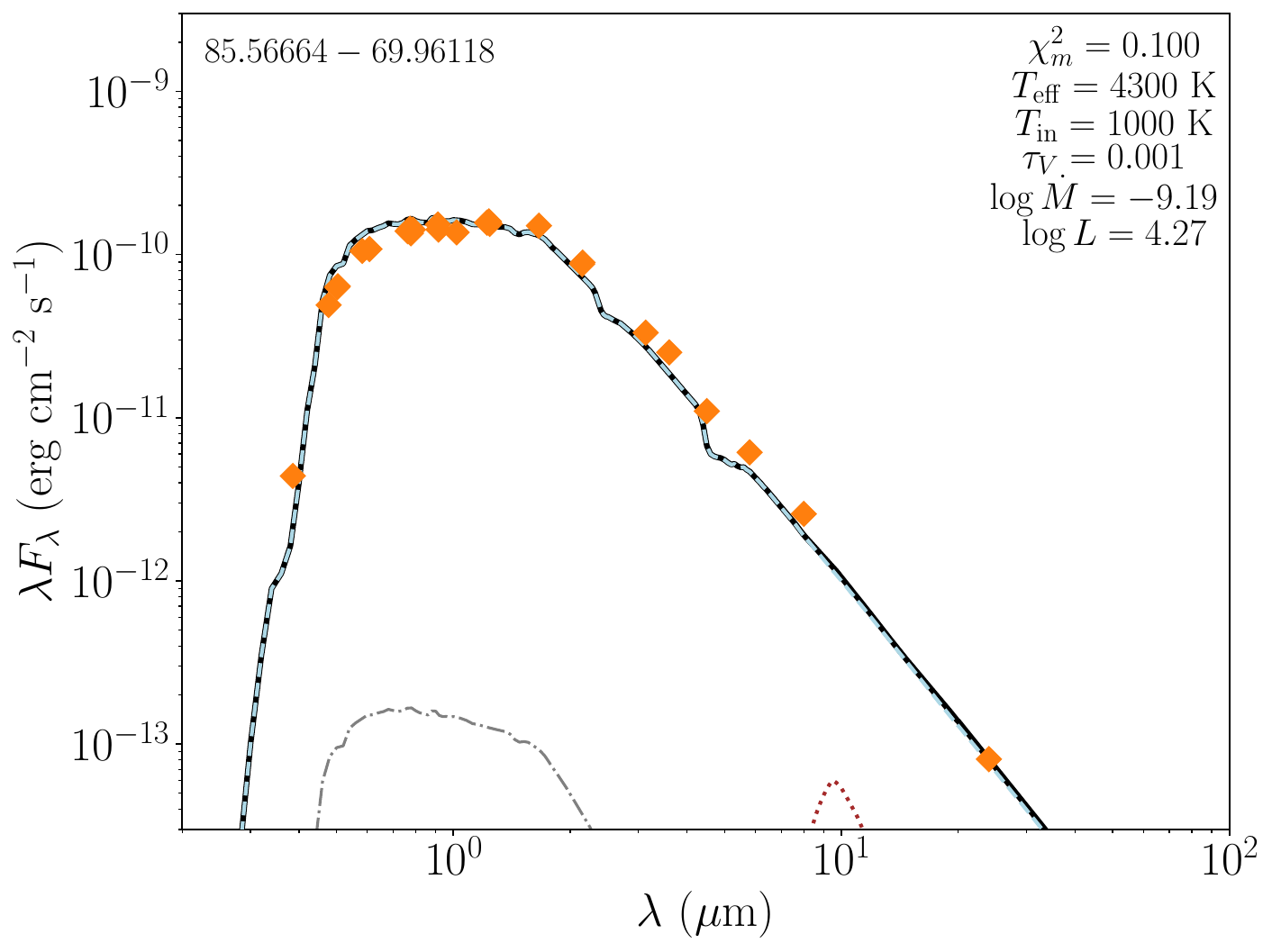}}
    \end{subfigure}
    \hfill
    \begin{subfigure}[t]{0.49\textwidth}
        \raisebox{-\height}{\includegraphics[width=\textwidth]{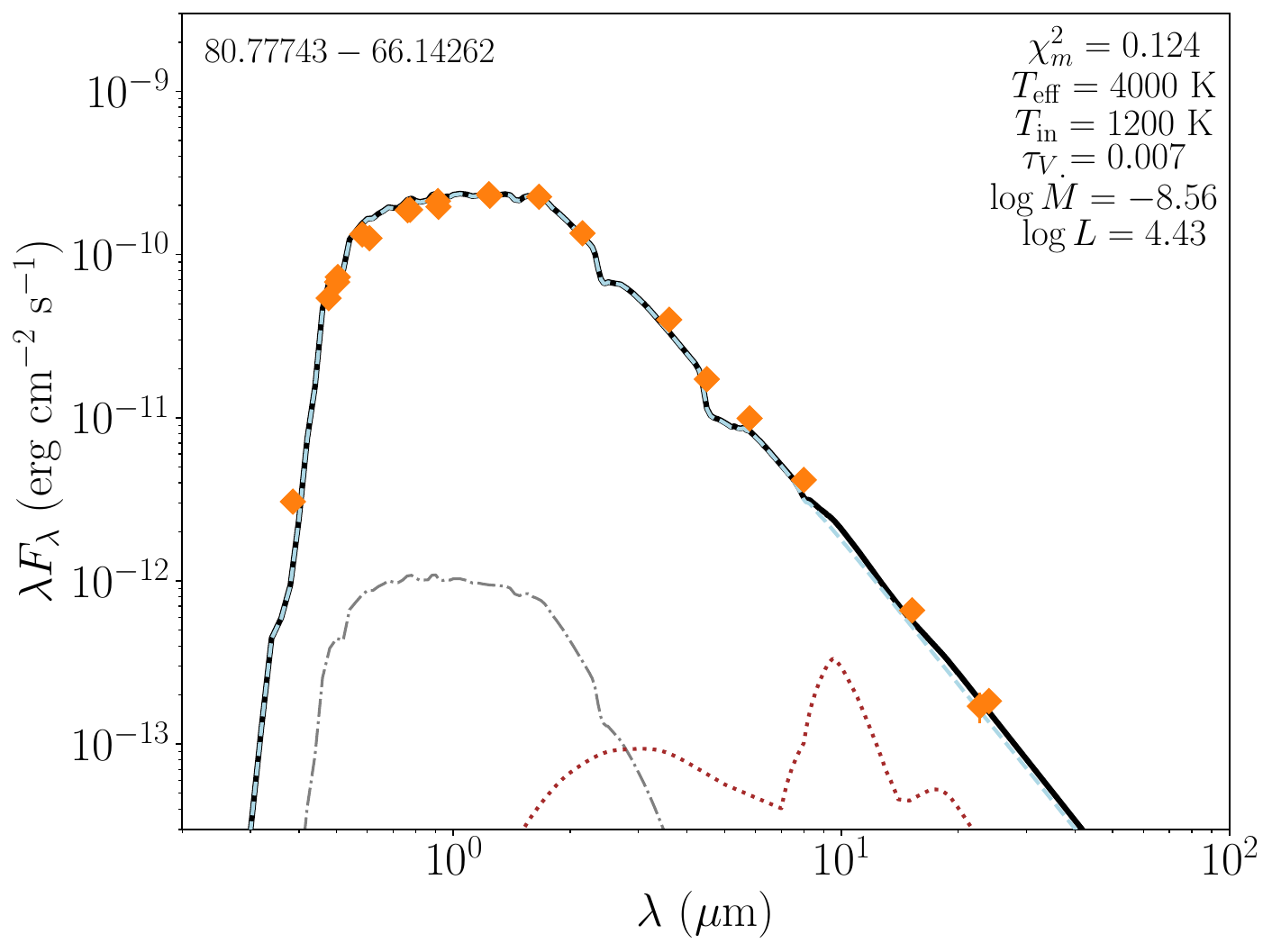}}
    \end{subfigure}
    \begin{subfigure}[t]{0.49\textwidth}
        \raisebox{-\height}{\includegraphics[width=\textwidth]{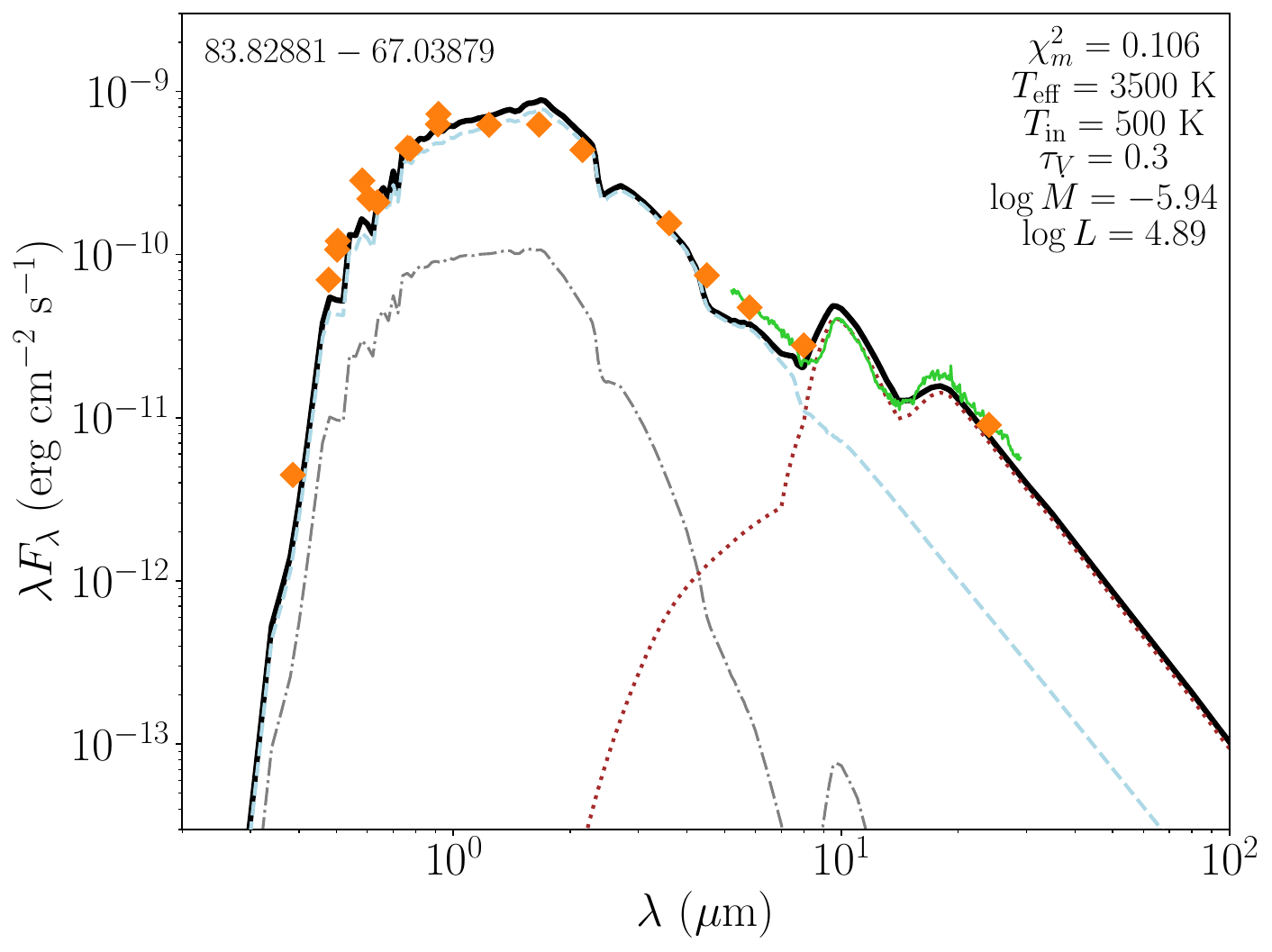}}
    \end{subfigure}
    \hfill
    \begin{subfigure}[t]{0.49\textwidth}
        \raisebox{-\height}{\includegraphics[width=\textwidth]{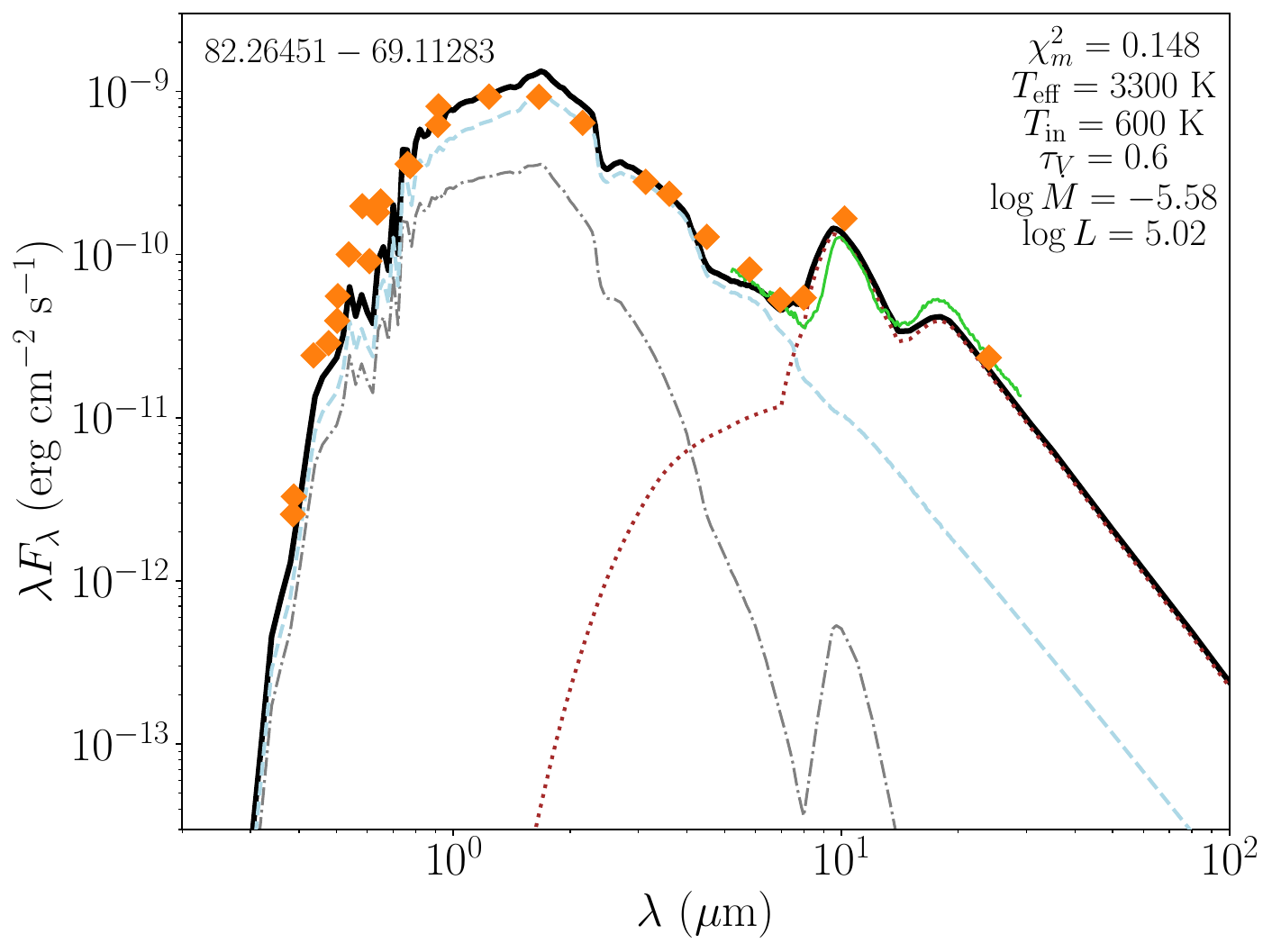}}
    \end{subfigure}
    \begin{subfigure}[t]{0.49\textwidth}
        \raisebox{-\height}{\includegraphics[width=\textwidth]{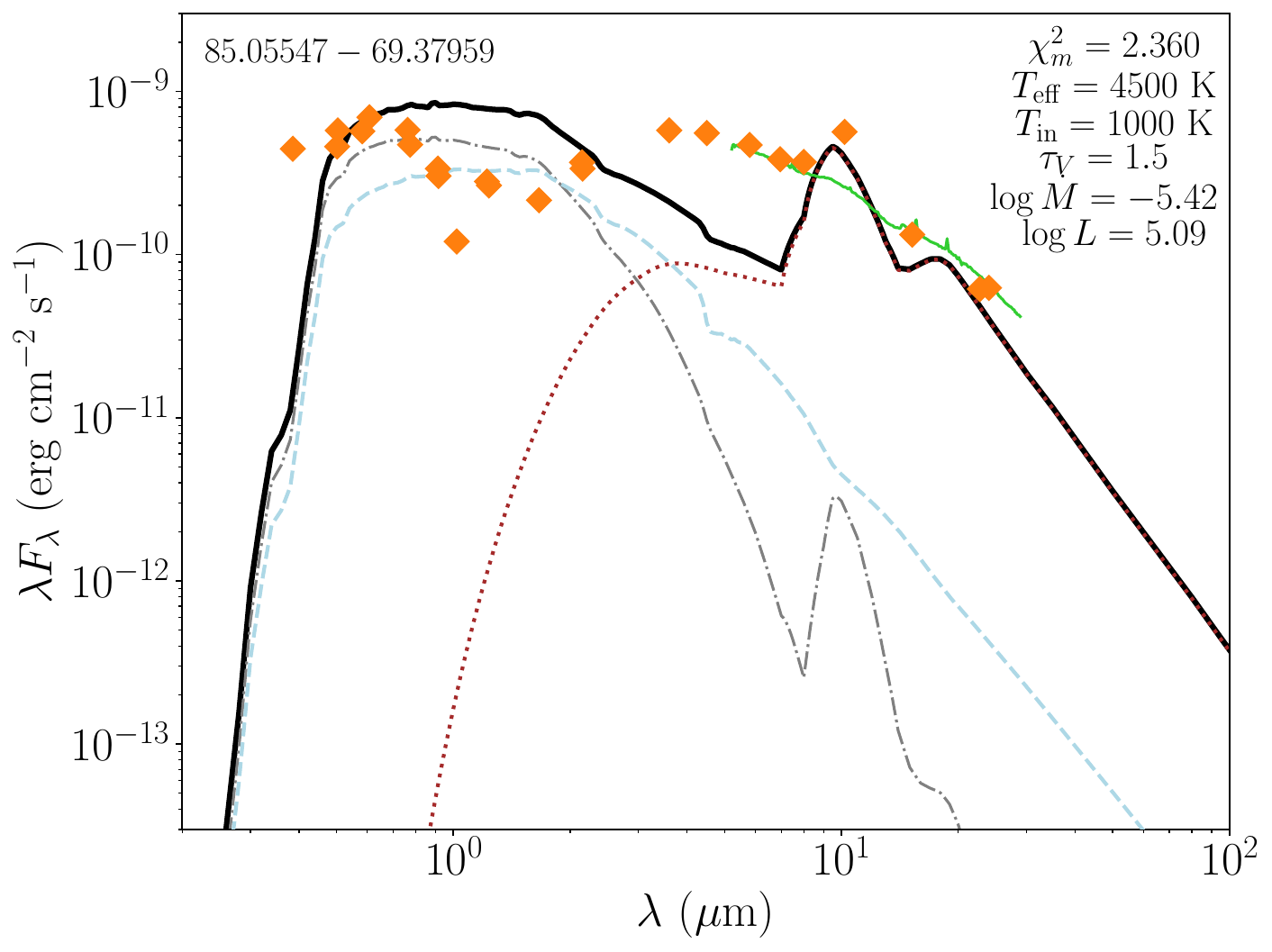}}
    \end{subfigure}
    \hfill
    \begin{subfigure}[t]{0.49\textwidth}
        \raisebox{-\height}{\includegraphics[width=\textwidth]{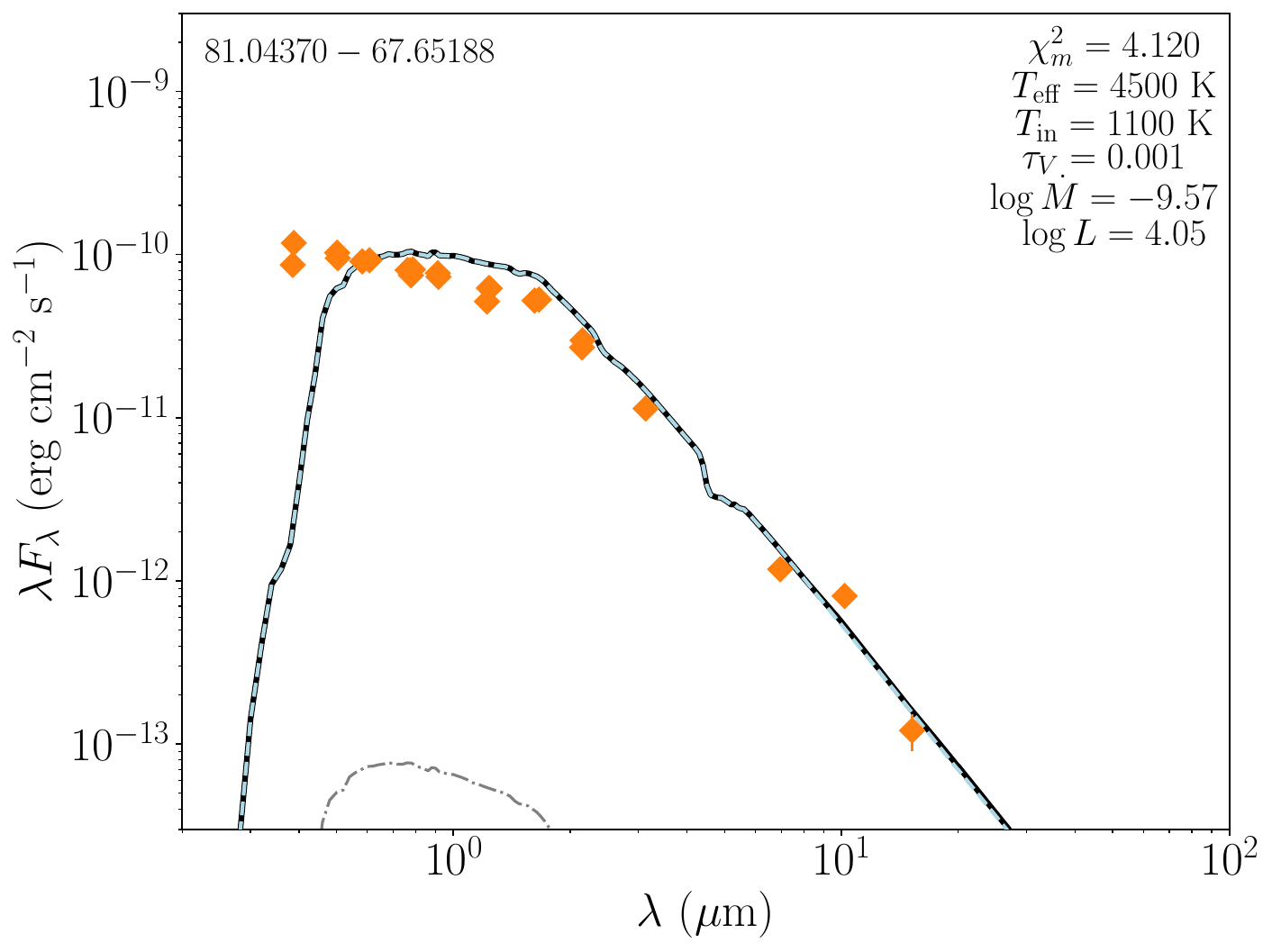}}
    \end{subfigure}

    \caption{Examples of SED fits of two optically thin (top row), two dusty (middle), and two peculiar (bottom row) RSGs. The orange diamonds show the observations, and the black line is the best-fit model consisting of the attenuated flux (dashed light blue), the scattered flux (dot-dashed grey) and dust emission (dotted brown). The green curve represents the \textit{Spitzer} IRS spectrum. The coordinates of each source are shown in the top left corner.} \label{fig:sed}
\end{figure*}

\autoref{fig:sed} shows six example SEDs. In the top row, we included two optically thin RSGs without IR excess, demonstrating the absence of a dust shell. In the middle row, the mid-IR excess shows the presence of a dust shell emitting in this wavelength range, and the bump at around $10 \ \mu$m reveals the presence of silicate dust. In the bottom row, we present some peculiar examples that could not be properly fit by a model with a significant IR excess or a UV excess, indicating a binary candidate. The bottom left source is a supergiant B[e] star (ID 225) that passed the photometric criteria along with two more stars (IDs 192, 706), found in the catalogue of \citet{Kraus_2019}. However, these stars are not included in our analysis of mass-loss rates because their model fits are poor, as we explain in the following section. We also found 57 matches of our sample with the \textit{Spitzer} InfraRed Spectrograph (IRS) Enhanced Products \citep{Houck_2004}, using a 1$\arcsec$ radius, to visually validate our SED fits; 3 of these matches are shown in Fig.~\ref{fig:sed}.


\section{Results} \label{sec:results}

\begin{table*}[h]
\centering
  \begin{threeparttable}
    \caption{Properties and derived parameters of RSG candidates in the LMC.}
    \renewcommand{\arraystretch}{1.3}
    \begin{tabular}{llllllllll}
    \hline\hline
ID &        R.A.  &        Dec.  &  $J_{\rm 2MASS}$ & $\sigma_{J\rm 2MASS}$  &  ... & $\log{\dot{M}}$                    & $T_\mathrm{eff}^{(J-K_s)_0}$ & $T_\mathrm{in}$            &     $\tau_V$    \\
 &        (deg) &        (deg) &  (mag)&  (mag) &  ... & $(\rm M_{\odot} \ \mathrm{yr}^{-1})$                    &  (K) &  (K)            &        \\
\hline 
1 &  77.681140 & $-$68.360497 &  12.084 &     0.022 &  ... & $-8.09_{-0.38}^{+0.90}$ &  $4061$  & $1200_{-500}^{+  0}$ &   $0.053_{-0.020}^{+0.020}$ \\
2 &  78.131344 & $-$68.802155 &  12.145 &     0.024 &  ... & $-8.27_{-0.39}^{+0.46}$ &  $4131$    & $1100_{-100}^{+100}$ &   $0.028_{-0.008}^{+0.005}$ \\
3 &  78.429980 & $-$68.964081 &  11.372 &     0.022 &  ... & $-7.93_{-0.17}^{+0.31}$ &  $4080$    & $1100_{-  0}^{+  0}$ &   $0.033_{-0.000}^{+0.000}$ \\
4 &  81.141740 & $-$66.431923 &  11.358 &     0.024 &  ... & $-7.13_{-1.25}^{+0.52}$ &  $3936$  & $ 400_{-  0}^{+800}$ &   $0.045_{-0.017}^{+0.028}$ \\
5 &  76.797033 & $-$69.191391 &  11.517 &     0.023 &  ... & $-9.44_{-0.43}^{+1.41}$ &  $3935$    & $ 900_{-500}^{+300}$ &   $0.001_{-0.000}^{+0.002}$ \\
    \vdots  &  \vdots & \vdots &  \vdots &    \vdots & \vdots  &  \vdots & \vdots & \vdots & \vdots 
    \\
    \hline
    \end{tabular}
    \label{tab:results} 
    \begin{tablenotes}
        \small
        \vspace{5pt}
        \item Notes: This table is available in its entirety in machine-readable format at the CDS. A portion of it is shown here for guidance regarding its form and content.
    \end{tablenotes}
  \end{threeparttable}
\end{table*}

\begin{figure*}[h]
    \centering
    \includegraphics[width=\linewidth]{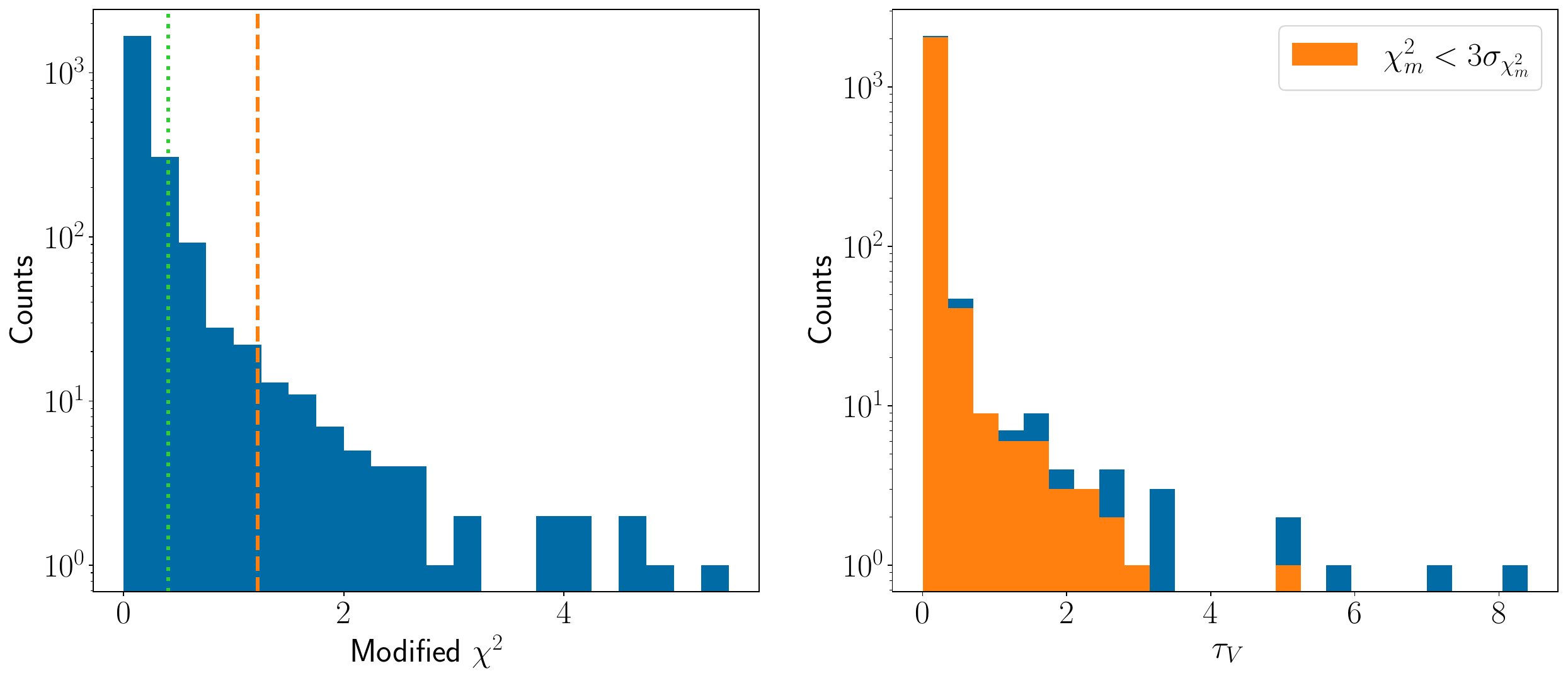}
    \caption{Distributions of SED fitting parameters. \textit{Left:} Distribution of the minimum $\chi^2_{mod}$. The vertical lines show the $\sigma_{\chi^2}$ (green dotted) and the $3\sigma_{\chi^2}$ (dashed orange) limits. \textit{Right:} Distribution of the best-fit optical depth, $\tau_V$. Orange indicates $\tau_V$ from models with $\chi^2_{mod}<3\sigma_{\chi^2}$.}
    \label{fig:chi2tau}
\end{figure*}

\begin{figure}[h]
    \includegraphics[width=\columnwidth]{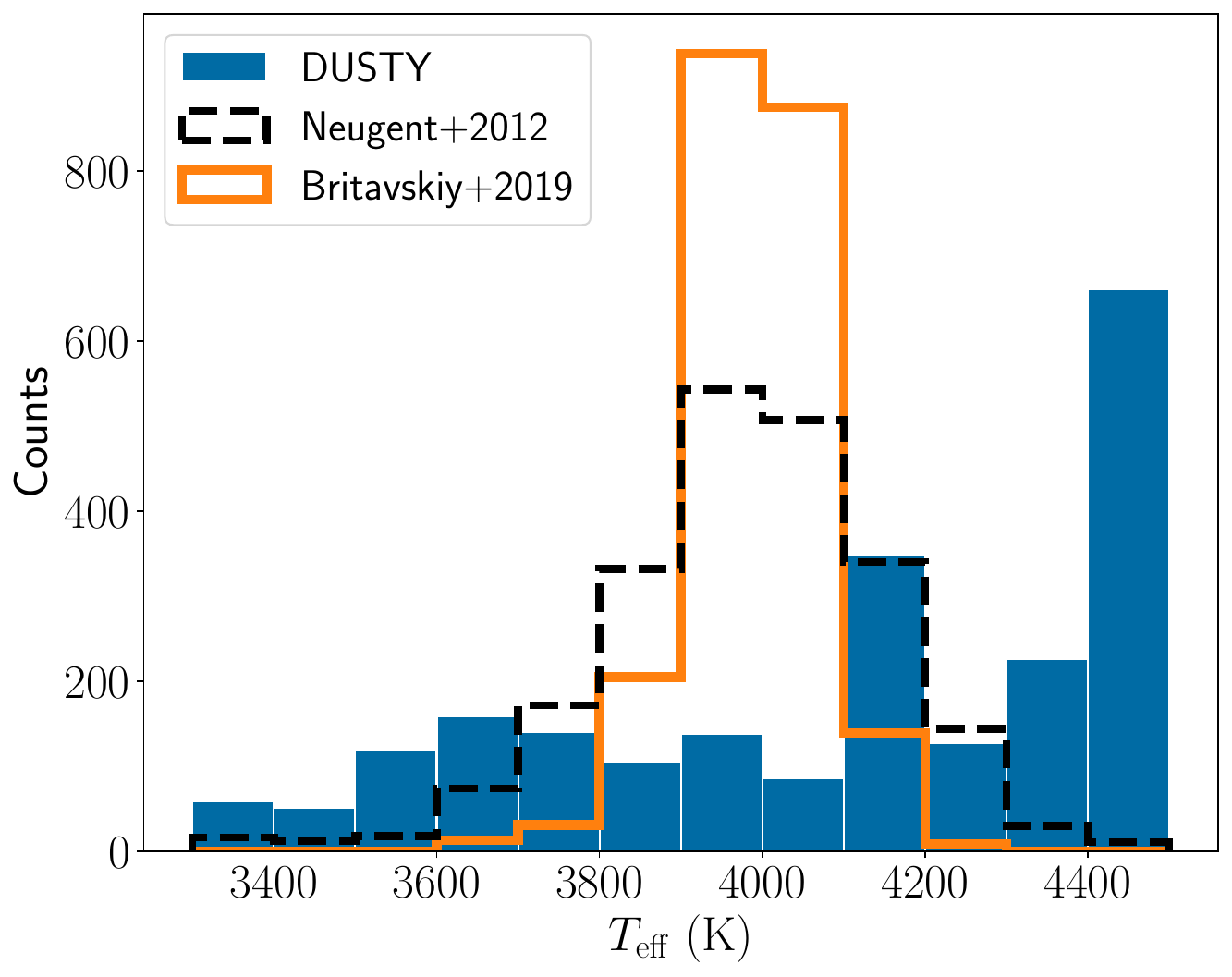}
    \caption{Distribution of the best-fit effective temperatures from \texttt{DUSTY}. The lines show the applied $J-K_s$ empirical prescriptions from \cite{Neugent_2012} (orange) and \cite{Britavskiy_2019} (dashed black) to our sample.
 }
    \label{fig:Teff}
\end{figure}

Fitting the observed SEDs with the \texttt{DUSTY} models, we obtained the dust shell parameters and mass-loss rates for each RSG. \autoref{fig:chi2tau} shows the histograms of the minimum $\chi^2_{mod}$ and the best-fit $\tau_V$. We consider models with $\chi^2_{mod}<3\sigma_{\chi^2}$ as a good fit, where the standard deviation is $\sigma_{\chi^2}=0.41$. Moreover, we do not present the parameters of 36 RSG candidates, as their SED could not be fit by any model of the grid. The average optical depth is $\tau_V = 0.09$. \autoref{fig:Teff} shows a histogram of the best-fit model effective temperatures on the left along with two colour-temperature, $T_\mathrm{eff}(J-K_s)$, LMC empirical prescriptions \citep{Neugent_2012, Britavskiy_2019} applied to our sample. In this study, we proceed with the relation from \cite{Britavskiy_2019} because it was calibrated on temperature measurements from $i$-band spectra, which are more reliable than TiO-band spectra \citep[for a discussion see][]{Davies_2013, deWit_2024}. The \texttt{DUSTY} $T_\mathrm{eff}$ reveals a discrepancy with those resulting from Eq.\ (\ref{eq:Teff}). One reason is that we used the modified $\chi^2$, which enhances the weight in the mid-IR part of the SED, and it looses some weight in the optical to the near-infrared, where the effective temperature affects the SED more strongly. This suggests that our derived $T_\mathrm{eff}$ from \texttt{DUSTY} are less reliable. \autoref{tab:results} includes our final catalogue of RSG candidates in the LMC with their coordinates, photometry, $Gaia$ astrometry, spectral classifications from the literature, and the parameters derived from this work.

\subsection{Mass-loss rates}

\begin{SCfigure*}[0.5]
    \includegraphics[width=1.5\linewidth]{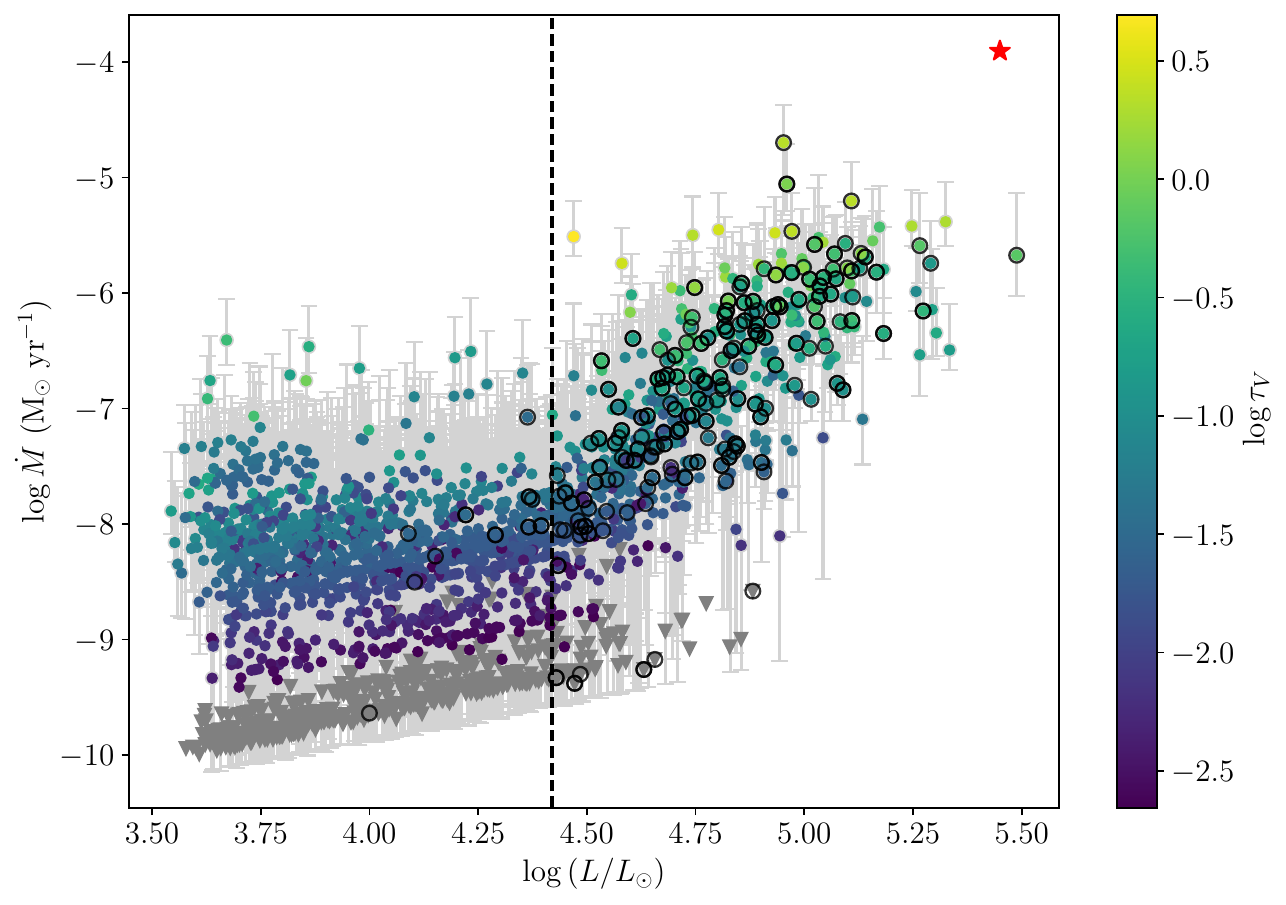}
    \caption{Derived mass-loss rate as a function of the luminosity of each RSG candidate in the LMC. The colour bar shows the best-fit $\tau_V$. The grey triangles represent the upper limits of targets fitted with the lowest values of the optical depth grid, and the open circles indicate the RSGs with spectral classifications. The dust-enshrouded RSG, WOH G64, is labelled with a red star. The position of the kink is shown with a dashed vertical line at $\log(L/L_\odot)=4.42$.}\label{fig:Mdot_L}
\end{SCfigure*}

We present the derived mass-loss rates versus the luminosities in Fig.~\ref{fig:Mdot_L}. We exclude 94 sources with best-fit models of $\chi^2_{mod}\geq3\sigma_{\chi^2}$, which suggests that their derived $\dot{M}$ may not be reliable. The colour bar shows the best-fit optical depth, $\tau_V$, of each source. The uncertainties are propagated from the error models defined in the previous section and the error on the gas-to-dust ratio. We note that the relative error becomes higher at $\log{(L/L_\odot)}<4.5$. The grey triangles show the upper limits for sources with $\tau_V \leq 0.002$. In this regime, the different model SEDs are indistinguishable and do not show any significant indications of dust in the CSM. We also mark the RSGs confirmed by spectral classification from the catalogues of \citet{Yang_2021lmc} and \citet{Gonzalez_2015}. The red star indicates WOH G64, a dust-enshrouded RSG. However, the grain size for this RSG should be lower than the range we considered in Sect.~\ref{sec:DSmodels}, resulting in a higher $\dot{M}$, as we discuss in \autoref{app:woh}. 

\sidecaptionvpos{figure}{c}

The average mass-loss rate is $\dot{M}\simeq 6.4\times10^{-7} \ M_{\odot}$ yr$^{-1}$, while for $\log{(L/L_\odot)}>4$, it is $\dot{M}\simeq 9.3\times10^{-7} \ M_{\odot}$ yr$^{-1}$. The latter corresponds to a mass of $\sim 0.1-1 \ M_{\odot}$ that is lost throughout the lifetime of an RSG, which is presumed to be $10^5-10^6$ yr, and an average dust-production rate of $\sim3.6\times10^{-9}\ M_{\odot}$ yr$^{-1}$. There is a noteworthy dispersion at $\log{(L/L_\odot)} \lesssim 4.1$, but these sources may not be RSGs, as previously mentioned. More significantly, we observe an enhancement of mass loss at $\log{(L/L_\odot)}\simeq4.4$ that appears as an increase in the slope of the $\dot{M}(L)$ relation or a kink, similar to recent findings in the SMC \citep{Yang_2023}. Finally, the mass loss additionally correlates with the effective temperature, the dust temperature, and variability among other properties of the RSG and CSM, which we further discuss in the following subsections.

\subsection{Variability}

\begin{figure}[h]
    \centering
    \includegraphics[width=\columnwidth]{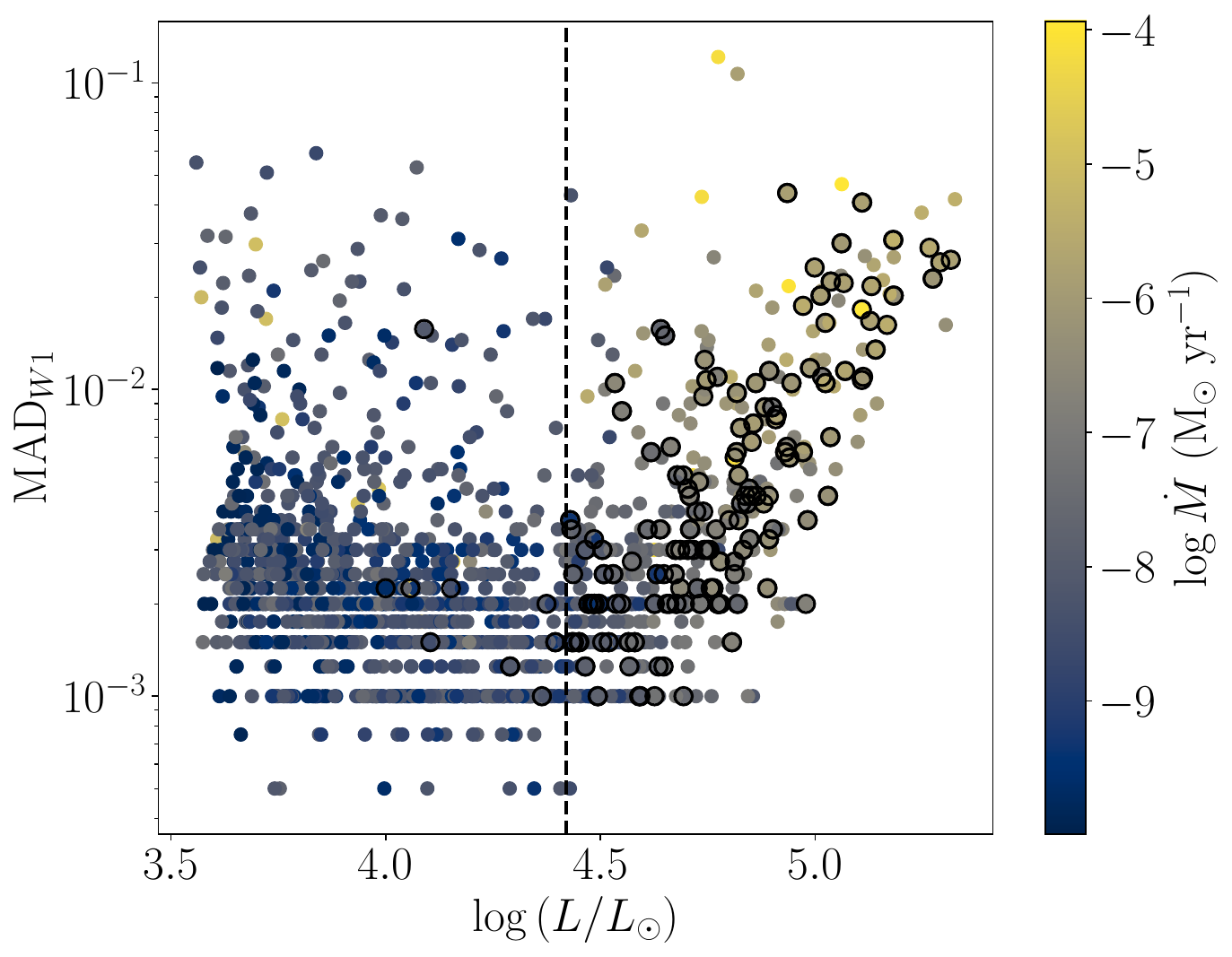}
    \caption{MAD of W1 band vs. luminosity diagram. The colour bar shows the corresponding mass-loss rate of each source. The open circles indicate the RSGs with spectral classifications, and the dashed line represents the position of the kink, as in Fig.\ \ref{fig:Mdot_L}.}
    \label{fig:mad}
\end{figure}

We used the epoch photometry from NEOWISE to calculate the median absolute deviation (MAD) of \textit{W}1 [3.4] band and examined the correlation with luminosity, as was done by \cite{Yang_2023}. We show this correlation in Fig.~\ref{fig:mad}.  The MAD value for \textit{W}1 [3.4] typically indicates the intrinsic variability of an RSG \citep{Yang_2018}. We observe that the trend of MAD with $L$ is nearly identical to the trend of the $\dot{M}(L)$. Thus, the variability appears to correlate with the enhancement of mass loss observed at the turning point around $\log{(L/L_\odot)}=4.4$. There is also high variability for some sources at lower luminosity than the turning point, which we discuss further in Sect.~\ref{sec:RHeB}. Finally, for the extreme assumption that RSGs could vary bolometrically up to 1 mag, it would induce a variation in the luminosity of up to 0.2 dex \citep{Beasor_2021}.

\subsection{Binarity}
\begin{figure}[h]
    \centering
    \includegraphics[width=\columnwidth]{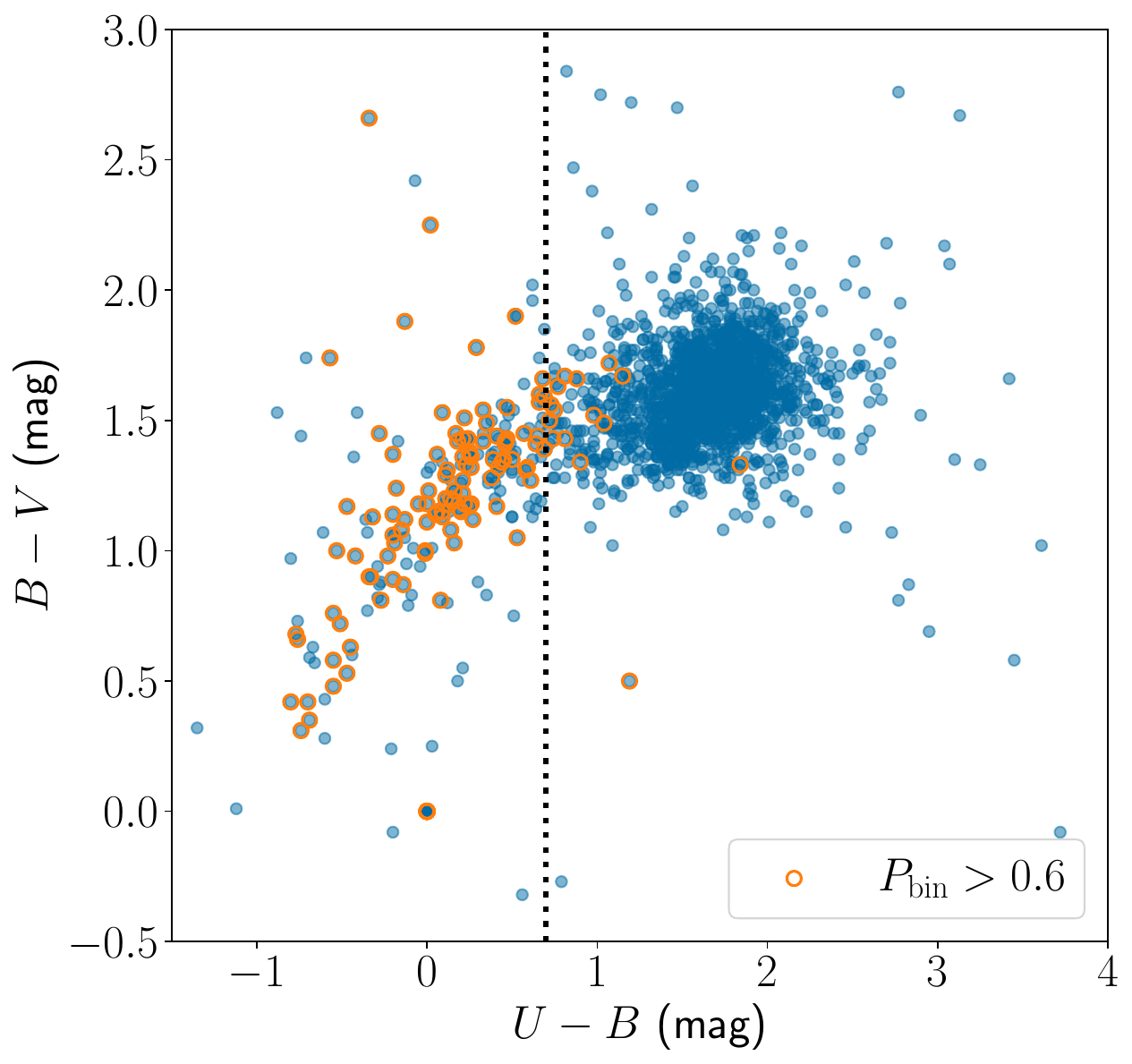}
    \caption{Colour-colour diagram of $U-B$ and $B-V$. The blue points show all sources with $UBV$ photometry. The orange circles indicate RSGs from \cite{Neugent_2020} with a probability of an OB companion $P_\mathrm{bin}>0.6$. The vertical dotted line corresponds to the limit applied to separate probable binary from single RSG.}
    \label{fig:binaries}
\end{figure}

\begin{figure}[h]
    \centering
    \includegraphics[width=\columnwidth]{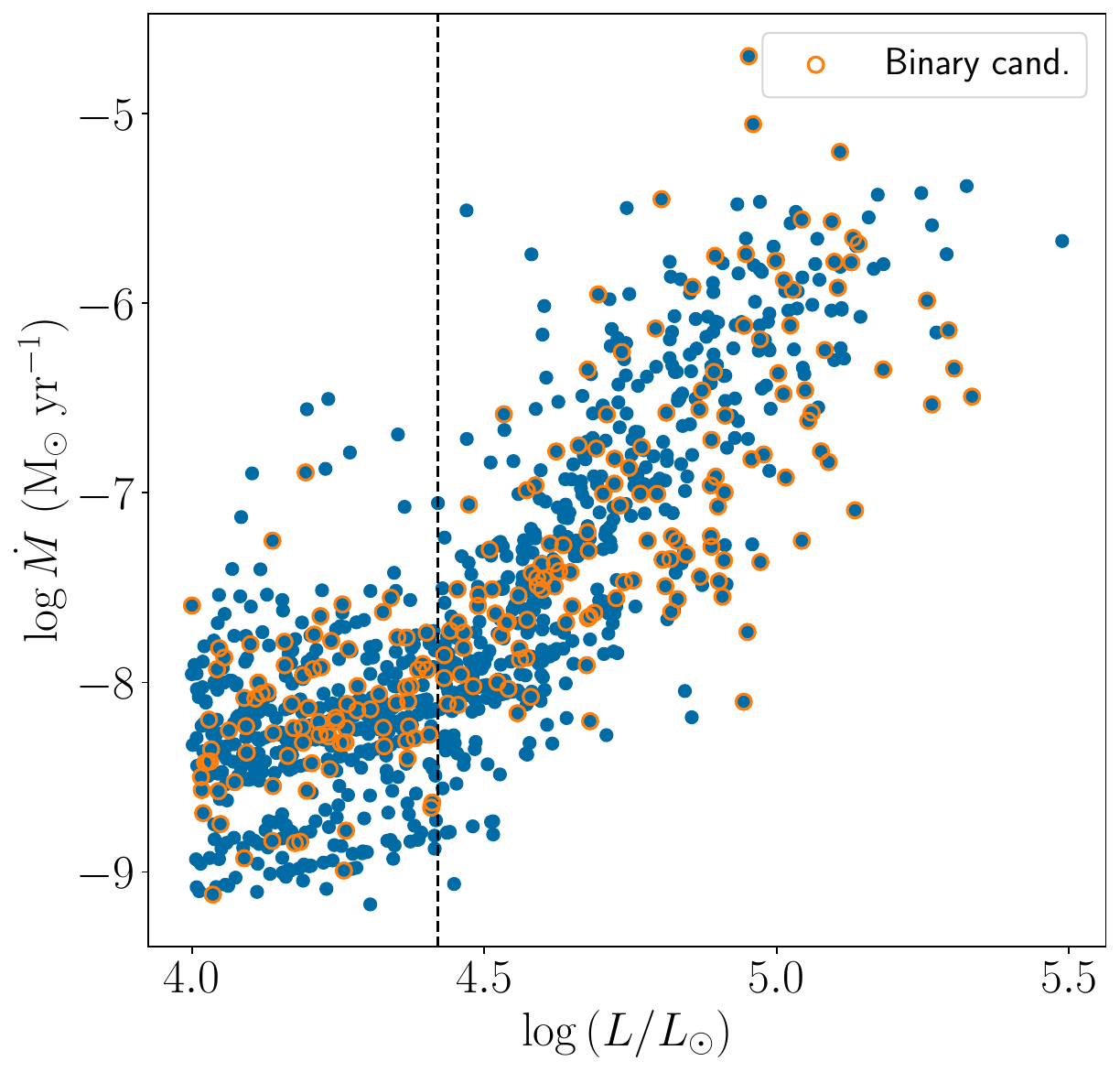}
    \caption{Same as Fig. \ref{fig:Mdot_L}, but for 
    $\log{(L/L_\odot)}>4$, indicating the binary candidates ($U-B<0.7$ mag) with orange circles.}
    \label{fig:mdot_binaries}
\end{figure}

Red supergiants can be part of a binary system with an OB companion \citep[e.g.][]{Neugent_2019}. Binary RSGs consist of around 20\% of the total RSG population \citep{Neugent_2020, Patrick_2022}. We further investigated this fraction in our sample following the $B-V$ versus $U-B$ colour diagram from \citet{Neugent_2020}, which distinguishes the binary population. We considered sources with $U-B<0.7$ mag as binary candidates. This resulted in a fraction of 21\% binary candidates in our complete sample of sources with $UBV$ photometry, and it strongly agrees with \citet{Neugent_2020}. \autoref{fig:binaries} shows the $B-V$ versus $U-B$ colour diagram of these sources, indicating those existing in the \citet{Neugent_2020} catalogue with a binary probability $P_\mathrm{bin}>0.6$. We demonstrate the $\dot{M}-L$ result, and we indicate the binary candidates with orange circles in Fig.~\ref{fig:mdot_binaries}. 

\subsection{$\dot{M}(L, T_\mathrm{eff})$}

\begin{figure*}[h]
    \includegraphics[width=0.51\linewidth]{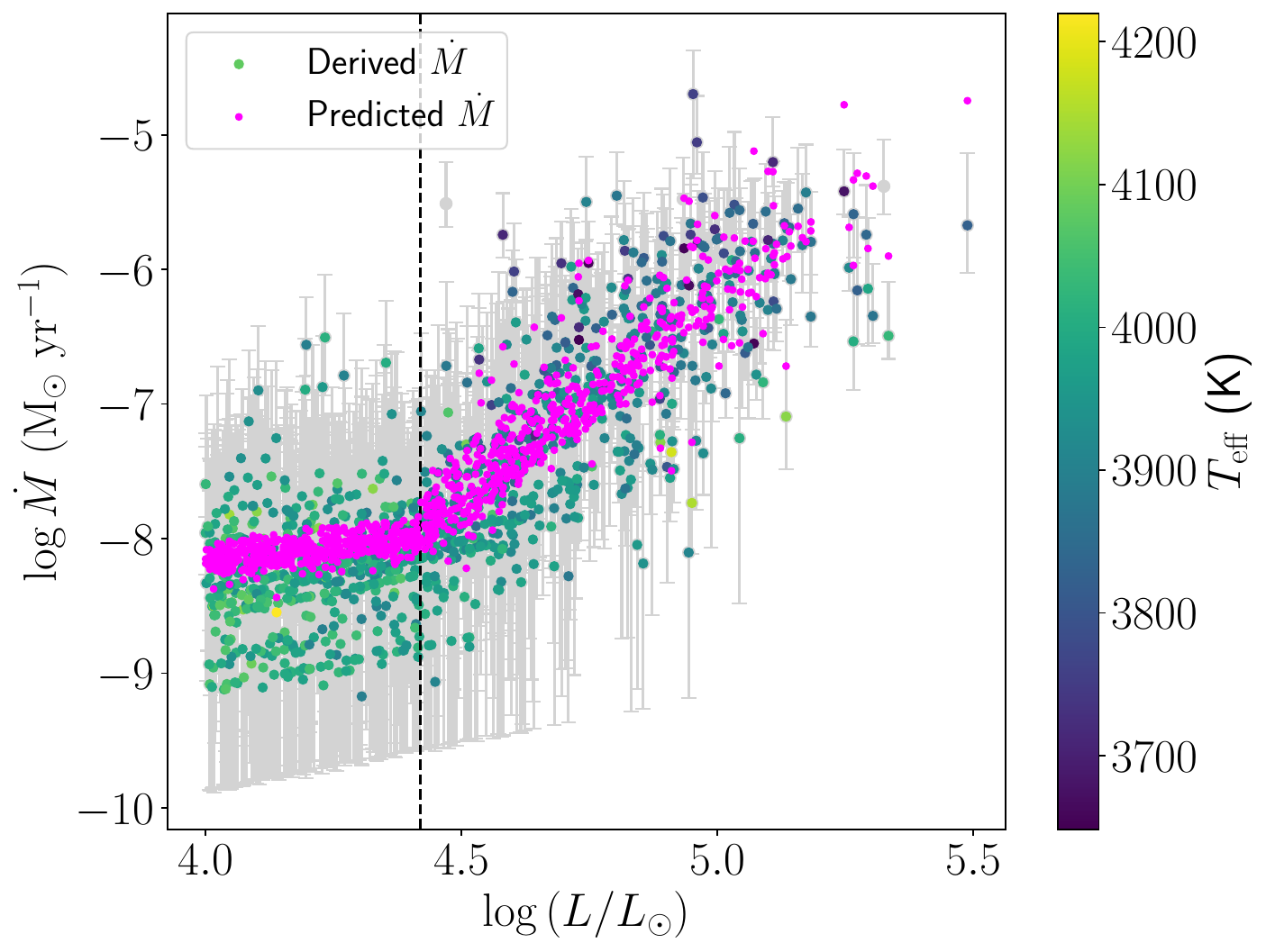}
    \includegraphics[width=0.49\linewidth]{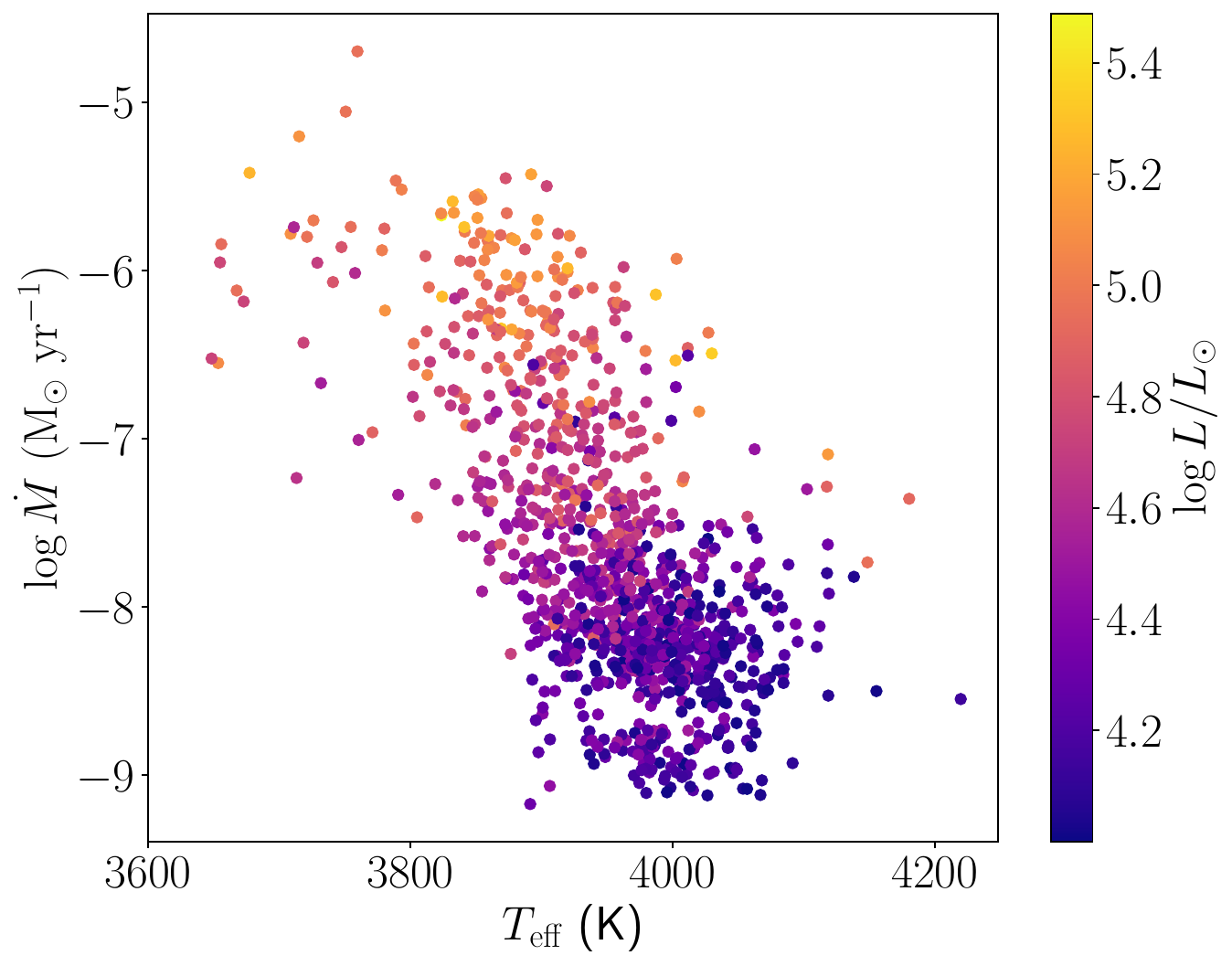}
    \caption{Mass-loss rates vs. luminosity (left) and effective temperature (right) calculated from Eq. (\ref{eq:Teff}). The magenta points correspond to the prediction of the derived $\dot{M}(L, T_\mathrm{eff})$ relation, Eq. (\ref{eq:Mdot}), showing the correlation of the mass-loss rate as a function of the luminosity and effective temperature.}
    \label{fig:Mdot_L_fit}
\end{figure*}

In addition to the luminosity, we examined the dependence of the mass-loss rate on the effective temperature calculated using Eq. (\ref{eq:Teff}) with an equal uncertainty of 140~K for each source \citep{Britavskiy_2019}. The right panel of Fig.~\ref{fig:Mdot_L_fit} shows that these two properties are anti-correlated, which agrees with the findings of \cite{deJager_1988} and \cite{vLoon_2005}. The $T_\mathrm{eff}(J-K_s)$ relation from \citet{Neugent_2012} based on synthetic models yields a similar correlation with $\dot{M}$, which is less steep since it predicts a broader range of $T_\mathrm{eff}$. However, we proceeded with the relation from \citet{Britavskiy_2019} because it is empirically derived rather than from synthetic models \citep[for a comparison between different $T_\mathrm{eff}$ prescriptions see][]{deWit_2024}. We fitted a broken relation dependent on both luminosity and effective temperature,
\begin{equation}
    \log{\dot{M}} = c_1 \log{L} + c_2\log{\left(\frac{T_\mathrm{eff}}{4000}\right)} + c_3 ,\label{eq:Mdot}
\end{equation}
with $\dot{M}$ the mass-loss rate in $M_{\odot} \ \mathrm{yr}^{-1}$, $L$ the luminosity in $L_\odot$, and $T_\mathrm{eff}$ the effective temperature in $\mathrm{K}$. We present the best-fit parameters in \autoref{tab:coef}. We derived the relation at $\log{L/L_\odot}>4$, roughly above the luminosity limit for RSGs in the LMC. The limit could be slightly higher, but regardless, the slope does not change up to the position of the kink. The left panel of Fig.~\ref{fig:Mdot_L_fit} shows the relation we applied to our sources compared with the derived $\dot{M}$.

\begin{table}[h]
    \centering
    \caption{Best-fit parameters of Eq. (\ref{eq:Mdot}).}
    \renewcommand{\arraystretch}{1.2}
    \begin{tabular}{c | c c c}
        \hline\hline
        $\log{L/L_\odot}$ & $c_1$   & $c_2$  & $c_3$ \\
         \hline 
        $<4.4$ & $0.26\pm0.2$  & $-14.19\pm4.48$  & $-9.17\pm0.86$ \\
        $\gtrsim4.4$ & $2.5\pm0.12$  & $-31.78\pm1.62$  & $-17.47\pm0.57$ \\
       \hline
    \end{tabular}
    \
    \label{tab:coef}
\end{table}

\subsection{Steady-state versus radiatively driven winds}

\begin{figure}[h]
    \includegraphics[width=\linewidth]{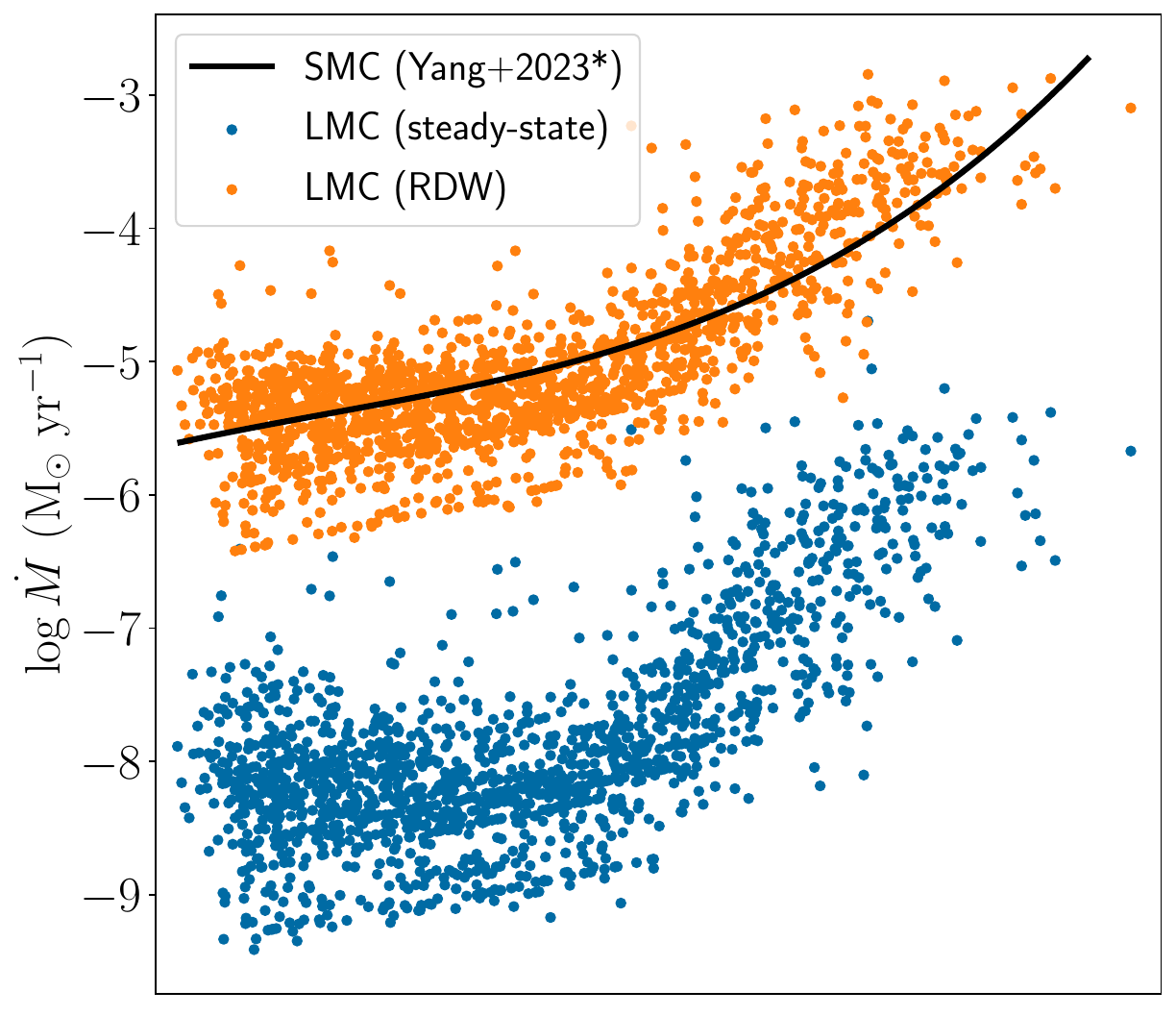}
    \includegraphics[width=\linewidth]{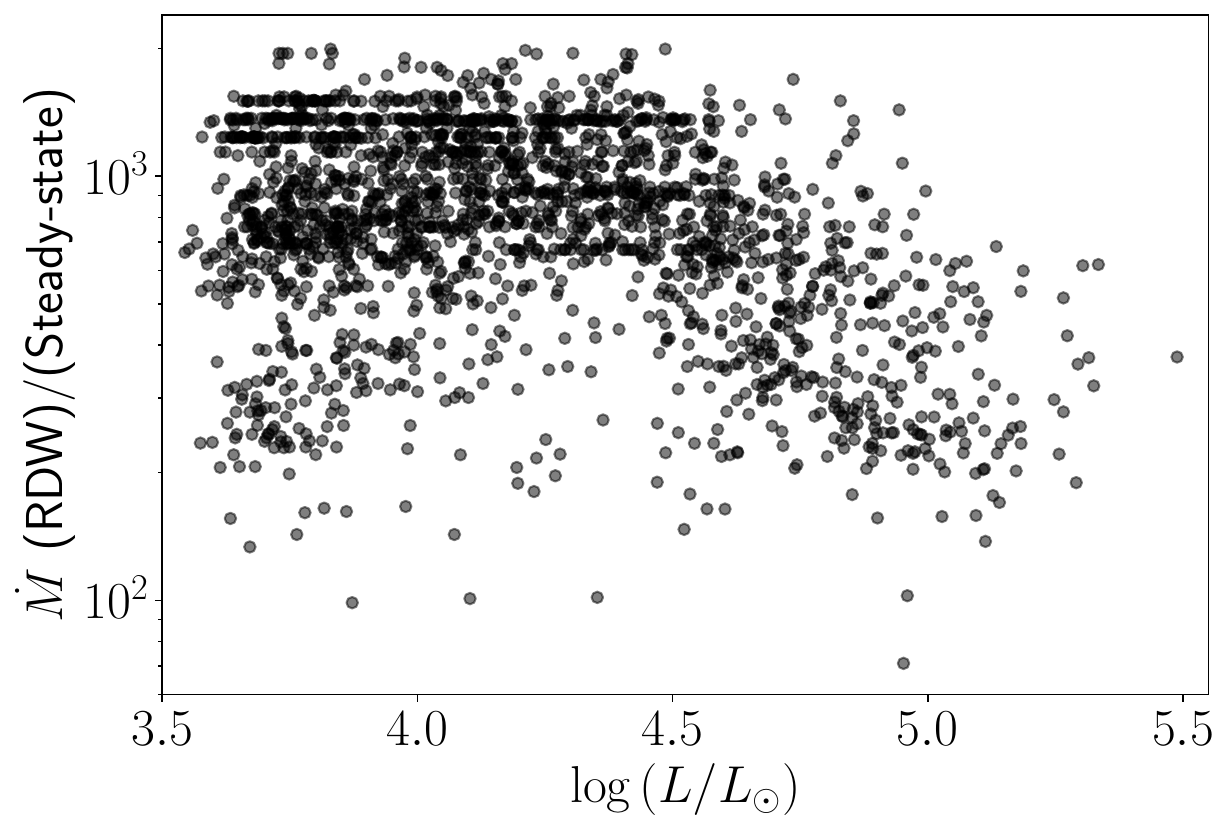}
    \caption{Comparison between the resulting mass-loss rates of steady-state winds and RDW. \textit{Top:} Mass-loss rate vs. luminosity. The results from steady-state winds and RDW are indicated with blue and orange, respectively. The solid line shows the relation by \cite{Yang_2023} in the SMC corrected by adding 0.7 dex. \textit{Bottom:} The ratio of the results of the two modes.}
    \label{fig:SSvsRDW}    
\end{figure}

Previous studies have shown a large discrepancy in their obtained $\dot{M}$ \citep[c.f.][]{vLoon_2005, Goldman_2017, Beasor_2020, Yang_2023}. In Fig.~\ref{fig:SSvsRDW}, we compare the resulting $\dot{M}$ of steady-state winds versus RDW as a function of $L$. The steady-state wind case has a slightly steeper slope than the RDW, mainly due to the different assumptions about the dependence of the outflow velocity on the luminosity. We corrected the $\dot{M}(L)$ relation from \cite{Yang_2023} by adding 0.7 dex to their $\log{\dot{M}}$, and we present it in the same figure with the black line. Their result of the SMC matches ours in the LMC, indicating that RSG winds are metallicity independent, apart from the luminosity limit of the kink (see Sect.~\ref{sec:mlr_discussion}). The bottom panel demonstrates the $\dot{M}$ ratio of the two modes. There is a significant difference in the results of the two assumptions by two to three orders of magnitude. The result from RDW would strip off most of the luminous RSGs and shift them to higher $T_\mathrm{eff}$ in the HR diagram, which is not confirmed by observations (Zapartas et al., in prep.).

\subsection{Grain size MRN [0.01, 1] versus [0.1, 1]} \label{sec:grain_size}

\begin{figure}[h]
    \centering
    \includegraphics[width=\columnwidth]{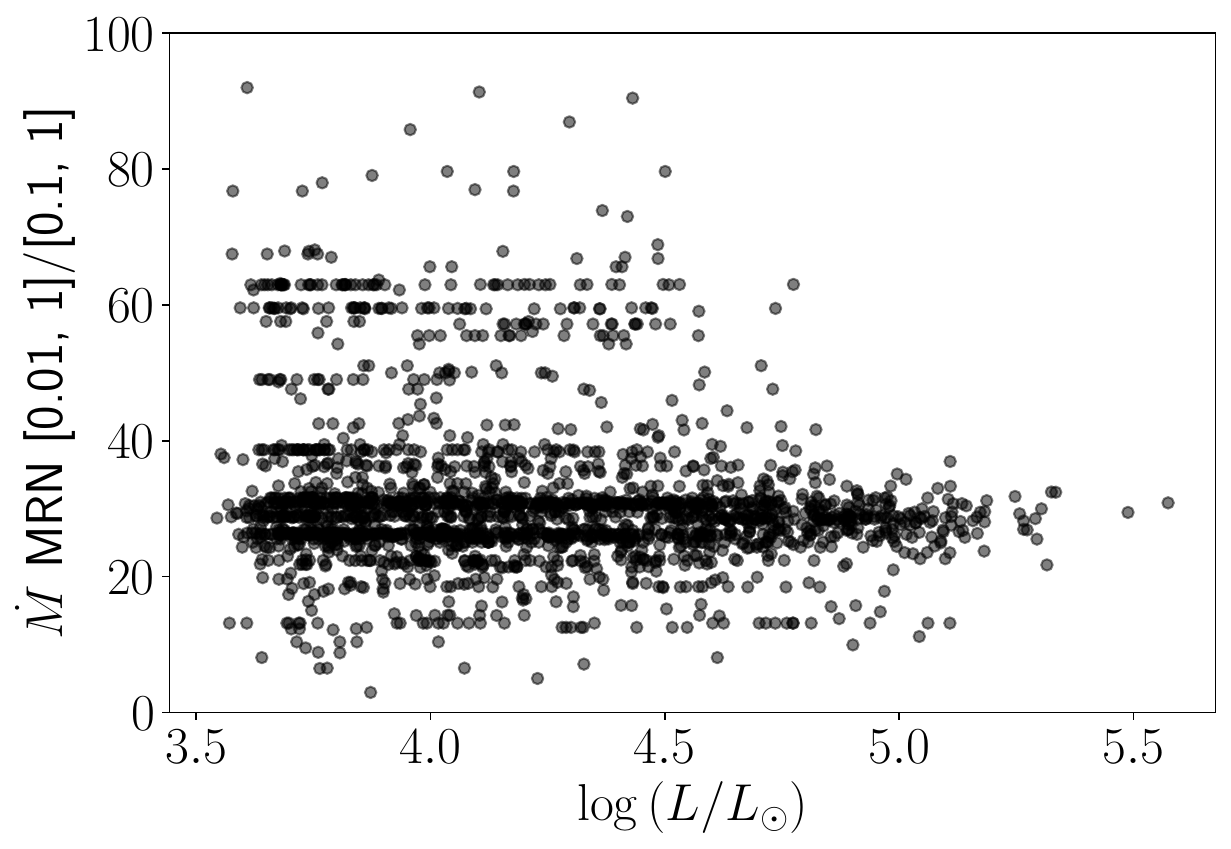}
    \includegraphics[width=\columnwidth]{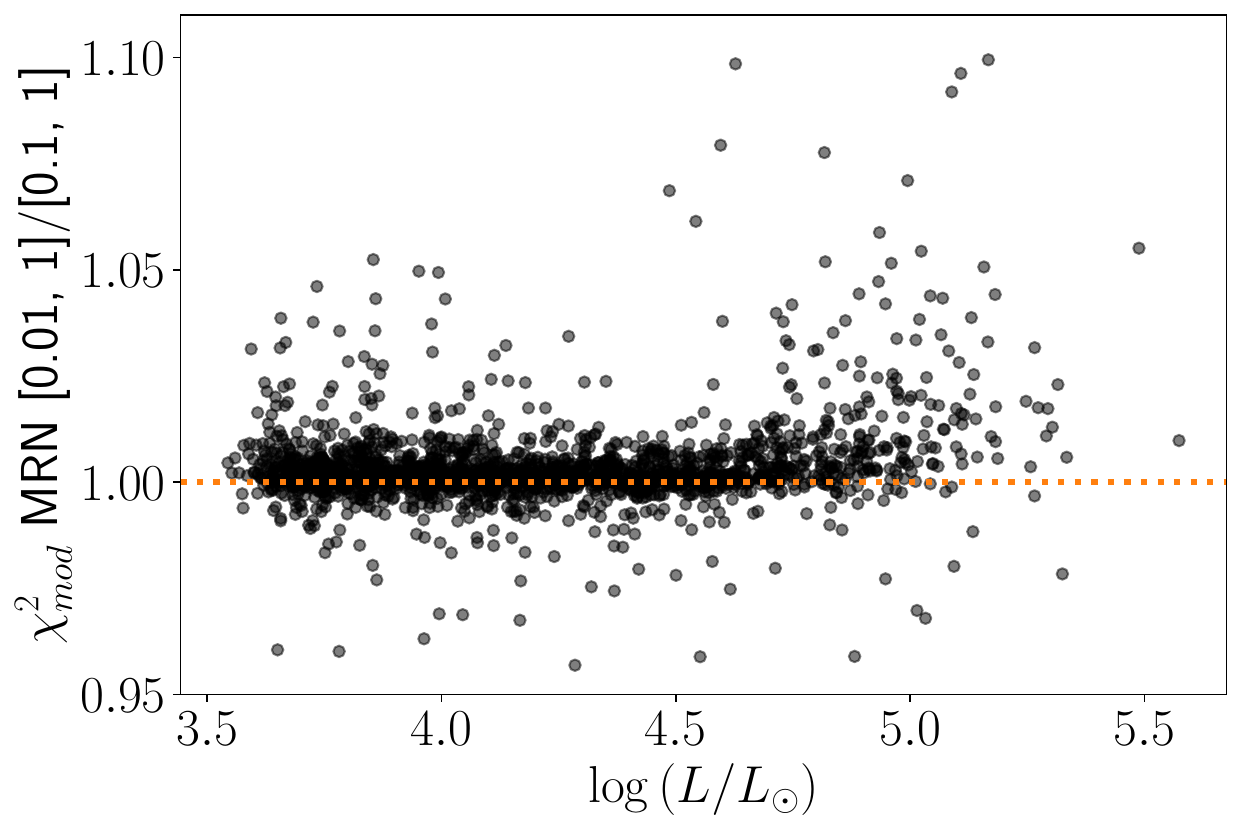}
    \caption{Effect of the grain size distribution on the mass-loss rate for a range of luminosities. \textit{Top:} Ratio of the mass-loss rates assuming an MRN grain size distribution in the range [0.01, 1] $\mu$m over the distribution in the range [0.1, 1] $\mu$m. \textit{Bottom:} Ratio of the minimum $\chi^2_{mod}$ between the two mentioned cases.}
    \label{fig:mrn_ratio}
\end{figure}

We used the modified MRN grain size distribution in the range of [0.01, 1] $\mu$m to compare with our assumption of [0.1, 1] $\mu$m. The lower size distribution has an average grain size of around 0.02 $\mu$m, and the average ratio $a/Q_V$ becomes higher by a factor of $\sim 20$ \citep[see Fig. 4 in][]{Wang_2021}, which linearly affects the $\dot{M}$. The top panel of Fig.~\ref{fig:mrn_ratio} shows the $\dot{M}$ ratio assuming a grain size in the range of [0.01, 1] over the range of [0.1, 1] $\mu$m. This assumption results in a higher $\dot{M}$ by a factor of around 25-30. The bottom panel of Fig.~\ref{fig:mrn_ratio} shows that the larger grain size appears to have slightly better-fit models for the highest luminosity sources. Despite this, the effect of grain sizes on the SED is indistinguishable and can therefore not be obtained with precision. We performed the same comparison using the RDW mode in \texttt{DUSTY}. We found an opposite correlation in this case; the smaller grain size yields an $\dot{M}$ lower by a factor of $\sim0.8$. This may have to do with the different dependence of the $\dot{M}$ from RDW on the extinction efficiency $Q$ \citep[see][]{Elitzur_2001}. We present the derived $\dot{M}(L, T_\mathrm{eff})$ relation for the mentioned assumptions in \autoref{app:rdw_a}.

\label{app:grain_size}


\section{Discussion} \label{sec:disc}

The most commonly used RSG $\dot{M}$ prescription is from \cite{deJager_1988}. Despite their limited sample size of 15 stars, it provides satisfactory results at solar metallicity \citep{Mauron_2011}. We derived an $\dot{M}(L, T_\mathrm{eff})$ relation with a similar correlation with luminosity and effective temperature as \cite{deJager_1988}. The anti-correlation with $T_\mathrm{eff}$ is expected because the RSGs expand their envelope as they evolve up the Hayashi track, where they steeply increase in luminosity and slightly drop in $T_\mathrm{eff}$.

It is noteworthy that we found a knee-like shape or kink of the mass-loss rate as a function of the luminosity, as presented in the previous section. Modelling an equally large, unbiased sample, \cite{Yang_2023} demonstrated that a similar turning point is correlated with the variability. Studying a sample of 42 RSGs, \citet{Humphreys_2020} found a transition in the $\dot{M}(L)$ relation at approximately $\log{(L/L_\odot)}=5$ and attributed it to instabilities in the atmosphere of evolved RSGs. Thus, the kink may indicate a change in the physical mechanism. \autoref{fig:mad} shows that the kink strongly correlates with the intrinsic photometric variability of the RSG. We could hypothesise that the variability is due to pulsations of the RSG in the stellar atmosphere, expelling gas to altitudes at which molecules can condense and form dust. In addition to this, increased radiation pressure on the dust shell could arise at a specific luminosity limit \citep{Vink_2023}. If turbulence is the dominant mass-loss mechanism \citep{Kee_2021}, measurements of the turbulent velocity versus the luminosity could provide better insight into the instabilities of the stellar atmosphere. It might be argued that sources with lower $L$ than the position of the kink are not RSGs and that these stars may be bound to a different physical driving mechanism. However, Fig.~\ref{fig:Mdot_L} demonstrates that there are spectroscopically confirmed RSGs below that point.

We observe a high dispersion in the mass-loss rates (see Fig.\ \ref{fig:Mdot_L}), which could be due to variability, or it could arise from different stellar parameters that affect the $\dot{M}$, such as different $T_\mathrm{eff}$. We showed in Fig.~\ref{fig:Mdot_L_fit} that $T_\mathrm{eff}$ partially compensates for the dispersion. In addition, different inner dust shell temperatures, which define the distance of the dust shell from the star, could contribute to it.  A lower dust temperature corresponds to higher $R_\mathrm{in}$ resulting in higher $\dot{M}$ (see Eq.~\ref{eq:dotM}). Moreover, \cite{Beasor_2020} found a correlation with the initial mass, as they targeted coeval RSGs in clusters. Different turbulent or outflow speeds could also affect this dispersion \citep[e.g.][]{Kee_2021}. Finally, the grain size and gas-to-dust ratio, which we have assumed to be uniform for our whole sample, can vary between RSGs, which might affect the $\dot{M}$ spread.

\subsection{Red helium-burning stars} \label{sec:RHeB}
At $\log(L/L_\odot)< 4.2$, we find enhanced variability for a fraction of sources ($\sim6\%$), mainly towards the lower luminosities of the sample. These sources are likely red helium-burning (RHeB) stars with masses $<8 \ M_{\odot}$. These stars are at the evolutionary point after the red giant phase and before becoming AGB stars when they perform a loop in the HR diagram on the horizontal branch. They undergo pulsations \citep{Kippenhahn_2013}, which could contribute to the observed variability. Although not certain, the wind mechanism could follow the RDW theory, similarly to the AGB phase. This potentially explains why our steady-wind assumption creates a high dispersion in the $\dot{M}$ results, in contrast to the RDW case, where they are more uniform (see Fig.~\ref{fig:SSvsRDW}). Finally, as mentioned previously, there may still be AGB star contamination despite the constraints applied to the initial sample.

\subsection{Luminous outliers in the $\dot{M}(L)$ relation}

We searched the literature for the sources that appear as outliers in Fig.~\ref{fig:Mdot_L}: \textit{i)} 5 sources with $\tau_V>2$ and $\log(L/L_\odot)>4.4$, and \textit{ii)} 11 sources with $\log(L/L_\odot)>5.2$. \textit{i)} ID 807 and ID 1348 are classified by photometry as long-period variables (LPVs) and O-rich AGB stars \citep{Soszynski_2009}, and ID 1139 is an M-type giant by spectroscopy \citep{deWit_2023}. The other 2 sources are M-type supergiants \citep{Wood_1983, Gonzalez_2015}. More specifically, we found that ID 2107, W61 7-8, with spectral type M3.5Ia \citep{Gonzalez_2015}, is a dust-enshrouded RSG according to criteria from \cite{Beasor_2022} with $K_s-W3=3.68$ mag. We did not have more RSGs with $K_s-W3>3.4$ mag, mainly because the CMD selection criteria from \cite{Neugent_2012} and \cite{Ren_2021} remove sources with high $J-K_s$ values, which rejected dust-enshrouded RSGs. \textit{ii)} Three of them (IDs 408, 1829, and 1866) do not have any classification, 2 are classified as M-type supergiants by photometry \citep{Westerlund_1981}, and the rest are classified as M-type supergiants by spectroscopy \citep{Wood_1983, Reid_1985, vLoon_2005, Gonzalez_2015, deWit_2023}. Five of these are binary candidates, as we indicate in Fig.~\ref{fig:mdot_binaries}. Finally, ID 1786, [W60] B90, is an extreme RSG, with evidence of a bow shock, undergoing episodic mass loss (Munoz-Sanchez, in prep.).

\subsection{Comparison with other works: Zoo of mass-loss rate prescriptions}
\label{sec:mlr_discussion}

There is a plethora of $\dot{M}$ prescriptions for RSGs with different assumptions and results. Our sample of RSGs with derived $\dot{M}$ is one of the largest, among others \citep{Wang_2021, Yang_2023, Wen_2024}. In \autoref{tab:mdot}, we present the prescriptions from the literature to which we compared our work. In addition to luminosity, some relations depend on the effective temperature, the stellar radius, the initial or current mass, the pulsation period $P$, or the gas-to-dust ratio. All are empirical relations, except for the analytic solution by \cite{Kee_2021}, who assumed that atmospheric turbulence is the dominant driving mechanism for mass loss and that it strongly depends on the turbulent speed, $v_\mathrm{turb}$. The luminosity $L$ and the mass $M$ are in solar units, and the temperature $T_\mathrm{eff}$ is in K. To compare them properly, we applied the average $T_\mathrm{eff}$ of our sample to all prescriptions and their suggested optimal parameters for the rest.

\begin{figure}[h]
    \centering
    \includegraphics[width=\columnwidth]{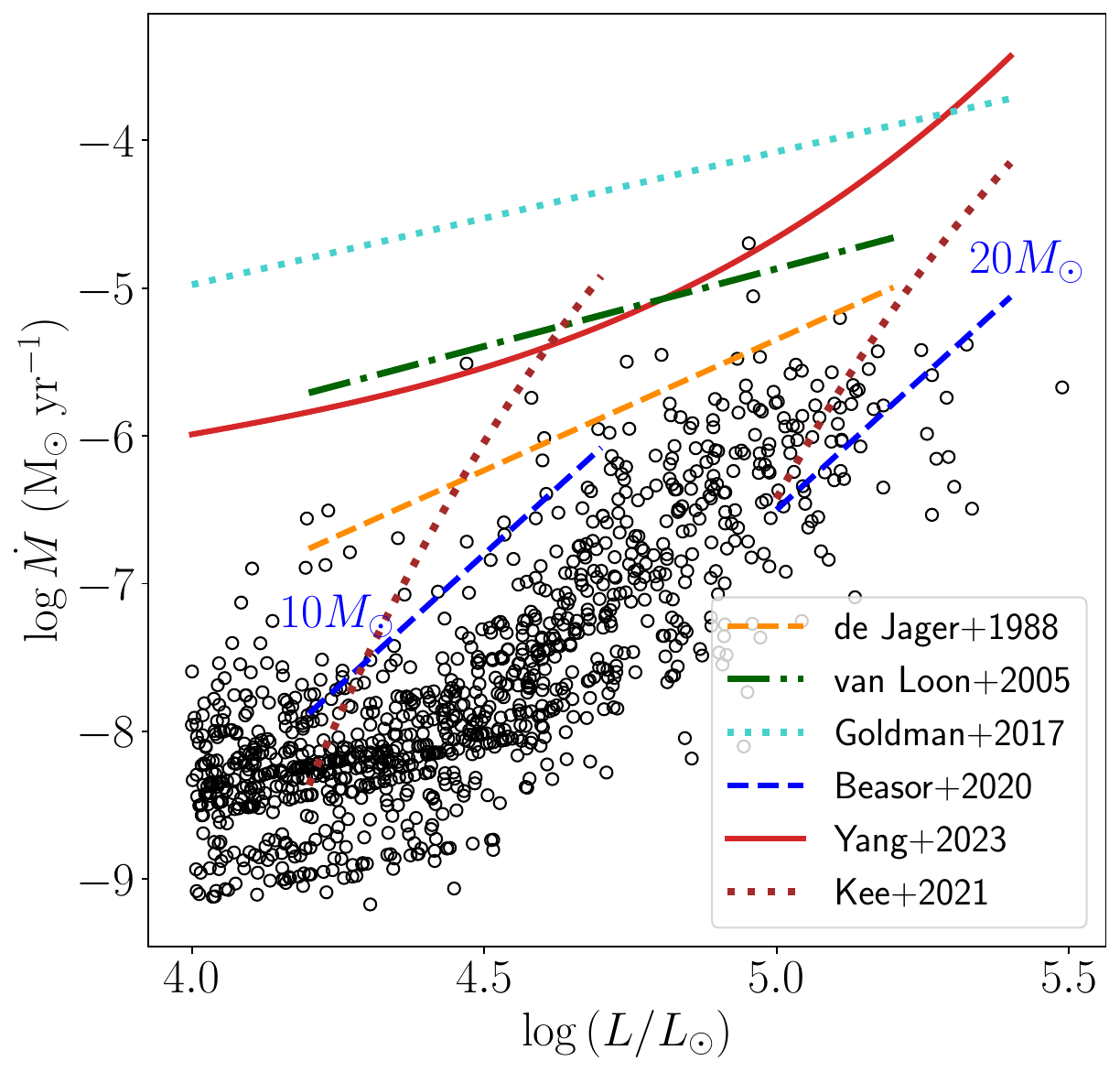}
    \caption{Comparison of our $\dot{M}(L)$ result (open circles) with prescriptions from previous works presented in \autoref{tab:mdot}. The two lines from \cite{Beasor_2020} correspond to an initial mass of 10 $M_{\odot}$ and 20 $M_{\odot}$, and the same holds for \cite{Kee_2021}, but represents the current mass.}
    \label{fig:Mdot_comp}
\end{figure}

In Fig.~\ref{fig:Mdot_comp} we compare the different relations with our results. Many uncertainties arise between different studies, for instance different samples, photometric bands used, metallicity, radiative-transport models, assumptions on the mass-loss mechanism, grain size, and gas-to-dust ratio. Our results agree better with \cite{Beasor_2023}, who used similar assumptions, although they had a smaller sample and luminosity range. The differences with \citep{Beasor_2023} lie within the errors and slight variations of the assumptions, such as the gas-to-dust ratio, the grain size and the outflow speed. The \cite{deJager_1988} relation overestimates the $\dot{M}$ at lower luminosities and is closer to ours at $\log{(L/L_\odot)}>4.8$, but less steep. \cite{Humphreys_2020} found slightly lower $\dot{M}$ than \cite{deJager_1988} at $4<\log{(L/L_\odot)}<5$, with an enhancement in the mass loss at $\log{(L/L_\odot)}>5$. \cite{Yang_2023}, \cite{vLoon_2005}, and especially \cite{Goldman_2017} found much higher $\dot{M}$ due to the RDW assumption in \texttt{DUSTY}, which results in two to three orders of magnitude higher $\dot{M}$ (see Fig.~\ref{fig:SSvsRDW}). More specifically, \cite{vLoon_2005} used the RDW analytic mode on \texttt{DUSTY}, which results in lower values than the fully numerical solution of RDW, as we verified. However, the analytic approximation is only applicable in the case of negligible drift \citep{Elitzur_2001}. \cite{Goldman_2017} and \cite{vLoon_2005} studied dust-enshrouded RSGs, and \cite{Goldman_2017} included AGB stars, which are expected to have high $\dot{M}$. Almost none of these studies found a kink in their results, mainly because their samples were small and biased towards high luminosities.

We found that the kink is located at $\log(L/L_\odot)\simeq4.4$, whereas in the SMC, it was found at $\log(L/L_\odot)=4.6$ \citep{Yang_2023}. We speculate that this luminosity shift is related to metallicity since the SMC is a lower-metallicity environment with $[Z]_\mathrm{SMC}=-0.53$ \citep{Davies_2015}. Metallicity could affect this physical procedure if the enhanced mass loss were due to instabilities in the atmosphere of RSGs. A lower $Z$ results in more compact stars during the main sequence, meaning a more massive core for the same initial mass and a higher $L$ for RSGs. This shifts the luminosity limit for RSGs and could explain the shift of the kink. In addition, lower $Z$ stars have weaker radiatively driven winds during the main sequence \citep{Vink_2001}, slowing down their rotation at a slower rate. The higher rotation results in enhanced mixing and a larger core \citep{Meynet_2006}, which again produces a more luminous RSG for the same initial mass.

Finally, our result agrees with the determined mass-loss rates of four RSGs from the observations of CO rotational line emission \citep{Decin_2023}, which provides confidence in our assumptions. The theoretical relation by \cite{Kee_2021} predicts higher $\dot{M}$ for low masses and has a similar slope as ours at high $L$ (see Fig.\ \ref{fig:Mdot_comp}). A further investigation of the $\dot{M}$ with observational and theoretical methods and robust samples could provide more solid support for the existing empirical relations.

\cite{Wen_2024} recently published a similar study of RSGs in the LMC. They used two-dimensional models from 2-DUST \citep{Ueta_2003} and steady-state winds, finding similar mass-loss rates. They also indicated the same position of the kink in the $\dot{M}-L$ relation. It is reassuring that two independent methods provide similar results. However, there are a few notable differences. Firstly, they only used the RSG catalogue from \citet{Ren_2021} and did not remove foreground sources existing in that catalogue. Secondly, they fit sources without photometry at $\lambda>8 \ \mu$m with low optical depth models, whereas we rejected them. Their empirical relation is therefore based on some erroneous mass-loss rate measurements. Thirdly, a significant difference is that we provide an $\dot{M}$ prescription dependent on both $L$ and $T_\mathrm{eff}$, explaining a large part of the dispersion in the results. Some other minor differences are that they used $r_\mathrm{gd}=500$ from \cite{Riebel_2012}, who used this abstract value without deriving it, but it is within our estimated errors, and $v_\infty=10$ km s$^{-1}$, which could be low for some extreme RSGs with $v_\infty=20-30$ km s$^{-1}$, as we discussed. Finally, they considered that sources with $\log(L/L_\odot)<4.5$ and high $\dot{M}$ are pseudo sources or are affected by nearby sources. This may be valid for some cases, but we showed that some of these have increased variability, indicating another type of star (e.g. RHeB giant). In summary, we provide a better-established $\dot{M}(L, T_\mathrm{eff})$ relation based on a cleaner RSG catalogue and an explanation for the scatter at both low $L$ and high $L$ sources.

\begin{table*}
    \centering
    \caption{Mass-loss rate prescriptions from different works.}
    \renewcommand{\arraystretch}{1.3}
    \begin{tabular}{l l l l}
        \hline\hline
       $\dot{M}$ relation $(\rm M_{\odot} \ \mathrm{yr}^{-1})$ & DUSTY mode & Reference & Sample \\
         \hline \\[-0.3cm]
       $\dot{M} = 10^{-8.158}L^{1.769}T_\mathrm{eff}^{1.676}$             &     -   &  \cite{deJager_1988} & Milky Way \\[0.1cm]
       $\dot{M}=10^{-5.56}\left(\frac{L}{10^4}\right)^{1.05}\left(\frac{T_{\text {eff }}}{3500}\right)^{-6.3}$  & RDW Analytic   &  \cite{vLoon_2005} & LMC \\[0.1cm]
       $\dot{M}=1.06 \times 10^{-5}\left(\frac{L}{10^4}\right)^{0.9}\left(\frac{P}{500 d}\right)^{0.75}\left(\frac{r_\mathrm{gd}}{200}\right)^{-0.03}$  & RDW   &  \cite{Goldman_2017} & LMC \\[0.1cm]
       Analytic solution (see reference)                                  &   -  &  \cite{Kee_2021} & - \\[0.1cm]
       $\log\dot{M} = -21.5-0.15M_\mathrm{ini}+3.6\log L$                 & Steady-state, $r^{-2}$   &  \cite{Beasor_2023} & Galactic clusters \\[0.1cm]
       $\log\dot{M} = 0.45(\log L)^3-5.26(\log L)^2+20.93\log L-34.56$    & RDW   &  \cite{Yang_2023} & SMC \\[0.1cm]
       $\log{\dot{M}} = 0.26 \log{L} - 14.19\log{\left(\frac{T_\mathrm{eff}}{4000}\right)} - 9.17, \ \log{L}<4.4$ & \multirow{2}{*}{Steady-state, $r^{-2}$}  & \multirow{2}{*}{This work} & \multirow{2}{*}{LMC} \\ [0.1cm]
       $\log{\dot{M}} = 2.5 \log{L} - 31.78\log{\left(\frac{T_\mathrm{eff}}{4000}\right)} - 17.47, \ \log{L}\gtrsim 4.4$ & & & \\

       \hline \\     
    \end{tabular}
    \label{tab:mdot}
\end{table*}

\subsection{Caveats}

Several caveats arise from \texttt{DUSTY} and the assumptions we made in our RSG sample. Deriving the mass loss from the circumstellar dust of RSGs, appearing in their SED as excess at $\lambda>3.6 \ \mu$m, constitutes a snapshot and has uncertainties. It neglects that RSGs can undergo episodic mass loss, forming a thick shell. However, these are rare events that occur in the late stages of RSGs. Furthermore, the gas-to-dust ratio is not uniform in every region of a galaxy. More luminous RSGs may have higher $r_\mathrm{gd}$ \citep{Goldman_2017}. However, it varies by less than an order of magnitude in the LMC, which would not affect the estimated $\dot{M}$ significantly, in contrast to the case of the SMC, in which the gas-to-dust ratio varies by a factor of $\sim20$ \citep{Clark_2023}.

The assumption of a spherically symmetric dust shell is not realistic. Dust around RSGs is clumped and asymmetric \citep{Smith_2001, Scicluna_2015}. However, clumping has a minor effect on the $\dot{M}$ \citep{Beasor_2016}. Luminosity measurements could be affected in the case of dense shells, resulting in higher or lower $L$, depending on the orientation of the clumps. This occurred in WOH G64, the luminosity of which is higher when integrating the SED than what \cite{Ohnaka_2008} inferred by modelling the dusty torus.

Binarity would affect the derived mass loss under the assumptions we made. The hotter OB companion could alter the shape of the dust shell or even heat and destroy part of the dust with its radiation field. We found that the impact of binaries on our derived $\dot{M}(L, T_\mathrm{eff})$ relation was negligible. We show this in \autoref{app:binaries}, and we provide a more conservative $\dot{M}(L, T_\mathrm{eff})$ relation by fitting only the probable single RSGs. We should mention that this fraction represents the current binary RSGs and does not exclude the binary history of the probable single RSGs with accretion of mass or merging \citep[e.g.][]{Zapartas_2019}. We observed that most of the high $L$ binary candidates appear to have lower $\dot{M}$ than the probable single ones for the same $L$ (see Fig.~\ref{fig:mdot_binaries}). This means that their SED exhibits a lower IR excess, which could originate from the destruction of dust by the OB companion. Finally, due to their UV excess, their derived $L$ is slightly higher, and they appear more to the right in the $\dot{M}-L$ plot than if they were single stars.

\section{Summary and conclusion} \label{sec:summary}

We have assembled and analysed a large, clean, and unbiased sample of 2,219 RSG candidates in the LMC, 1,376 of which have $\log(L/L_\odot)>4$. We used photometry in 49 filters ranging from the UV to the mid-IR to construct SEDs, from which we measured luminosities and effective temperatures, and we derived the mass-loss rates for these sources using the radiative code \texttt{DUSTY}. The average mass-loss rate is $6.4\times10^{-7} \ M_{\odot}$ yr$^{-1}$, while for $\log{(L/L_\odot)}>4$, it is $9.3\times10^{-7} \ M_{\odot}$ yr$^{-1}$. The latter corresponds to a mass of $\sim 0.1-1 \ M_{\odot}$ lost throughout the lifetime of an RSG, presumed to be $10^5-10^6$ yr, and to an average dust-production rate of $\sim3.6\times10^{-9} \ M_{\odot}$ yr$^{-1}$.

We presented an accurate $\dot{M}$ relation as a function of both $L$ and $T_\mathrm{eff}$, finding it to be proportional to $L$ and inversely proportional to the $T_\mathrm{eff}$. We found a turning point or kink in the $\dot{M}-L$ relation similar to the one in the SMC \citep{Yang_2023}, but at lower luminosity, $\log(L/L_\odot)\simeq4.4$, as in \cite{Wen_2024}. The kink indicates an enhancement of mass loss beyond this luminosity limit, and we showed that this tightly correlates with variability. This indicates a physical mechanism that starts to dominate the RSG wind beyond this limit. Furthermore, the varying location of the kink between the SMC and LMC suggests that this enhancement depends on the metallicity.

We investigated the effect of the assumptions made in \texttt{DUSTY} on the resulting mass-loss rates. We found that the density distribution of the wind profile affects the RSG mass-loss rates most and that the RDW assumption causes the discrepancy of two to three orders of magnitude in the RSG mass-loss rate measurements in the literature. The RDW assumption produces unrealistically high values for quiescent mass loss in RSGs, whereas the assumption of a steady-state wind yields results that are confirmed by independent methods (e.g. measurement of gas), which have been measured for a small number of RSGs. We derived mass-loss rates for our sample using RDW that closely match the SMC rates measured by \citep{Yang_2023} under this assumption. This indicates that apart from the position of the kink, RSG mass-loss rates can be independent of metallicity. Additionally, we tested the effect of the grain size in the case of a steady-state wind and found that a grain size of $a<0.1\ \mu$m increases $\dot{M}$ up to a factor of 25-30. We found a binarity fraction of 21\% based on UV excess and showed that when these binary RSG are included, the derived $\dot{M}$ relation is not significantly affected.

The current investigation provided an improved method and a well-established $\dot{M}(L, T_\mathrm{eff})$ relation compared to previous works. Repeating this exercise for the SMC, M31 and M33 following the same method and assumptions will confirm whether mass loss in RSGs depends on the metallicity. Further research on the theoretical derivation of the $\dot{M}-L$ relation, as in \cite{Kee_2021}, and future measurements of gas emission lines \citep[e.g. CO;][]{Decin_2023} provide independent ways of confirming the estimated mass-loss rates. 


\begin{acknowledgements}
KA, AZB, SdW, EZ, GMS, and GM acknowledge funding support from the European Research Council (ERC) under the European Union’s Horizon 2020 research and innovation program (Grant agreement No. 772086. EZ also acknowledges support from the Hellenic Foundation for Research and Innovation (H.F.R.I.) under the ``3rd Call for H.F.R.I. Research Projects to Support Post-Doctoral Researchers'' (Project No: 7933). KA thanks Ming Yang for helping with the use of \texttt{DUSTY} and Emma Beasor for valuable discussions regarding \texttt{DUSTY} and the RSG wind properties. KA would also like to thank the participants of the VLT-FLAMES Tarantula Survey (VFTS) meeting in 2023 who provided useful questions and comments on our preliminary results.
\end{acknowledgements}

\bibliographystyle{aa}
\bibliography{ref}

\begin{thebibliography}{105}
\expandafter\ifx\csname natexlab\endcsname\relax\def\natexlab#1{#1}\fi

\bibitem[{{Arroyo-Torres} {et~al.}(2015){Arroyo-Torres}, {Wittkowski}, {Chiavassa}, {Scholz}, {Freytag}, {Marcaide}, {Hauschildt}, {Wood}, \& {Abellan}}]{Arroyo-Torres_2015}
{Arroyo-Torres}, B., {Wittkowski}, M., {Chiavassa}, A., {et~al.} 2015, \aap, 575, A50

\bibitem[{{Beasor} \& {Davies}(2016)}]{Beasor_2016}
{Beasor}, E.~R. \& {Davies}, B. 2016, \mnras, 463, 1269

\bibitem[{{Beasor} {et~al.}(2021){Beasor}, {Davies}, {Smith}, {Gehrz}, \& {Figer}}]{Beasor_2021}
{Beasor}, E.~R., {Davies}, B., {Smith}, N., {Gehrz}, R.~D., \& {Figer}, D.~F. 2021, \apj, 912, 16

\bibitem[{{Beasor} {et~al.}(2020){Beasor}, {Davies}, {Smith}, {van Loon}, {Gehrz}, \& {Figer}}]{Beasor_2020}
{Beasor}, E.~R., {Davies}, B., {Smith}, N., {et~al.} 2020, \mnras, 492, 5994

\bibitem[{{Beasor} {et~al.}(2023){Beasor}, {Davies}, {Smith}, {van Loon}, {Gehrz}, \& {Figer}}]{Beasor_2023}
{Beasor}, E.~R., {Davies}, B., {Smith}, N., {et~al.} 2023, \mnras, 524, 2460

\bibitem[{{Beasor} \& {Smith}(2022)}]{Beasor_2022}
{Beasor}, E.~R. \& {Smith}, N. 2022, \apj, 933, 41

\bibitem[{{Bianchi} {et~al.}(2017){Bianchi}, {Shiao}, \& {Thilker}}]{galex}
{Bianchi}, L., {Shiao}, B., \& {Thilker}, D. 2017, \apjs, 230, 24

\bibitem[{{Boyer} {et~al.}(2011){Boyer}, {Srinivasan}, {van Loon}, {McDonald}, {Meixner}, {Zaritsky}, {Gordon}, {Kemper}, {Babler}, {Block}, {Bracker}, {Engelbracht}, {Hora}, {Indebetouw}, {Meade}, {Misselt}, {Robitaille}, {Sewi{\l}o}, {Shiao}, \& {Whitney}}]{sage_lmc}
{Boyer}, M.~L., {Srinivasan}, S., {van Loon}, J.~T., {et~al.} 2011, \aj, 142, 103

\bibitem[{{Britavskiy} {et~al.}(2019){Britavskiy}, {Lennon}, {Patrick}, {Evans}, {Herrero}, {Langer}, {van Loon}, {Clark}, {Schneider}, {Almeida}, {Sana}, {de Koter}, \& {Taylor}}]{Britavskiy_2019}
{Britavskiy}, N., {Lennon}, D.~J., {Patrick}, L.~R., {et~al.} 2019, \aap, 624, A128

\bibitem[{{Cioni} {et~al.}(2011){Cioni}, {Clementini}, {Girardi}, {Guandalini}, {Gullieuszik}, {Miszalski}, {Moretti}, {Ripepi}, {Rubele}, {Bagheri}, {Bekki}, {Cross}, {de Blok}, {de Grijs}, {Emerson}, {Evans}, {Gibson}, {Gonzales-Solares}, {Groenewegen}, {Irwin}, {Ivanov}, {Lewis}, {Marconi}, {Marquette}, {Mastropietro}, {Moore}, {Napiwotzki}, {Naylor}, {Oliveira}, {Read}, {Sutorius}, {van Loon}, {Wilkinson}, \& {Wood}}]{vmc}
{Cioni}, M. R.~L., {Clementini}, G., {Girardi}, L., {et~al.} 2011, \aap, 527, A116

\bibitem[{{Clark} {et~al.}(2023){Clark}, {Roman-Duval}, {Gordon}, {Bot}, {Smith}, \& {Hagen}}]{Clark_2023}
{Clark}, C. J.~R., {Roman-Duval}, J.~C., {Gordon}, K.~D., {et~al.} 2023, \apj, 946, 42

\bibitem[{{Cutri} {et~al.}(2021){Cutri}, {Wright}, {Conrow}, {Fowler}, {Eisenhardt}, {Grillmair}, {Kirkpatrick}, {Masci}, {McCallon}, {Wheelock}, {Fajardo-Acosta}, {Yan}, {Benford}, {Harbut}, {Jarrett}, {Lake}, {Leisawitz}, {Ressler}, {Stanford}, {Tsai}, {Liu}, {Helou}, {Mainzer}, {Gettngs}, {Gonzalez}, {Hoffman}, {Marsh}, {Padgett}, {Skrutskie}, {Beck}, {Papin}, \& {Wittman}}]{allwise}
{Cutri}, R.~M., {Wright}, E.~L., {Conrow}, T., {et~al.} 2021, VizieR Online Data Catalog, II/328

\bibitem[{{Davies} \& {Beasor}(2020)}]{Davies_2020}
{Davies}, B. \& {Beasor}, E.~R. 2020, \mnras, 493, 468

\bibitem[{{Davies} {et~al.}(2015){Davies}, {Kudritzki}, {Gazak}, {Plez}, {Bergemann}, {Evans}, \& {Patrick}}]{Davies_2015}
{Davies}, B., {Kudritzki}, R.-P., {Gazak}, Z., {et~al.} 2015, \apj, 806, 21

\bibitem[{{Davies} {et~al.}(2013){Davies}, {Kudritzki}, {Plez}, {Trager}, {Lan{\c{c}}on}, {Gazak}, {Bergemann}, {Evans}, \& {Chiavassa}}]{Davies_2013}
{Davies}, B., {Kudritzki}, R.-P., {Plez}, B., {et~al.} 2013, \apj, 767, 3

\bibitem[{{de Jager} {et~al.}(1988){de Jager}, {Nieuwenhuijzen}, \& {van der Hucht}}]{deJager_1988}
{de Jager}, C., {Nieuwenhuijzen}, H., \& {van der Hucht}, K.~A. 1988, \aaps, 72, 259

\bibitem[{{de Wit} {et~al.}(2024){de Wit}, {Bonanos}, {Antoniadis}, {Zapartas}, {Ruiz}, {Britavskiy}, {Christodoulou}, {De}, {Maravelias}, {Munoz-Sanchez}, \& {Tsopela}}]{deWit_2024}
{de Wit}, S., {Bonanos}, A.~Z., {Antoniadis}, K., {et~al.} 2024, arXiv e-prints, arXiv:2402.12442

\bibitem[{{de Wit} {et~al.}(2023){de Wit}, {Bonanos}, {Tramper}, {Yang}, {Maravelias}, {Boutsia}, {Britavskiy}, \& {Zapartas}}]{deWit_2023}
{de Wit}, S., {Bonanos}, A.~Z., {Tramper}, F., {et~al.} 2023, \aap, 669, A86

\bibitem[{{Decin} {et~al.}(2006){Decin}, {Hony}, {de Koter}, {Justtanont}, {Tielens}, \& {Waters}}]{Decin_2006}
{Decin}, L., {Hony}, S., {de Koter}, A., {et~al.} 2006, \aap, 456, 549

\bibitem[{{Decin} {et~al.}(2024){Decin}, {Richards}, {Marchant}, \& {Sana}}]{Decin_2023}
{Decin}, L., {Richards}, A.~M.~S., {Marchant}, P., \& {Sana}, H. 2024, \aap, 681, A17

\bibitem[{{Draine} \& {Lee}(1984)}]{DraineLee_1984}
{Draine}, B.~T. \& {Lee}, H.~M. 1984, \apj, 285, 89

\bibitem[{{Dupree} {et~al.}(2022){Dupree}, {Strassmeier}, {Calderwood}, {Granzer}, {Weber}, {Kravchenko}, {Matthews}, {Montarg{\`e}s}, {Tappin}, \& {Thompson}}]{Dupree_2022}
{Dupree}, A.~K., {Strassmeier}, K.~G., {Calderwood}, T., {et~al.} 2022, \apj, 936, 18

\bibitem[{{Eldridge} {et~al.}(2013){Eldridge}, {Fraser}, {Smartt}, {et~al.}}]{Eldridge_2013}
{Eldridge}, J.~J., {Fraser}, M., {Smartt}, S.~J., {et~al.} 2013, \mnras, 436, 774

\bibitem[{{Elitzur} \& {Ivezi{\'c}}(2001)}]{Elitzur_2001}
{Elitzur}, M. \& {Ivezi{\'c}}, {\v{Z}}. 2001, \mnras, 327, 403

\bibitem[{{Gaia Collaboration} {et~al.}(2016){Gaia Collaboration}, {Prusti}, {de Bruijne}, {Brown}, {Vallenari}, {Babusiaux}, {Bailer-Jones}, {Bastian}, {Biermann}, {Evans}, {Eyer}, {Jansen}, {Jordi}, {Klioner}, {Lammers}, {Lindegren}, {Luri}, {Mignard}, {Milligan}, {Panem}, {Poinsignon}, {Pourbaix}, {Randich}, {Sarri}, {Sartoretti}, {Siddiqui}, {Soubiran}, {Valette}, {van Leeuwen}, {Walton}, {Aerts}, {Arenou}, {Cropper}, {Drimmel}, {H{\o}g}, {Katz}, {Lattanzi}, {O'Mullane}, {Grebel}, {Holland}, {Huc}, {Passot}, {Bramante}, {Cacciari}, {Casta{\~n}eda}, {Chaoul}, {Cheek}, {De Angeli}, {Fabricius}, {Guerra}, {Hern{\'a}ndez}, {Jean-Antoine-Piccolo}, {Masana}, {Messineo}, {Mowlavi}, {Nienartowicz}, {Ord{\'o}{\~n}ez-Blanco}, {Panuzzo}, {Portell}, {Richards}, {Riello}, {Seabroke}, {Tanga}, {Th{\'e}venin}, {Torra}, {Els}, {Gracia-Abril}, {Comoretto}, {Garcia-Reinaldos}, {Lock}, {Mercier}, {Altmann}, {Andrae}, {Astraatmadja}, {Bellas-Velidis}, {Benson}, {Berthier}, {Blomme}, {Busso}, {Carry}, {Cellino}, {Clementini},
  {Cowell}, {Creevey}, {Cuypers}, {Davidson}, {De Ridder}, {de Torres}, {Delchambre}, {Dell'Oro}, {Ducourant}, {Fr{\'e}mat}, {Garc{\'\i}a-Torres}, {Gosset}, {Halbwachs}, {Hambly}, {Harrison}, {Hauser}, {Hestroffer}, {Hodgkin}, {Huckle}, {Hutton}, {Jasniewicz}, {Jordan}, {Kontizas}, {Korn}, {Lanzafame}, {Manteiga}, {Moitinho}, {Muinonen}, {Osinde}, {Pancino}, {Pauwels}, {Petit}, {Recio-Blanco}, {Robin}, {Sarro}, {Siopis}, {Smith}, {Smith}, {Sozzetti}, {Thuillot}, {van Reeven}, {Viala}, {Abbas}, {Abreu Aramburu}, {Accart}, {Aguado}, {Allan}, {Allasia}, {Altavilla}, {{\'A}lvarez}, {Alves}, {Anderson}, {Andrei}, {Anglada Varela}, {Antiche}, {Antoja}, {Ant{\'o}n}, {Arcay}, {Atzei}, {Ayache}, {Bach}, {Baker}, {Balaguer-N{\'u}{\~n}ez}, {Barache}, {Barata}, {Barbier}, {Barblan}, {Baroni}, {Barrado y Navascu{\'e}s}, {Barros}, {Barstow}, {Becciani}, {Bellazzini}, {Bellei}, {Bello Garc{\'\i}a}, {Belokurov}, {Bendjoya}, {Berihuete}, {Bianchi}, {Bienaym{\'e}}, {Billebaud}, {Blagorodnova}, {Blanco-Cuaresma}, {Boch},
  {Bombrun}, {Borrachero}, {Bouquillon}, {Bourda}, {Bouy}, {Bragaglia}, {Breddels}, {Brouillet}, {Br{\"u}semeister}, {Bucciarelli}, {Budnik}, {Burgess}, {Burgon}, {Burlacu}, {Busonero}, {Buzzi}, {Caffau}, {Cambras}, {Campbell}, {Cancelliere}, {Cantat-Gaudin}, {Carlucci}, {Carrasco}, {Castellani}, {Charlot}, {Charnas}, {Charvet}, {Chassat}, {Chiavassa}, {Clotet}, {Cocozza}, {Collins}, {Collins}, {Costigan}, {Crifo}, {Cross}, {Crosta}, {Crowley}, {Dafonte}, {Damerdji}, {Dapergolas}, {David}, {David}, {De Cat}, {de Felice}, {de Laverny}, {De Luise}, {De March}, {de Martino}, {de Souza}, {Debosscher}, {del Pozo}, {Delbo}, {Delgado}, {Delgado}, {di Marco}, {Di Matteo}, {Diakite}, {Distefano}, {Dolding}, {Dos Anjos}, {Drazinos}, {Dur{\'a}n}, {Dzigan}, {Ecale}, {Edvardsson}, {Enke}, {Erdmann}, {Escolar}, {Espina}, {Evans}, {Eynard Bontemps}, {Fabre}, {Fabrizio}, {Faigler}, {Falc{\~a}o}, {Farr{\`a}s Casas}, {Faye}, {Federici}, {Fedorets}, {Fern{\'a}ndez-Hern{\'a}ndez}, {Fernique}, {Fienga}, {Figueras}, {Filippi},
  {Findeisen}, {Fonti}, {Fouesneau}, {Fraile}, {Fraser}, {Fuchs}, {Furnell}, {Gai}, {Galleti}, {Galluccio}, {Garabato}, {Garc{\'\i}a-Sedano}, {Gar{\'e}}, {Garofalo}, {Garralda}, {Gavras}, {Gerssen}, {Geyer}, {Gilmore}, {Girona}, {Giuffrida}, {Gomes}, {Gonz{\'a}lez-Marcos}, {Gonz{\'a}lez-N{\'u}{\~n}ez}, {Gonz{\'a}lez-Vidal}, {Granvik}, {Guerrier}, {Guillout}, {Guiraud}, {G{\'u}rpide}, {Guti{\'e}rrez-S{\'a}nchez}, {Guy}, {Haigron}, {Hatzidimitriou}, {Haywood}, {Heiter}, {Helmi}, {Hobbs}, {Hofmann}, {Holl}, {Holland}, {Hunt}, {Hypki}, {Icardi}, {Irwin}, {Jevardat de Fombelle}, {Jofr{\'e}}, {Jonker}, {Jorissen}, {Julbe}, {Karampelas}, {Kochoska}, {Kohley}, {Kolenberg}, {Kontizas}, {Koposov}, {Kordopatis}, {Koubsky}, {Kowalczyk}, {Krone-Martins}, {Kudryashova}, {Kull}, {Bachchan}, {Lacoste-Seris}, {Lanza}, {Lavigne}, {Le Poncin-Lafitte}, {Lebreton}, {Lebzelter}, {Leccia}, {Leclerc}, {Lecoeur-Taibi}, {Lemaitre}, {Lenhardt}, {Leroux}, {Liao}, {Licata}, {Lindstr{\o}m}, {Lister}, {Livanou}, {Lobel}, {L{\"o}ffler},
  {L{\'o}pez}, {Lopez-Lozano}, {Lorenz}, {Loureiro}, {MacDonald}, {Magalh{\~a}es Fernandes}, {Managau}, {Mann}, {Mantelet}, {Marchal}, {Marchant}, {Marconi}, {Marie}, {Marinoni}, {Marrese}, {Marschalk{\'o}}, {Marshall}, {Mart{\'\i}n-Fleitas}, {Martino}, {Mary}, {Matijevi{\v{c}}}, {Mazeh}, {McMillan}, {Messina}, {Mestre}, {Michalik}, {Millar}, {Miranda}, {Molina}, {Molinaro}, {Molinaro}, {Moln{\'a}r}, {Moniez}, {Montegriffo}, {Monteiro}, {Mor}, {Mora}, {Morbidelli}, {Morel}, {Morgenthaler}, {Morley}, {Morris}, {Mulone}, {Muraveva}, {Musella}, {Narbonne}, {Nelemans}, {Nicastro}, {Noval}, {Ord{\'e}novic}, {Ordieres-Mer{\'e}}, {Osborne}, {Pagani}, {Pagano}, {Pailler}, {Palacin}, {Palaversa}, {Parsons}, {Paulsen}, {Pecoraro}, {Pedrosa}, {Pentik{\"a}inen}, {Pereira}, {Pichon}, {Piersimoni}, {Pineau}, {Plachy}, {Plum}, {Poujoulet}, {Pr{\v{s}}a}, {Pulone}, {Ragaini}, {Rago}, {Rambaux}, {Ramos-Lerate}, {Ranalli}, {Rauw}, {Read}, {Regibo}, {Renk}, {Reyl{\'e}}, {Ribeiro}, {Rimoldini}, {Ripepi}, {Riva}, {Rixon},
  {Roelens}, {Romero-G{\'o}mez}, {Rowell}, {Royer}, {Rudolph}, {Ruiz-Dern}, {Sadowski}, {Sagrist{\`a} Sell{\'e}s}, {Sahlmann}, {Salgado}, {Salguero}, {Sarasso}, {Savietto}, {Schnorhk}, {Schultheis}, {Sciacca}, {Segol}, {Segovia}, {Segransan}, {Serpell}, {Shih}, {Smareglia}, {Smart}, {Smith}, {Solano}, {Solitro}, {Sordo}, {Soria Nieto}, {Souchay}, {Spagna}, {Spoto}, {Stampa}, {Steele}, {Steidelm{\"u}ller}, {Stephenson}, {Stoev}, {Suess}, {S{\"u}veges}, {Surdej}, {Szabados}, {Szegedi-Elek}, {Tapiador}, {Taris}, {Tauran}, {Taylor}, {Teixeira}, {Terrett}, {Tingley}, {Trager}, {Turon}, {Ulla}, {Utrilla}, {Valentini}, {van Elteren}, {Van Hemelryck}, {van Leeuwen}, {Varadi}, {Vecchiato}, {Veljanoski}, {Via}, {Vicente}, {Vogt}, {Voss}, {Votruba}, {Voutsinas}, {Walmsley}, {Weiler}, {Weingrill}, {Werner}, {Wevers}, {Whitehead}, {Wyrzykowski}, {Yoldas}, {{\v{Z}}erjal}, {Zucker}, {Zurbach}, {Zwitter}, {Alecu}, {Allen}, {Allende Prieto}, {Amorim}, {Anglada-Escud{\'e}}, {Arsenijevic}, {Azaz}, {Balm}, {Beck}, {Bernstein},
  {Bigot}, {Bijaoui}, {Blasco}, {Bonfigli}, {Bono}, {Boudreault}, {Bressan}, {Brown}, {Brunet}, {Bunclark}, {Buonanno}, {Butkevich}, {Carret}, {Carrion}, {Chemin}, {Ch{\'e}reau}, {Corcione}, {Darmigny}, {de Boer}, {de Teodoro}, {de Zeeuw}, {Delle Luche}, {Domingues}, {Dubath}, {Fodor}, {Fr{\'e}zouls}, {Fries}, {Fustes}, {Fyfe}, {Gallardo}, {Gallegos}, {Gardiol}, {Gebran}, {Gomboc}, {G{\'o}mez}, {Grux}, {Gueguen}, {Heyrovsky}, {Hoar}, {Iannicola}, {Isasi Parache}, {Janotto}, {Joliet}, {Jonckheere}, {Keil}, {Kim}, {Klagyivik}, {Klar}, {Knude}, {Kochukhov}, {Kolka}, {Kos}, {Kutka}, {Lainey}, {LeBouquin}, {Liu}, {Loreggia}, {Makarov}, {Marseille}, {Martayan}, {Martinez-Rubi}, {Massart}, {Meynadier}, {Mignot}, {Munari}, {Nguyen}, {Nordlander}, {Ocvirk}, {O'Flaherty}, {Olias Sanz}, {Ortiz}, {Osorio}, {Oszkiewicz}, {Ouzounis}, {Palmer}, {Park}, {Pasquato}, {Peltzer}, {Peralta}, {P{\'e}turaud}, {Pieniluoma}, {Pigozzi}, {Poels}, {Prat}, {Prod'homme}, {Raison}, {Rebordao}, {Risquez}, {Rocca-Volmerange}, {Rosen},
  {Ruiz-Fuertes}, {Russo}, {Sembay}, {Serraller Vizcaino}, {Short}, {Siebert}, {Silva}, {Sinachopoulos}, {Slezak}, {Soffel}, {Sosnowska}, {Strai{\v{z}}ys}, {ter Linden}, {Terrell}, {Theil}, {Tiede}, {Troisi}, {Tsalmantza}, {Tur}, {Vaccari}, {Vachier}, {Valles}, {Van Hamme}, {Veltz}, {Virtanen}, {Wallut}, {Wichmann}, {Wilkinson}, {Ziaeepour}, \& {Zschocke}}]{gaia1}
{Gaia Collaboration}, {Prusti}, T., {de Bruijne}, J.~H.~J., {et~al.} 2016, \aap, 595, A1

\bibitem[{{Gaia Collaboration} {et~al.}(2023){Gaia Collaboration}, {Vallenari}, {Brown}, {Prusti}, {de Bruijne}, {Arenou}, {Babusiaux}, {Biermann}, {Creevey}, {Ducourant}, {Evans}, {Eyer}, {Guerra}, {Hutton}, {Jordi}, {Klioner}, {Lammers}, {Lindegren}, {Luri}, {Mignard}, {Panem}, {Pourbaix}, {Randich}, {Sartoretti}, {Soubiran}, {Tanga}, {Walton}, {Bailer-Jones}, {Bastian}, {Drimmel}, {Jansen}, {Katz}, {Lattanzi}, {van Leeuwen}, {Bakker}, {Cacciari}, {Casta{\~n}eda}, {De Angeli}, {Fabricius}, {Fouesneau}, {Fr{\'e}mat}, {Galluccio}, {Guerrier}, {Heiter}, {Masana}, {Messineo}, {Mowlavi}, {Nicolas}, {Nienartowicz}, {Pailler}, {Panuzzo}, {Riclet}, {Roux}, {Seabroke}, {Sordo}, {Th{\'e}venin}, {Gracia-Abril}, {Portell}, {Teyssier}, {Altmann}, {Andrae}, {Audard}, {Bellas-Velidis}, {Benson}, {Berthier}, {Blomme}, {Burgess}, {Busonero}, {Busso}, {C{\'a}novas}, {Carry}, {Cellino}, {Cheek}, {Clementini}, {Damerdji}, {Davidson}, {de Teodoro}, {Nu{\~n}ez Campos}, {Delchambre}, {Dell'Oro}, {Esquej},
  {Fern{\'a}ndez-Hern{\'a}ndez}, {Fraile}, {Garabato}, {Garc{\'\i}a-Lario}, {Gosset}, {Haigron}, {Halbwachs}, {Hambly}, {Harrison}, {Hern{\'a}ndez}, {Hestroffer}, {Hodgkin}, {Holl}, {Jan{\ss}en}, {Jevardat de Fombelle}, {Jordan}, {Krone-Martins}, {Lanzafame}, {L{\"o}ffler}, {Marchal}, {Marrese}, {Moitinho}, {Muinonen}, {Osborne}, {Pancino}, {Pauwels}, {Recio-Blanco}, {Reyl{\'e}}, {Riello}, {Rimoldini}, {Roegiers}, {Rybizki}, {Sarro}, {Siopis}, {Smith}, {Sozzetti}, {Utrilla}, {van Leeuwen}, {Abbas}, {{\'A}brah{\'a}m}, {Abreu Aramburu}, {Aerts}, {Aguado}, {Ajaj}, {Aldea-Montero}, {Altavilla}, {{\'A}lvarez}, {Alves}, {Anders}, {Anderson}, {Anglada Varela}, {Antoja}, {Baines}, {Baker}, {Balaguer-N{\'u}{\~n}ez}, {Balbinot}, {Balog}, {Barache}, {Barbato}, {Barros}, {Barstow}, {Bartolom{\'e}}, {Bassilana}, {Bauchet}, {Becciani}, {Bellazzini}, {Berihuete}, {Bernet}, {Bertone}, {Bianchi}, {Binnenfeld}, {Blanco-Cuaresma}, {Blazere}, {Boch}, {Bombrun}, {Bossini}, {Bouquillon}, {Bragaglia}, {Bramante}, {Breedt},
  {Bressan}, {Brouillet}, {Brugaletta}, {Bucciarelli}, {Burlacu}, {Butkevich}, {Buzzi}, {Caffau}, {Cancelliere}, {Cantat-Gaudin}, {Carballo}, {Carlucci}, {Carnerero}, {Carrasco}, {Casamiquela}, {Castellani}, {Castro-Ginard}, {Chaoul}, {Charlot}, {Chemin}, {Chiaramida}, {Chiavassa}, {Chornay}, {Comoretto}, {Contursi}, {Cooper}, {Cornez}, {Cowell}, {Crifo}, {Cropper}, {Crosta}, {Crowley}, {Dafonte}, {Dapergolas}, {David}, {David}, {de Laverny}, {De Luise}, {De March}, {De Ridder}, {de Souza}, {de Torres}, {del Peloso}, {del Pozo}, {Delbo}, {Delgado}, {Delisle}, {Demouchy}, {Dharmawardena}, {Di Matteo}, {Diakite}, {Diener}, {Distefano}, {Dolding}, {Edvardsson}, {Enke}, {Fabre}, {Fabrizio}, {Faigler}, {Fedorets}, {Fernique}, {Fienga}, {Figueras}, {Fournier}, {Fouron}, {Fragkoudi}, {Gai}, {Garcia-Gutierrez}, {Garcia-Reinaldos}, {Garc{\'\i}a-Torres}, {Garofalo}, {Gavel}, {Gavras}, {Gerlach}, {Geyer}, {Giacobbe}, {Gilmore}, {Girona}, {Giuffrida}, {Gomel}, {Gomez}, {Gonz{\'a}lez-N{\'u}{\~n}ez},
  {Gonz{\'a}lez-Santamar{\'\i}a}, {Gonz{\'a}lez-Vidal}, {Granvik}, {Guillout}, {Guiraud}, {Guti{\'e}rrez-S{\'a}nchez}, {Guy}, {Hatzidimitriou}, {Hauser}, {Haywood}, {Helmer}, {Helmi}, {Sarmiento}, {Hidalgo}, {Hilger}, {H{\l}adczuk}, {Hobbs}, {Holland}, {Huckle}, {Jardine}, {Jasniewicz}, {Jean-Antoine Piccolo}, {Jim{\'e}nez-Arranz}, {Jorissen}, {Juaristi Campillo}, {Julbe}, {Karbevska}, {Kervella}, {Khanna}, {Kontizas}, {Kordopatis}, {Korn}, {K{\'o}sp{\'a}l}, {Kostrzewa-Rutkowska}, {Kruszy{\'n}ska}, {Kun}, {Laizeau}, {Lambert}, {Lanza}, {Lasne}, {Le Campion}, {Lebreton}, {Lebzelter}, {Leccia}, {Leclerc}, {Lecoeur-Taibi}, {Liao}, {Licata}, {Lindstr{\o}m}, {Lister}, {Livanou}, {Lobel}, {Lorca}, {Loup}, {Madrero Pardo}, {Magdaleno Romeo}, {Managau}, {Mann}, {Manteiga}, {Marchant}, {Marconi}, {Marcos}, {Marcos Santos}, {Mar{\'\i}n Pina}, {Marinoni}, {Marocco}, {Marshall}, {Martin Polo}, {Mart{\'\i}n-Fleitas}, {Marton}, {Mary}, {Masip}, {Massari}, {Mastrobuono-Battisti}, {Mazeh}, {McMillan}, {Messina}, {Michalik},
  {Millar}, {Mints}, {Molina}, {Molinaro}, {Moln{\'a}r}, {Monari}, {Mongui{\'o}}, {Montegriffo}, {Montero}, {Mor}, {Mora}, {Morbidelli}, {Morel}, {Morris}, {Muraveva}, {Murphy}, {Musella}, {Nagy}, {Noval}, {Oca{\~n}a}, {Ogden}, {Ordenovic}, {Osinde}, {Pagani}, {Pagano}, {Palaversa}, {Palicio}, {Pallas-Quintela}, {Panahi}, {Payne-Wardenaar}, {Pe{\~n}alosa Esteller}, {Penttil{\"a}}, {Pichon}, {Piersimoni}, {Pineau}, {Plachy}, {Plum}, {Poggio}, {Pr{\v{s}}a}, {Pulone}, {Racero}, {Ragaini}, {Rainer}, {Raiteri}, {Rambaux}, {Ramos}, {Ramos-Lerate}, {Re Fiorentin}, {Regibo}, {Richards}, {Rios Diaz}, {Ripepi}, {Riva}, {Rix}, {Rixon}, {Robichon}, {Robin}, {Robin}, {Roelens}, {Rogues}, {Rohrbasser}, {Romero-G{\'o}mez}, {Rowell}, {Royer}, {Ruz Mieres}, {Rybicki}, {Sadowski}, {S{\'a}ez N{\'u}{\~n}ez}, {Sagrist{\`a} Sell{\'e}s}, {Sahlmann}, {Salguero}, {Samaras}, {Sanchez Gimenez}, {Sanna}, {Santove{\~n}a}, {Sarasso}, {Schultheis}, {Sciacca}, {Segol}, {Segovia}, {S{\'e}gransan}, {Semeux}, {Shahaf}, {Siddiqui}, {Siebert},
  {Siltala}, {Silvelo}, {Slezak}, {Slezak}, {Smart}, {Snaith}, {Solano}, {Solitro}, {Souami}, {Souchay}, {Spagna}, {Spina}, {Spoto}, {Steele}, {Steidelm{\"u}ller}, {Stephenson}, {S{\"u}veges}, {Surdej}, {Szabados}, {Szegedi-Elek}, {Taris}, {Taylor}, {Teixeira}, {Tolomei}, {Tonello}, {Torra}, {Torra}, {Torralba Elipe}, {Trabucchi}, {Tsounis}, {Turon}, {Ulla}, {Unger}, {Vaillant}, {van Dillen}, {van Reeven}, {Vanel}, {Vecchiato}, {Viala}, {Vicente}, {Voutsinas}, {Weiler}, {Wevers}, {Wyrzykowski}, {Yoldas}, {Yvard}, {Zhao}, {Zorec}, {Zucker}, \& {Zwitter}}]{gaia2}
{Gaia Collaboration}, {Vallenari}, A., {Brown}, A.~G.~A., {et~al.} 2023, \aap, 674, A1

\bibitem[{{Gail} {et~al.}(1984){Gail}, {Keller}, \& {Sedlmayr}}]{Gail_1984}
{Gail}, H.~P., {Keller}, R., \& {Sedlmayr}, E. 1984, \aap, 133, 320

\bibitem[{{Gail} \& {Sedlmayr}(1999)}]{Gail_1999}
{Gail}, H.~P. \& {Sedlmayr}, E. 1999, \aap, 347, 594

\bibitem[{{Gail} {et~al.}(2020){Gail}, {Tamanai}, {Pucci}, \& {Dohmen}}]{Gail_2020}
{Gail}, H.-P., {Tamanai}, A., {Pucci}, A., \& {Dohmen}, R. 2020, \aap, 644, A139

\bibitem[{{Goldman} {et~al.}(2017){Goldman}, {van Loon}, {Zijlstra}, {Green}, {Wood}, {Nanni}, {Imai}, {Whitelock}, {Matsuura}, {Groenewegen}, \& {G{\'o}mez}}]{Goldman_2017}
{Goldman}, S.~R., {van Loon}, J.~T., {Zijlstra}, A.~A., {et~al.} 2017, \mnras, 465, 403

\bibitem[{{Gonz{\'a}lez-Fern{\'a}ndez} {et~al.}(2015){Gonz{\'a}lez-Fern{\'a}ndez}, {Dorda}, {Negueruela}, \& {Marco}}]{Gonzalez_2015}
{Gonz{\'a}lez-Fern{\'a}ndez}, C., {Dorda}, R., {Negueruela}, I., \& {Marco}, A. 2015, \aap, 578, A3

\bibitem[{{Gordon} {et~al.}(2018){Gordon}, {Humphreys}, {Jones}, {Shenoy}, {Gehrz}, {Helton}, {Marengo}, {Hinz}, \& {Hoffmann}}]{Gordon_2018}
{Gordon}, M.~S., {Humphreys}, R.~M., {Jones}, T.~J., {et~al.} 2018, \aj, 155, 212

\bibitem[{{Groenewegen} \& {Sloan}(2018)}]{Groenewegen_2018}
{Groenewegen}, M.~A.~T. \& {Sloan}, G.~C. 2018, \aap, 609, A114

\bibitem[{{Groenewegen} {et~al.}(2009){Groenewegen}, {Sloan}, {Soszy{\'n}ski}, {et~al.}}]{Groenewegen_2009}
{Groenewegen}, M.~A.~T., {Sloan}, G.~C., {Soszy{\'n}ski}, I., {et~al.} 2009, \aap, 506, 1277

\bibitem[{{Gustafsson} {et~al.}(2008){Gustafsson}, {Edvardsson}, {Eriksson}, {J{\o}rgensen}, {Nordlund}, \& {Plez}}]{Gustafsson_2008}
{Gustafsson}, B., {Edvardsson}, B., {Eriksson}, K., {et~al.} 2008, \aap, 486, 951

\bibitem[{{Haubois} {et~al.}(2019){Haubois}, {Norris}, {Tuthill}, {Pinte}, {Kervella}, {Girard}, {Kostogryz}, {Berdyugina}, {Perrin}, {Lacour}, {Chiavassa}, \& {Ridgway}}]{Haubois_2019}
{Haubois}, X., {Norris}, B., {Tuthill}, P.~G., {et~al.} 2019, \aap, 628, A101

\bibitem[{{Heger} {et~al.}(2003){Heger}, {Fryer}, {Woosley}, {et~al.}}]{Heger_2003}
{Heger}, A., {Fryer}, C.~L., {Woosley}, S.~E., {et~al.} 2003, \apj, 591, 288

\bibitem[{{H{\"o}fner} {et~al.}(2003){H{\"o}fner}, {Gautschy-Loidl}, {Aringer}, \& {J{\o}rgensen}}]{Hofner_2003}
{H{\"o}fner}, S., {Gautschy-Loidl}, R., {Aringer}, B., \& {J{\o}rgensen}, U.~G. 2003, \aap, 399, 589

\bibitem[{{Houck} {et~al.}(2004){Houck}, {Roellig}, {Van Cleve}, {Forrest}, {Herter}, {Lawrence}, {Matthews}, {Reitsema}, {Soifer}, {Watson}, {Weedman}, {Huisjen}, {Troeltzsch}, {Barry}, {Bernard-Salas}, {Blacken}, {Brandl}, {Charmandaris}, {Devost}, {Gull}, {Hall}, {Henderson}, {Higdon}, {Pirger}, {Schoenwald}, {Sloan}, {Uchida}, {Appleton}, {Armus}, {Burgdorf}, {Fajardo-Acosta}, {Grillmair}, {Ingalls}, {Morris}, \& {Teplitz}}]{Houck_2004}
{Houck}, J.~R., {Roellig}, T.~L., {Van Cleve}, J., {et~al.} 2004, in Society of Photo-Optical Instrumentation Engineers (SPIE) Conference Series, Vol. 5487, Optical, Infrared, and Millimeter Space Telescopes, ed. J.~C. {Mather}, 62--76

\bibitem[{{Humphreys} {et~al.}(2020){Humphreys}, {Helmel}, {Jones}, \& {Gordon}}]{Humphreys_2020}
{Humphreys}, R.~M., {Helmel}, G., {Jones}, T.~J., \& {Gordon}, M.~S. 2020, \aj, 160, 145

\bibitem[{{Ivezic} \& {Elitzur}(1997)}]{Ivezic_1997}
{Ivezic}, Z. \& {Elitzur}, M. 1997, \mnras, 287, 799

\bibitem[{{Ivezić} \& {Elitzur}(1995)}]{Ivezic_1995}
{Ivezić}, Z. \& {Elitzur}, M. 1995, \apj, 445, 415

\bibitem[{{Jim{\'e}nez-Arranz} {et~al.}(2023){Jim{\'e}nez-Arranz}, {Romero-G{\'o}mez}, {Luri}, {McMillan}, {Antoja}, {Chemin}, {Roca-F{\`a}brega}, {Masana}, \& {Muros}}]{Gaia_LMC}
{Jim{\'e}nez-Arranz}, {\'O}., {Romero-G{\'o}mez}, M., {Luri}, X., {et~al.} 2023, \aap, 669, A91

\bibitem[{{Josselin} \& {Plez}(2007)}]{Josselin_2007}
{Josselin}, E. \& {Plez}, B. 2007, \aap, 469, 671

\bibitem[{{Kato} {et~al.}(2012){Kato}, {Ita}, {Onaka}, {Tanab{\'e}}, {Shimonishi}, {Sakon}, {Kaneda}, {Kawamura}, {Wada}, {Usui}, {Koo}, {Matsuura}, \& {Takahashi}}]{akari}
{Kato}, D., {Ita}, Y., {Onaka}, T., {et~al.} 2012, \aj, 144, 179

\bibitem[{{Kato} {et~al.}(2007){Kato}, {Nagashima}, {Nagayama}, {Kurita}, {Koerwer}, {Kawai}, {Yamamuro}, {Zenno}, {Nishiyama}, {Baba}, {Kadowaki}, {Haba}, {Hatano}, {Shimizu}, {Nishimura}, {Nagata}, {Sato}, {Murai}, {Kawazu}, {Nakajima}, {Nakaya}, {Kandori}, {Kusakabe}, {Ishihara}, {Kaneyasu}, {Hashimoto}, {Tamura}, {Tanab{\'e}}, {Ita}, {Matsunaga}, {Nakada}, {Sugitani}, {Wakamatsu}, {Glass}, {Feast}, {Menzies}, {Whitelock}, {Fourie}, {Stoffels}, {Evans}, \& {Hasegawa}}]{IRSF}
{Kato}, D., {Nagashima}, C., {Nagayama}, T., {et~al.} 2007, \pasj, 59, 615

\bibitem[{{Kee} {et~al.}(2021){Kee}, {Sundqvist}, {Decin}, {et~al.}}]{Kee_2021}
{Kee}, N.~D., {Sundqvist}, J.~O., {Decin}, L., {et~al.} 2021, \aap, 646, A180

\bibitem[{{Kippenhahn} {et~al.}(2013){Kippenhahn}, {Weigert}, \& {Weiss}}]{Kippenhahn_2013}
{Kippenhahn}, R., {Weigert}, A., \& {Weiss}, A. 2013, {Stellar Structure and Evolution}

\bibitem[{{Kochanek} {et~al.}(2012){Kochanek}, {Khan}, \& {Dai}}]{Kochanek_2012}
{Kochanek}, C.~S., {Khan}, R., \& {Dai}, X. 2012, \apj, 759, 20

\bibitem[{{Kraus}(2019)}]{Kraus_2019}
{Kraus}, M. 2019, Galaxies, 7, 83

\bibitem[{{Levesque} {et~al.}(2009){Levesque}, {Massey}, {Plez}, \& {Olsen}}]{Levesque_2009}
{Levesque}, E.~M., {Massey}, P., {Plez}, B., \& {Olsen}, K. A.~G. 2009, \aj, 137, 4744

\bibitem[{{Liu} {et~al.}(2017){Liu}, {Jiang}, {Li}, \& {Gao}}]{Liu_2017}
{Liu}, J., {Jiang}, B.~W., {Li}, A., \& {Gao}, J. 2017, \mnras, 466, 1963

\bibitem[{{Mainzer} {et~al.}(2014){Mainzer}, {Bauer}, {Cutri}, {Grav}, {Masiero}, {Beck}, {Clarkson}, {Conrow}, {Dailey}, {Eisenhardt}, {Fabinsky}, {Fajardo-Acosta}, {Fowler}, {Gelino}, {Grillmair}, {Heinrichsen}, {Kendall}, {Kirkpatrick}, {Liu}, {Masci}, {McCallon}, {Nugent}, {Papin}, {Rice}, {Royer}, {Ryan}, {Sevilla}, {Sonnett}, {Stevenson}, {Thompson}, {Wheelock}, {Wiemer}, {Wittman}, {Wright}, \& {Yan}}]{Mainzer_2014}
{Mainzer}, A., {Bauer}, J., {Cutri}, R.~M., {et~al.} 2014, \apj, 792, 30

\bibitem[{{Mainzer} {et~al.}(2011){Mainzer}, {Bauer}, {Grav}, {Masiero}, {Cutri}, {Dailey}, {Eisenhardt}, {McMillan}, {Wright}, {Walker}, {Jedicke}, {Spahr}, {Tholen}, {Alles}, {Beck}, {Brandenburg}, {Conrow}, {Evans}, {Fowler}, {Jarrett}, {Marsh}, {Masci}, {McCallon}, {Wheelock}, {Wittman}, {Wyatt}, {DeBaun}, {Elliott}, {Elsbury}, {Gautier}, {Gomillion}, {Leisawitz}, {Maleszewski}, {Micheli}, \& {Wilkins}}]{Mainzer_2011}
{Mainzer}, A., {Bauer}, J., {Grav}, T., {et~al.} 2011, \apj, 731, 53

\bibitem[{{Marshall} {et~al.}(2004){Marshall}, {van Loon}, {Matsuura}, {Wood}, {Zijlstra}, \& {Whitelock}}]{Marshall_2004}
{Marshall}, J.~R., {van Loon}, J.~T., {Matsuura}, M., {et~al.} 2004, \mnras, 355, 1348

\bibitem[{{Massey}(2002)}]{Massey_2002}
{Massey}, P. 2002, \apjs, 141, 81

\bibitem[{{Massey} {et~al.}(2023){Massey}, {Neugent}, {Ekstr{\"o}m}, {et~al.}}]{Massey_2023}
{Massey}, P., {Neugent}, K.~F., {Ekstr{\"o}m}, S., {et~al.} 2023, \apj, 942, 69

\bibitem[{{Massey} {et~al.}(2007){Massey}, {Olsen}, {Hodge}, {Jacoby}, {McNeill}, {Smith}, \& {Strong}}]{Massey_2007}
{Massey}, P., {Olsen}, K.~A.~G., {Hodge}, P.~W., {et~al.} 2007, \aj, 133, 2393

\bibitem[{{Mathis} {et~al.}(1977){Mathis}, {Rumpl}, \& {Nordsieck}}]{MRN}
{Mathis}, J.~S., {Rumpl}, W., \& {Nordsieck}, K.~H. 1977, \apj, 217, 425

\bibitem[{{Mauron} \& {Josselin}(2011)}]{Mauron_2011}
{Mauron}, N. \& {Josselin}, E. 2011, \aap, 526, A156

\bibitem[{{McDonald} {et~al.}(2011){McDonald}, {Boyer}, {van Loon}, \& {Zijlstra}}]{McDonald_2011}
{McDonald}, I., {Boyer}, M.~L., {van Loon}, J.~T., \& {Zijlstra}, A.~A. 2011, \apj, 730, 71

\bibitem[{{Meynet} {et~al.}(2006){Meynet}, {Ekstr{\"o}m}, \& {Maeder}}]{Meynet_2006}
{Meynet}, G., {Ekstr{\"o}m}, S., \& {Maeder}, A. 2006, \aap, 447, 623

\bibitem[{{Meynet} \& {Maeder}(2003)}]{Meynet_2003}
{Meynet}, G. \& {Maeder}, A. 2003, \aap, 404, 975

\bibitem[{{Montarg{\`e}s} {et~al.}(2021){Montarg{\`e}s}, {Cannon}, {Lagadec}, {de Koter}, {Kervella}, {Sanchez-Bermudez}, {Paladini}, {Cantalloube}, {Decin}, {Scicluna}, {Kravchenko}, {Dupree}, {Ridgway}, {Wittkowski}, {Anugu}, {Norris}, {Rau}, {Perrin}, {Chiavassa}, {Kraus}, {Monnier}, {Millour}, {Le Bouquin}, {Haubois}, {Lopez}, {Stee}, \& {Danchi}}]{Montarges_2021}
{Montarg{\`e}s}, M., {Cannon}, E., {Lagadec}, E., {et~al.} 2021, \nat, 594, 365

\bibitem[{{Neilson} \& {Lester}(2008)}]{Neilson_2008}
{Neilson}, H.~R. \& {Lester}, J.~B. 2008, \apj, 684, 569

\bibitem[{{Neugent} {et~al.}(2019){Neugent}, {Levesque}, {Massey}, {et~al.}}]{Neugent_2019}
{Neugent}, K.~F., {Levesque}, E.~M., {Massey}, P., {et~al.} 2019, \apj, 875, 124

\bibitem[{{Neugent} {et~al.}(2020){Neugent}, {Levesque}, {Massey}, {et~al.}}]{Neugent_2020}
{Neugent}, K.~F., {Levesque}, E.~M., {Massey}, P., {et~al.} 2020, \apj, 900, 118

\bibitem[{{Neugent} {et~al.}(2012){Neugent}, {Massey}, {Skiff}, \& {Meynet}}]{Neugent_2012}
{Neugent}, K.~F., {Massey}, P., {Skiff}, B., \& {Meynet}, G. 2012, \apj, 749, 177

\bibitem[{{Nidever} {et~al.}(2021){Nidever}, {Dey}, {Fasbender}, {Juneau}, {Meisner}, {Wishart}, {Scott}, {Matt}, {Nikutta}, \& {Pucha}}]{Nidever_2021}
{Nidever}, D.~L., {Dey}, A., {Fasbender}, K., {et~al.} 2021, \aj, 161, 192

\bibitem[{{Ohnaka} {et~al.}(2008){Ohnaka}, {Driebe}, {Hofmann}, {Weigelt}, \& {Wittkowski}}]{Ohnaka_2008}
{Ohnaka}, K., {Driebe}, T., {Hofmann}, K.~H., {Weigelt}, G., \& {Wittkowski}, M. 2008, \aap, 484, 371

\bibitem[{{Ohnaka} {et~al.}(2017){Ohnaka}, {Weigelt}, \& {Hofmann}}]{Ohnaka_2017}
{Ohnaka}, K., {Weigelt}, G., \& {Hofmann}, K.~H. 2017, \nat, 548, 310

\bibitem[{{Onken} {et~al.}(2019){Onken}, {Wolf}, {Bessell}, {Chang}, {Da Costa}, {Luvaul}, {Mackey}, {Schmidt}, \& {Shao}}]{Onken_2019}
{Onken}, C.~A., {Wolf}, C., {Bessell}, M.~S., {et~al.} 2019, \pasa, 36, e033

\bibitem[{{Patrick} {et~al.}(2022){Patrick}, {Thilker}, {Lennon}, {Bianchi}, {Schootemeijer}, {Dorda}, {Langer}, \& {Negueruela}}]{Patrick_2022}
{Patrick}, L.~R., {Thilker}, D., {Lennon}, D.~J., {et~al.} 2022, \mnras, 513, 5847

\bibitem[{{Pietrzy{\'n}ski} {et~al.}(2019){Pietrzy{\'n}ski}, {Graczyk}, {Gallenne}, {Gieren}, {Thompson}, {Pilecki}, {Karczmarek}, {G{\'o}rski}, {Suchomska}, {Taormina}, {Zgirski}, {Wielg{\'o}rski}, {Ko{\l}aczkowski}, {Konorski}, {Villanova}, {Nardetto}, {Kervella}, {Bresolin}, {Kudritzki}, {Storm}, {Smolec}, \& {Narloch}}]{lmc_distance}
{Pietrzy{\'n}ski}, G., {Graczyk}, D., {Gallenne}, A., {et~al.} 2019, \nat, 567, 200

\bibitem[{{Reid} \& {Mould}(1985)}]{Reid_1985}
{Reid}, N. \& {Mould}, J. 1985, \apj, 299, 236

\bibitem[{{Ren} {et~al.}(2021){Ren}, {Jiang}, {Yang}, {Wang}, \& {Ren}}]{Ren_2021}
{Ren}, Y., {Jiang}, B., {Yang}, M., {Wang}, T., \& {Ren}, T. 2021, \apj, 923, 232

\bibitem[{{Richards} {et~al.}(1998){Richards}, {Yates}, \& {Cohen}}]{Richards&Yates_1998}
{Richards}, A.~M.~S., {Yates}, J.~A., \& {Cohen}, R.~J. 1998, \mnras, 299, 319

\bibitem[{{Riebel} {et~al.}(2012){Riebel}, {Srinivasan}, {Sargent}, \& {Meixner}}]{Riebel_2012}
{Riebel}, D., {Srinivasan}, S., {Sargent}, B., \& {Meixner}, M. 2012, \apj, 753, 71

\bibitem[{{Sargent} {et~al.}(2011){Sargent}, {Srinivasan}, \& {Meixner}}]{Sargent_2011}
{Sargent}, B.~A., {Srinivasan}, S., \& {Meixner}, M. 2011, \apj, 728, 93

\bibitem[{{Scicluna} {et~al.}(2015){Scicluna}, {Siebenmorgen}, {Wesson}, {Blommaert}, {Kasper}, {Voshchinnikov}, \& {Wolf}}]{Scicluna_2015}
{Scicluna}, P., {Siebenmorgen}, R., {Wesson}, R., {et~al.} 2015, \aap, 584, L10

\bibitem[{{Shenoy} {et~al.}(2016){Shenoy}, {Humphreys}, {Jones}, {Marengo}, {Gehrz}, {Helton}, {Hoffmann}, {Skemer}, \& {Hinz}}]{Shenoy_2016}
{Shenoy}, D., {Humphreys}, R.~M., {Jones}, T.~J., {et~al.} 2016, \aj, 151, 51

\bibitem[{{Smartt}(2009)}]{Smartt_2009}
{Smartt}, S.~J. 2009, \araa, 47, 63

\bibitem[{{Smith} {et~al.}(2001){Smith}, {Humphreys}, {Davidson}, {Gehrz}, {Schuster}, \& {Krautter}}]{Smith_2001}
{Smith}, N., {Humphreys}, R.~M., {Davidson}, K., {et~al.} 2001, \aj, 121, 1111

\bibitem[{{Soszy{\'n}ski} {et~al.}(2009){Soszy{\'n}ski}, {Udalski}, {Szyma{\'n}ski}, {Kubiak}, {Pietrzy{\'n}ski}, {Wyrzykowski}, {Szewczyk}, {Ulaczyk}, \& {Poleski}}]{Soszynski_2009}
{Soszy{\'n}ski}, I., {Udalski}, A., {Szyma{\'n}ski}, M.~K., {et~al.} 2009, \actaa, 59, 239

\bibitem[{{Ueta} \& {Meixner}(2003)}]{Ueta_2003}
{Ueta}, T. \& {Meixner}, M. 2003, \apj, 586, 1338

\bibitem[{{Ulaczyk} {et~al.}(2012){Ulaczyk}, {Szyma{\'n}ski}, {Udalski}, {Kubiak}, {Pietrzy{\'n}ski}, {Soszy{\'n}ski}, {Wyrzykowski}, {Poleski}, {Gieren}, {Walker}, \& {Garcia-Varela}}]{ogleIII}
{Ulaczyk}, K., {Szyma{\'n}ski}, M.~K., {Udalski}, A., {et~al.} 2012, \actaa, 62, 247

\bibitem[{{van Loon}(2000)}]{vLoon_2000}
{van Loon}, J.~T. 2000, \aap, 354, 125

\bibitem[{{van Loon} {et~al.}(2005){van Loon}, {Cioni}, {Zijlstra}, \& {Loup}}]{vLoon_2005}
{van Loon}, J.~T., {Cioni}, M. R.~L., {Zijlstra}, A.~A., \& {Loup}, C. 2005, \aap, 438, 273

\bibitem[{{van Loon} {et~al.}(2001){van Loon}, {Zijlstra}, {Bujarrabal}, \& {Nyman}}]{vLoon_2001}
{van Loon}, J.~T., {Zijlstra}, A.~A., {Bujarrabal}, V., \& {Nyman}, L.~{\r{A}}. 2001, \aap, 368, 950

\bibitem[{{Verhoelst} {et~al.}(2009){Verhoelst}, {Waters}, {Verhoeff}, {Dijkstra}, {van Winckel}, {Pel}, \& {Peletier}}]{Verhoelst_2009}
{Verhoelst}, T., {Waters}, L.~B.~F.~M., {Verhoeff}, A., {et~al.} 2009, \aap, 503, 837

\bibitem[{{Vink} {et~al.}(2001){Vink}, {de Koter}, \& {Lamers}}]{Vink_2001}
{Vink}, J.~S., {de Koter}, A., \& {Lamers}, H.~J.~G.~L.~M. 2001, \aap, 369, 574

\bibitem[{{Vink} \& {Sabhahit}(2023)}]{Vink_2023}
{Vink}, J.~S. \& {Sabhahit}, G.~N. 2023, \aap, 678, L3

\bibitem[{{Wang} \& {Chen}(2019)}]{Wang_2019}
{Wang}, S. \& {Chen}, X. 2019, \apj, 877, 116

\bibitem[{{Wang} {et~al.}(2021){Wang}, {Jiang}, {Ren}, {Yang}, \& {Li}}]{Wang_2021}
{Wang}, T., {Jiang}, B., {Ren}, Y., {Yang}, M., \& {Li}, J. 2021, \apj, 912, 112

\bibitem[{{Wen} {et~al.}(2024){Wen}, {Gao}, {Yang}, {Chen}, {Ren}, {Wang}, \& {Jiang}}]{Wen_2024}
{Wen}, J., {Gao}, J., {Yang}, M., {et~al.} 2024, \aj, 167, 51

\bibitem[{{Westerlund} {et~al.}(1981){Westerlund}, {Olander}, \& {Hedin}}]{Westerlund_1981}
{Westerlund}, B.~E., {Olander}, N., \& {Hedin}, B. 1981, \aaps, 43, 267

\bibitem[{{Wood} {et~al.}(1983){Wood}, {Bessell}, \& {Fox}}]{Wood_1983}
{Wood}, P.~R., {Bessell}, M.~S., \& {Fox}, M.~W. 1983, \apj, 272, 99

\bibitem[{{Yang} {et~al.}(2021){Yang}, {Bonanos}, {Jiang}, {Gao}, {Gavras}, {Maravelias}, {Wang}, {Chen}, {Lam}, {Ren}, {Tramper}, \& {Spetsieri}}]{Yang_2021lmc}
{Yang}, M., {Bonanos}, A.~Z., {Jiang}, B., {et~al.} 2021, \aap, 646, A141

\bibitem[{{Yang} {et~al.}(2023){Yang}, {Bonanos}, {Jiang}, {Zapartas}, {Gao}, {Ren}, {Lam}, {Wang}, {Maravelias}, {Gavras}, {Wang}, {Chen}, {Tramper}, {de Wit}, {Chen}, {Wen}, {Liu}, {Tian}, {Antoniadis}, \& {Luo}}]{Yang_2023}
{Yang}, M., {Bonanos}, A.~Z., {Jiang}, B., {et~al.} 2023, \aap, 676, A84

\bibitem[{{Yang} {et~al.}(2019){Yang}, {Bonanos}, {Jiang}, {Gao}, {Gavras}, {Maravelias}, {Ren}, {Wang}, {Xue}, {Tramper}, {Spetsieri}, \& {Pouliasis}}]{Yang_2019}
{Yang}, M., {Bonanos}, A.~Z., {Jiang}, B.-W., {et~al.} 2019, \aap, 629, A91

\bibitem[{{Yang} {et~al.}(2018){Yang}, {Bonanos}, {Jiang}, {Gao}, {Xue}, {Wang}, {Lam}, {Spetsieri}, {Ren}, \& {Gavras}}]{Yang_2018}
{Yang}, M., {Bonanos}, A.~Z., {Jiang}, B.-W., {et~al.} 2018, \aap, 616, A175

\bibitem[{{Yoon} \& {Cantiello}(2010)}]{Yoon_2010}
{Yoon}, S.-C. \& {Cantiello}, M. 2010, \apjl, 717, L62

\bibitem[{{Yoon} {et~al.}(2017){Yoon}, {Dessart}, \& {Clocchiatti}}]{Yoon_2017}
{Yoon}, S.-C., {Dessart}, L., \& {Clocchiatti}, A. 2017, \apj, 840, 10

\bibitem[{{Zapartas} {et~al.}(2019){Zapartas}, {de Mink}, {Justham}, {Smith}, {de Koter}, {Renzo}, {Arcavi}, {Farmer}, {G{\"o}tberg}, \& {Toonen}}]{Zapartas_2019}
{Zapartas}, E., {de Mink}, S.~E., {Justham}, S., {et~al.} 2019, \aap, 631, A5

\bibitem[{{Zapartas} {et~al.}(2017){Zapartas}, {de Mink}, {Van Dyk}, {Fox}, {Smith}, {Bostroem}, {de Koter}, {Filippenko}, {Izzard}, {Kelly}, {Neijssel}, {Renzo}, \& {Ryder}}]{Zapartas_2017}
{Zapartas}, E., {de Mink}, S.~E., {Van Dyk}, S.~D., {et~al.} 2017, \apj, 842, 125

\end{thebibliography}

\begin{appendix} 

\section{Dust-enshrouded RSGs} \label{app:woh}
\begin{figure}[h]
    \includegraphics[width=\linewidth]{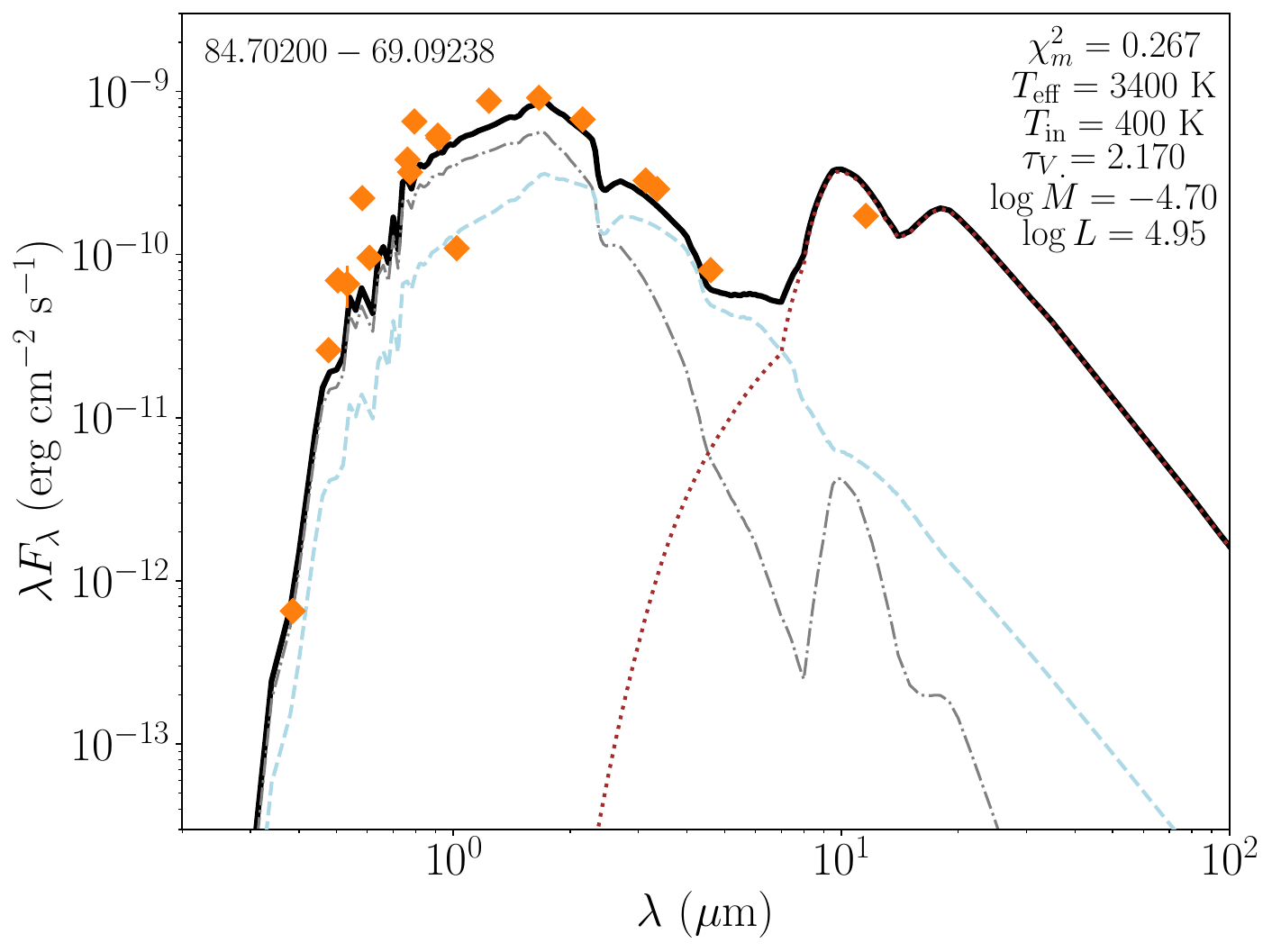}
    \caption{SED of W61 7-6 and its derived parameters. The lines and symbols represent the same as in Fig.~\ref{fig:sed}.}
    \label{fig:deRSG}    
\end{figure}

We report two dust-enshrouded RSGs in our sample, W61 7-8 and WOH G64. We found that W61 7-8 (ID 2107) has $K_s-W3=3.68$ mag, satisfying the criteria of \citet{Beasor_2022} to constitute a dust-enshrouded RSG. We present its SED, best-fit model, and derived parameters in Fig.~\ref{fig:deRSG}.

\begin{figure}[h]
    \includegraphics[width=\linewidth]{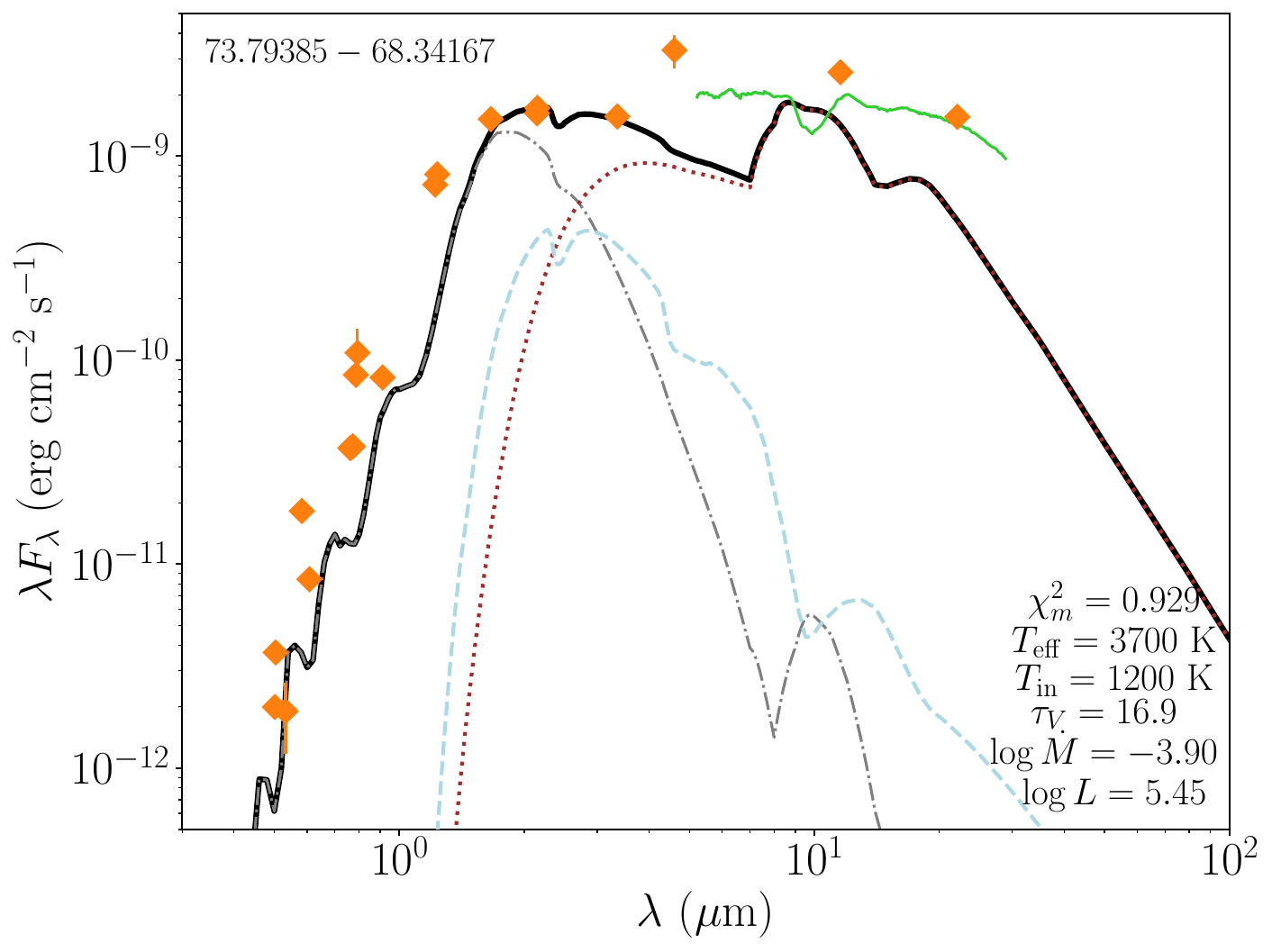}
    \includegraphics[width=\linewidth]{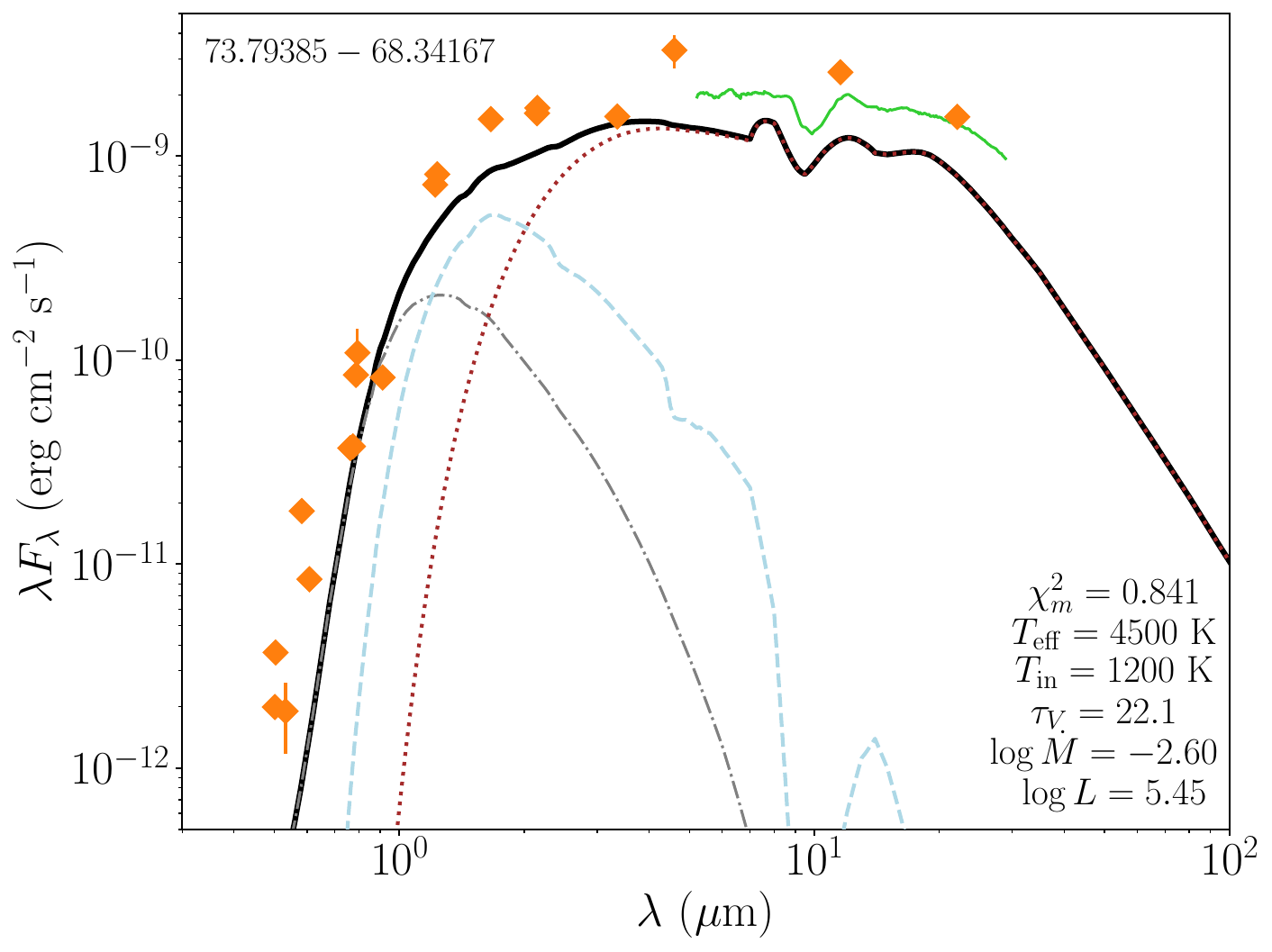}
    \caption{Two \texttt{DUSTY} SED models of WOH G64 with different grain size assumptions. The lines and symbols represent the same as in Fig.~\ref{fig:sed}. \textit{Top:} Grain size with $a=0.3 \ \mu$m. \textit{Bottom:} MRN distribution with $0.01\leq a \leq 0.15 \ \mu$m.}
    \label{fig:woh}    
\end{figure}

WOH G64 is a well-studied dust-enshrouded RSG with an optically thick dust shell \citep{Ohnaka_2008, Levesque_2009, Beasor_2022}. It lacks a spherically symmetric shell, but has a dust torus. This was modelled by \cite{Ohnaka_2008}, who inferred a luminosity of $\log{(L/L_\odot)}=5.45$. Integrating the SED would overestimate its luminosity because of the asymmetric shell. The grid of optical depths we used in this case was in the range of [14, 26] $\mu$m. In Fig.~\ref{fig:woh}, we present the best-fit models of WOH G64 with two different assumptions for the grain size: $a=0.3 \ \mu$m following \cite{Beasor_2022} and the MRN distribution with $0.01\leq a \leq 0.15 \ \mu$m, as inferred by \cite{Ohnaka_2008}. In the first case, the result is similar to that of \cite{Beasor_2022}, with a difference in the mass-loss rate that could be attributed to the different photometry and fitting method, $\log{\dot{M}/(M_{\odot} \ \mathrm{yr}^{-1})}=-3.9^{+0.31}_{-0.17}$. In the second case, we found a better fit that also reproduces the shape of the IRS spectrum. This means that a smaller grain size is better, and it also results in a higher mass-loss rate, $\log{\dot{M}/(M_{\odot} \ \mathrm{yr}^{-1})}=-2.6^{+0.31}_{-0.17}$. We should mention that the WISE photometry is not consistent with the rest of the SED because the source is brighter than the instrument saturation limits. However, it passes the quality criteria we defined in Section~\ref{sec:sample}.

\clearpage


\section{Mass-loss rate relations for radiatively driven winds and different grain size distribution} \label{app:rdw_a}

\begin{figure}[h]
    \centering
    \includegraphics[width=\columnwidth]{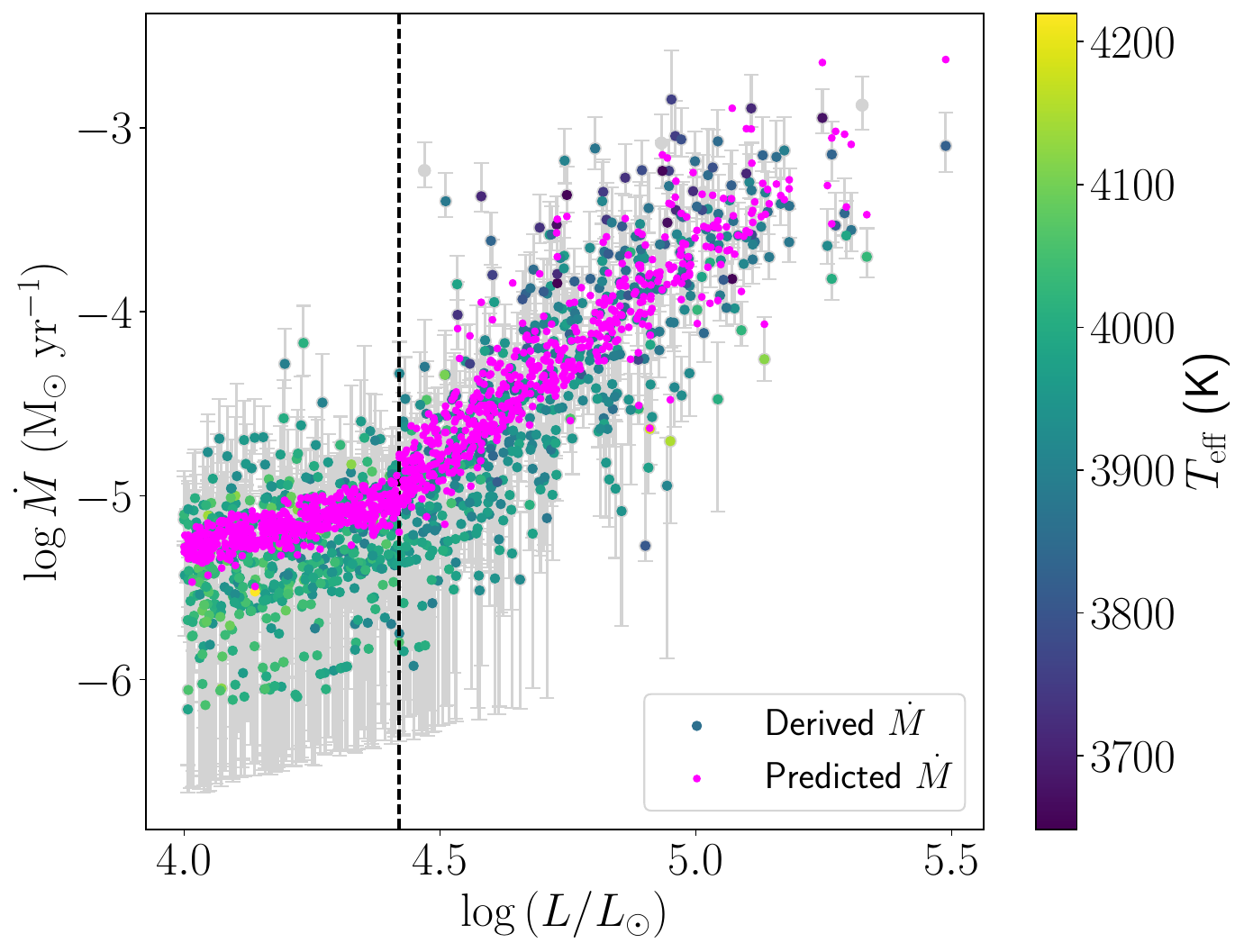}
    \caption{Same as Fig. \ref{fig:Mdot_L_fit} (left), but for RDW.}
    \label{fig:Mdot_L_rdw}
\end{figure}

\begin{figure}[h]
    \centering
    \includegraphics[width=\columnwidth]{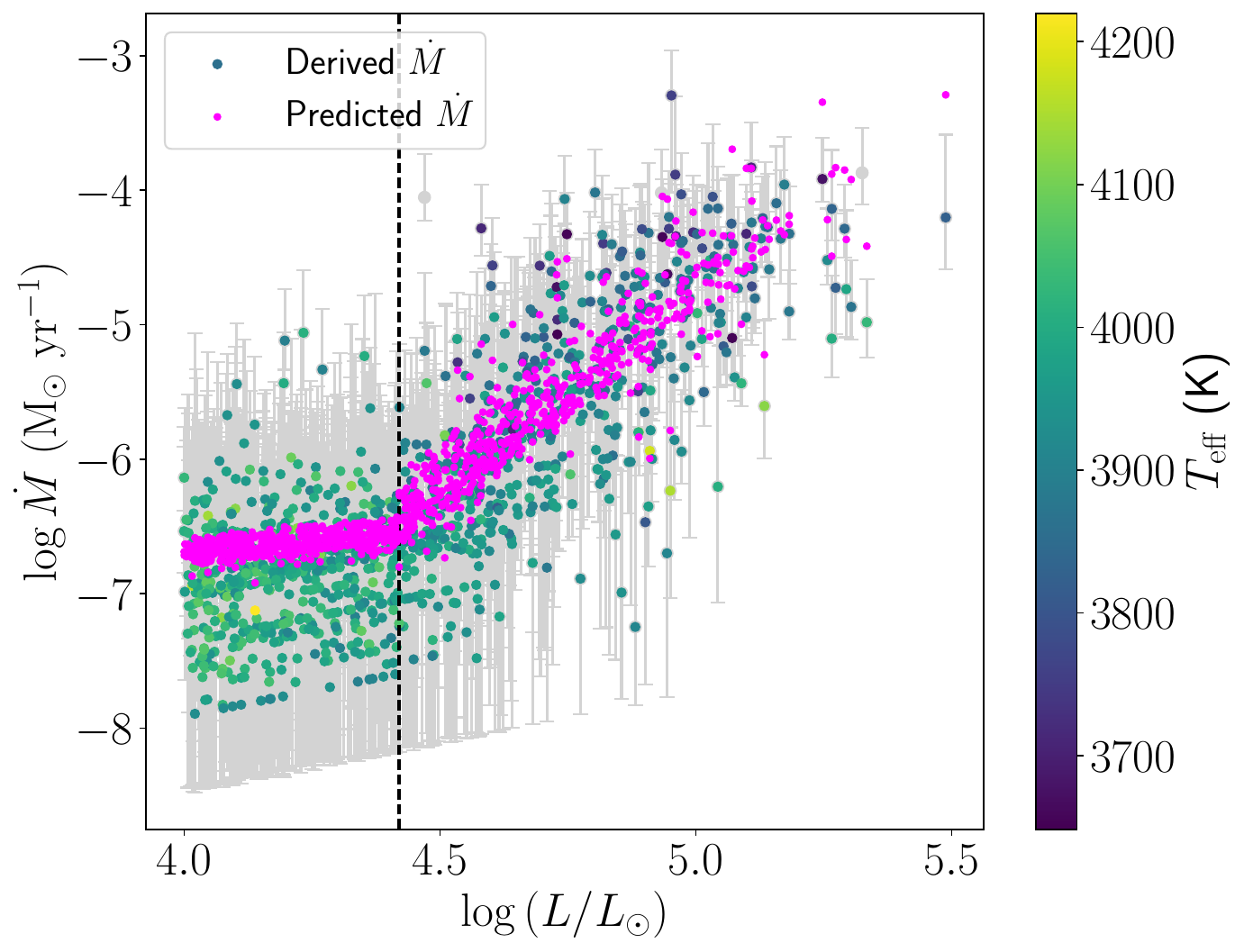}
    \caption{Same as Fig. \ref{fig:Mdot_L_fit} (left), but for an MRN grain size distribution in the range [0.01, 1] $\mu$m.}
    \label{fig:Mdot_ss_mrn001}
\end{figure}

\begin{figure}[h]
    \centering
    \includegraphics[width=\columnwidth]{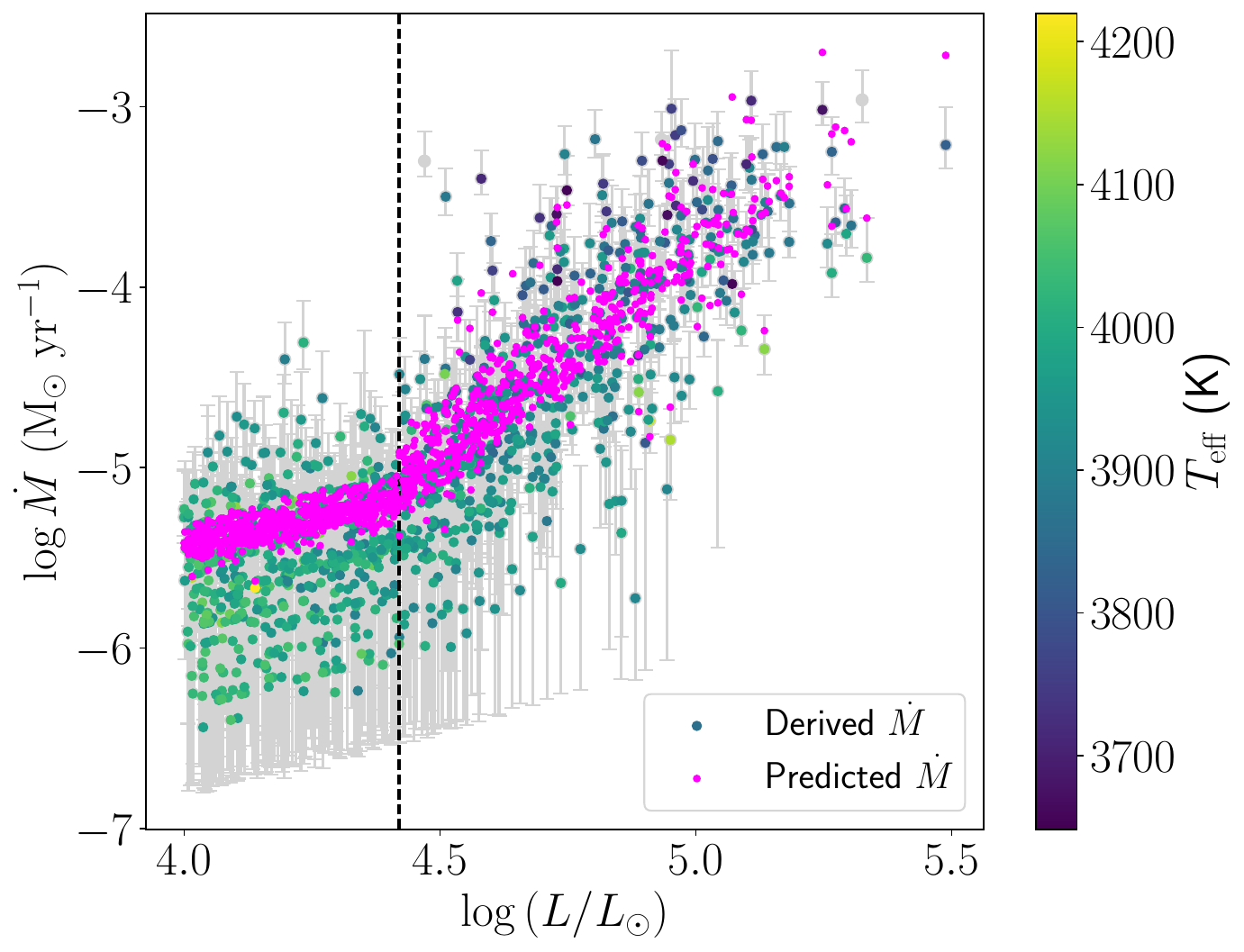}
    \caption{Same as Fig. \ref{fig:Mdot_L_fit} (left), but for RDW and an MRN grain size distribution in the range [0.01, 1] $\mu$m.}
    \label{fig:Mdot_rdw_mrn001}
\end{figure}

The best-fit parameters of each wind and grain size assumption are presented in \autoref{tab:coef_rdw}. We show the derived $\dot{M}(L, T_\mathrm{eff})$ relations in Fig.~\ref{fig:Mdot_L_rdw}, \ref{fig:Mdot_ss_mrn001}, and \ref{fig:Mdot_rdw_mrn001} for RDW and an MRN grain size distribution [0.1, 1] $\mu$m, steady-state winds and  MRN grain size distribution [0.01, 1] $\mu$m, and RDW and MRN grain size distribution [0.01, 1] $\mu$m, respectively.

\begin{table}[h]
    \centering
    \caption{Best-fit parameters of Eq. (\ref{eq:Mdot}) for RDW, steady-state winds with a smaller grain size and only single RSGs.}
    \renewcommand{\arraystretch}{1.2}
    \begin{tabular}{c | c c c}
        \hline\hline
        $\log{L/L_\odot}$ & $c_1$   & $c_2$  & $c_3$ \\
         \hline
        \rowcolor{lightgray!50} \multicolumn{4}{c}{RDW, MRN [0.1, 1]} \\
        $<4.4$ & $0.43\pm0.15$  & $-14.42\pm3.42$  & $-6.97\pm0.62$ \\
        $\gtrsim4.4$ & $1.68\pm0.09$  & $-26.88\pm1.65$  & $-12.46\pm0.41$ \\
       \hline
        \rowcolor{lightgray!50} \multicolumn{4}{c}{RDW, MRN [0.01, 1]} \\
        $<4.4$ & $0.48\pm0.07$  & $-13.48\pm1.66$  & $-7.15\pm0.31$ \\
        $\gtrsim4.4$ & $1.65\pm0.04$  & $-25.32\pm0.79$  & $-12.31\pm0.2$ \\
       \hline
       \rowcolor{lightgray!50} \multicolumn{4}{c}{Steady-state, MRN [0.01, 1]} \\
        $<4.4$ & $0.21\pm0.16$  & $-10.32\pm3.27$  & $-7.5\pm0.67$ \\
        $\gtrsim4.4$ & $2.46\pm0.09$  & $-31.78\pm1.62$  & $-17.47\pm0.43$ \\
        \hline
       \rowcolor{lightgray!50} \multicolumn{4}{c}{Steady-state, MRN [0.1, 1], "single" RSGs} \\
        $<4.4$ & $0.16\pm0.24$  & $-24.93\pm6.07$  & $-8.8\pm1$ \\
        $\gtrsim4.4$ & $2.77\pm0.15$  & $-33.08\pm2.97$  & $-20.39\pm0.7$ \\
        \hline
    \end{tabular}
    \
    \label{tab:coef_rdw}
\end{table}
\clearpage


\section{Binary candidate RSGs}  \label{app:binaries}

\begin{figure}[h]
    \centering
    \includegraphics[width=\columnwidth]{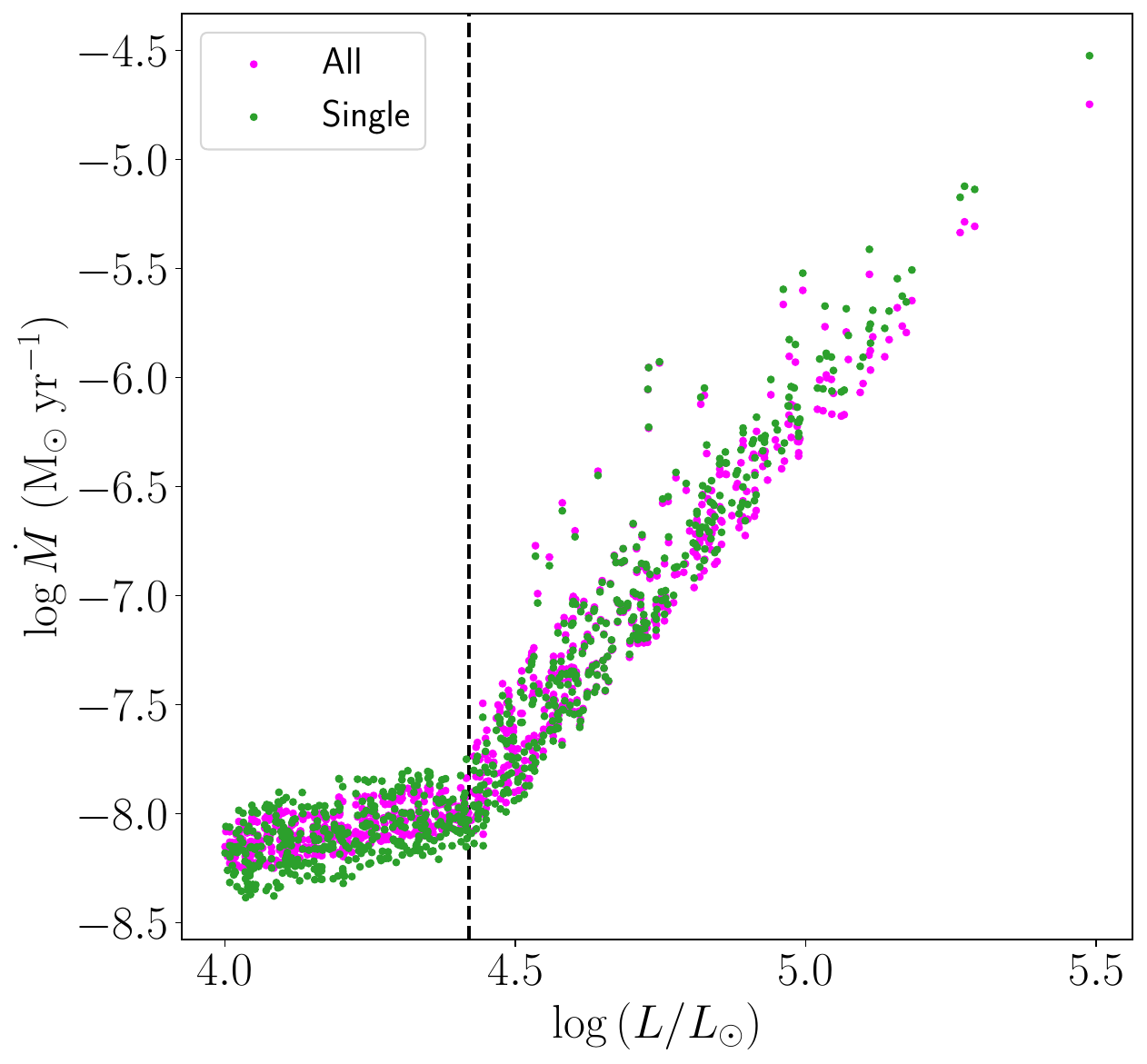}
    \caption{Derived $\dot{M}(L, T_\mathrm{eff})$ from the complete RSG sample (magenta), as in Fig. \ref{fig:Mdot_L_fit}, and from the single RSG sample with $U-B>0.7$ mag (green). Both relations were applied to the single RSG sample for a one-to-one comparison.}
    \label{fig:mdot_bin_presc}
\end{figure}

We removed the binary candidates and fitted Eq. (\ref{eq:Mdot}) in the new sample. We demonstrate the new relation of the probable single RSGs with green points in Fig. \ref{fig:mdot_bin_presc}. The magenta points show the same relation as in Fig. \ref{fig:Mdot_L_fit}. The two relations differ marginally and mainly in the change of the slope at high $L$ because a significant fraction of binary candidates occupies the lower $\dot{M}$ part at high $L$, as shown in Fig. \ref{fig:mdot_binaries}. If these sources are indeed binaries, then we estimated the mass-loss rates under uncertain assumptions, such as the spherical symmetry of the dust shell and the luminosity of the star. We present the best-fit parameters of Eq. (\ref{eq:Mdot}) for the probably single RSGs in \autoref{tab:coef_rdw}.

\end{appendix}

\end{document}